\documentclass{mn2e}

\usepackage[dvips]{graphics}

\newif\ifAMStwofonts

\title{Evolutionary SED diagnostics of starburst galaxies: 
signature of bimodality}

\author[T. Takagi, N. Arimoto and H. Hanami]
 {T. Takagi$^{1, 2, 3}$\thanks{E-mail: t.takagi@ic.ac.uk}
 , N. Arimoto$^{4, 5}$ and H. Hanami$^6$ \\ 
$^1$ Blackett Laboratory, Imperial College, 
Prince Consort Road, London SW7 2BZ, UK \\
$^2$The Institute of Space and Astronautical Science, 3-1-1 Yoshinodai, 
Sagamihara, Kanagawa
229-8510, Japan \\
$^3$Department of Physics, Rikkyo University, 3-34-1 Nishi-Ikebukuro,
Toshima-ku, Tokyo 171-8501, Japan\\ 
$^4$Institute of Astronomy, School of Science, University of Tokyo,
2-21-1 Osawa, Mitaka, Tokyo 181-0015, Japan \\
$^5$National Astronomical Observatory,
2-21-1 Osawa, Mitaka, Tokyo 181-8588, Japan \\
$^6$Physics Section, Faculty of Humanities and Social Sciences, Iwate
University, Morioka, 020-8550, Japan
}

\date{}

\begin{document}

\maketitle

\begin{abstract}
We construct an evolutionary spectral energy distribution 
(SED) model of a starburst region, from the ultraviolet 
to submillimetre wavelengths. 
This model allows us to derive the star formation rate, 
optical depth by dust and apparent effective radius of starburst 
regions at various wavelengths; as a result, the intrinsic 
surface brightness of starburst regions can be derived.   
Using this SED model, we analyse 16 UV-selected starburst 
galaxies and 10 ultraluminous infrared galaxies. 
The derived star formation rates and optical depths are compared with 
emission line measurements and found to be consistent. 
The derived apparent effective 
radii are also consistent with observations. 
From the SED analysis, we find a bimodal property 
of the star formation rate with the optical depth and the 
compactness of stellar distributions. 
While mild starbursts have a limiting intrinsic surface brightness
$ L_{bol} r_e^{-2} \simeq 10^{12}$ L$_{\odot} $kpc$^{-2}$,
intense starbursts tend to be more
heavily obscured and concentrated within a characteristic
scale of $ r_e \simeq 0.3 $ kpc. 
We suggest that the mild starbursts can be triggered by a 
self-gravitating disc instability in which feedback is effective 
in the shallow gravitational potential. 
On the other hand, the intense starbursts can be induced via 
an external dynamical perturbation like galaxy merging, in which
feedback is less effective due to the deep gravitational potential
attained by the large gas concentration within the central starburst
region. 
\end{abstract}


\section{Introduction}

Starbursts are an intensive mode of star formation in galaxies
(Storchi-Bergmann, Calzetti \& Kinney 1994; Kennicutt 1998).
Ultraluminous infrared galaxies (ULIRGs) with $L_{\rm IR} \ge
10^{12}$ L$_{\odot}$ are probably the most active and luminous 
starburst galaxies.
Indeed recent ISO observations suggest that many ULIRGs are
powered by starbursts with little contribution from AGN 
(Genzel et al. 1998; Lutz et al. 1998).  On the other hand,
starburst galaxies of lower activity are HII galaxies and blue compact dwarf
galaxies, known as UV-selected starburst galaxies (UVSBGs;
Kinney et al. 1993; Storch-Bergmann et al. 1994;
Calzetti, Kinney \& Storchi-Bergmann 1994; 
McQuade, Calzetti \& Kinney 1995; Gordon, Calzetti \& Witt 1997;
Takagi, Arimoto \& Vansevi\v cius 1999).

Recently, an upper limit of `bolometric surface brightness'
was given by Meurer et al. (1997) for
a sample of starburst galaxies observed in the rest frame 1) UV, 
2) far infrared (FIR) and H$\alpha$,
and 3) 21 cm radio continuum emission. This limit seems to be
physically associated with
an instability in gaseous discs (e.g. Toomre 1964; Quirk 1972)
as Kennicutt (1989) had already suggested for the star
formation in normal disc galaxies.
In their study, the effective radii are determined from 
observations at the different wavelengths for each galaxy.
However,  if a galaxy is optically thick 
the observed effective radius strongly depends on the
degree of extinction, since the
observed light comes out solely from the region in which the optical
depth is less than unity.
Moreover, many ULIRGs have multiple starburst regions
(e.g. Scoville et al. 2000; Surace \& Sanders 2000),
while most of UVSBGs show only one major central starburst region.
These effects suggest that the panchromatic
intensity limit on starbursts proposed by Meurer et al. (1997)
may not be appropriate for ULIRGs.

To determine the intensity limit of starbursts with various levels of 
activity ranging from UVSBGs to ULIRGs, it is necessary to estimate
both star formation rates (SFRs) and effective radii in a unified manner,
whatever the degree of extinction.
While typical $V$-band optical depths in UVSBGs are
$\tau_V \sim 0.3 - 2$ (Takagi, Arimoto, \& Vansevi\v cius 1999;
Storchi-Bergmann et al. 1994),
those in ULIRGs reach $\tau_V \sim 5-50$ (Genzel et al. 1998).
Such a wide variation in $\tau_V$ causes a serious problem in
deriving SFRs, since
none of the currently used SFR indicators, such as
H$\alpha$ (or other Balmer lines), UV continuum, or FIR
luminosities, are commonly applicable 
for starbursts ranging from UVSBGs to ULIRGs.
The FIR luminosity is not a good measure of bolometric luminosity
when starburst regions are 
optically thin, while H$\alpha$ luminosity is difficult to use if
the mean surface gas density becomes larger
than 50 M$_\odot$ pc$^{-2}$, corresponding to
$\tau_V \sim 2$, causing a significant extinction
at H$\alpha$ (Kennicutt 1998).
What is worse, by using Br$\gamma$ and FIR luminosities,
Kennicutt (1998) has shown 
that the FIR luminosities give systematically higher SFRs
(by a factor of $\sim 2$) than the Br$\gamma$s.
Sullivan et al. (2000) compared the H$\alpha$- and UV-derived
SFRs of nearby starbursts to find that starburst galaxies are
typically over-luminous in the UV for a given H$\alpha$ luminosity. 
The effect is strongest in the less luminous galaxies.

When comparing ULIRGs and UVSBGs, 
the bolometric surface brightness should be derived for
each individual starburst region in ULIRGs. 
The wealth of ground-based low-resolution imaging 
surveys have provided little information on the stars in the starburst region
near the confusion limit $ < 1''$.  Even the
{\em Hubble Space Telescope} ({\it HST}) 
with its higher resolution $ <0.1''$ fails to
go deep enough into the starburst regions, which are obscured
by dust at optical wavelengths.  The Near-Infrared Camera and
Multi-Object Spectrometer (NICMOS) on {\it HST} can
observe the obscured starburst regions more directly at
near infrared (NIR) wavelengths where the effects of dust extinction
are reduced significantly compared to visual wavelengths.
Scoville et al. (2000)
studied the morphologies of 24 ULIRGs with NICMOS and found that
light profiles of nine of them were well fit by an $r^{1/4}$ law, 
rather than an exponential profile.
The apparent effective radii of ULIRGs tend to be more compact than
the extent of gas observed with high-resolution
imaging of the CO emission (Bryant $\&$
Scoville 1999).  This suggests that the distribution of stars is 
more concentrated than that of dust and gas in the starburst regions.
Thus, the dimming of light due to the surrounding dust
can be so large that the real distribution of stars
deviates from the apparent distribution of the light even the longest
observable wavelength with NICMOS of 2.2 $\mu$m.

It is thus crucial to establish a measure of both SFRs and
the spatial distribution of stars 
which can be applied to starbursts of any optical depth.
Clearly, a new recipe to derive both SFRs and the stellar distributions
is required to investigate the physical properties of
starbursts comprehensively.

In this paper, we will use spectral energy distributions (SEDs)
from the far ultraviolet (FUV) to submillimetre (submm) 
to derive SFRs of starburst regions. The SED
from the UV to submm can represent the bolometric luminosity,
irrespective of the dust extinction. We construct an `evolutionary'
SED model for starburst regions from a simple but realistic point of
view.  We show that our SED model can explain a wide
variety of SEDs from optically thin starbursts (UVSBGs)
to optically thick ones (ULIRGs).
It is important to note that our SED model can derive not only the
SFR, but also the other starburst properties like the optical depth.
Moreover, we can derive the apparent effective radius of 
starbursts at various wavelengths for a given geometry, 
since we properly take into account the radiative transfer in 
a dusty medium. 
Therefore, the evolutionary stages of starburst galaxies can be
investigated in a unified manner for a whole range of activity, which
can then be compared with the observations of the SED and the effective
radius, directly. The systematic estimation
of the SFRs and the effective radius can give an answer to the question:
is the panchromatic limit of starburst
intensity reported by Meurer et al. (1997) also valid for ULIRGs?

This paper is organized as follows.
In Section 2, we describe our evolutionary SED model for starburst
galaxies. In Section 3, we summarize the model properties.  In Section 4,
we apply our model to a sample of UVSBGs and ULIRGs in the local Universe 
and describe the resulting properties of nearby starburst galaxies derived
from our SED model fitting. Sections 5 and 6 give discussions and
conclusions, respectively.

\section{Evolutionary SED model for Starburst Regions}


\subsection {Star formation and chemical evolution}

We consider a model for the chemical enrichment
of a starburst region into which gas is supplied continuously.
When a large amount of gas is supplied into the central region of a
galaxy, a starburst is triggered. Since the details of the gas
supplying mechanism are yet to be investigated, it would not be outrageous to
suppose that star formation and chemical evolution in the starburst regions
are approximately described by an infall model of chemical evolution 
(Arimoto, et al. 1992). 
Ubiquitous galactic outflows from the starbursts have been 
observed with the X-ray emission from the hot gas driving the flow, 
optical line emission produced by the warm gas, and the interstellar 
absorption lines 
(e.g. Heckman et al. 2000; Lehnert $\&$ Heckman 1996). 
Heckman et al. (2000) suggested that most of the outflow gas consists of 
ambient material which has been 'mass-loaded' into the hot gas. 
Therefore, we assume that the amount of gas 
in the outflow from a starburst region has negligible impact on the 
chemical evolution as a whole. This is true if the mass loss 
rate from a starburst region is less than 10\% of the SFR. 
Under this simple evolutionary picture, a starburst is characterized by 
the rates of gas infall and star formation.

The time variation of gas mass $M_g(t)$, total stellar mass $M_*(t)$, 
and gas metallicity 
$Z_g(t)$ are given by the equations:
\begin{eqnarray}
\frac{dM_g(t)}{dt} &=& -\psi (t) + E(t)+ \xi_i (t), \\
\frac{dM_*(t)}{dt} &=& \psi (t) - E(t), \\
\frac{d(Z_g M_g)}{dt} &=& -\psi(t) Z_g(t) + E_Z(t)+\xi_i(t) Z_i,
\end{eqnarray}
where $E(t)$, $E_Z(t)$, and $Z_i$ are the ejection rates of the gas and
the metals from dying stars, and the metallicity of
the supplied gas, respectively. We set the initial condition as 
$M_g(0)=M_*(0)=0$ for all the calculations in this paper. 
Although some amount of 
gas and stars unrelated to the starbursts are initially expected 
in the starburst region, we assume that the amount of initial gas and 
stars are negligible in both the chemical and photometric evolution of 
starburst region. For $Z_i$, the star formation history before the 
starburst event is important. We will explicitly note the adopted 
values of $Z_i$ later. We assume $Z_i$ is constant during 
the starburst event. 

The SFR $\psi (t)$ is given by 
\begin{equation}
\psi (t)= \frac{1}{t_*} M_g (t),
\end{equation}
where $t_*$ and $M_g (t)$ are the star formation time-scale
and the gas mass in the starburst region, respectively.
The gas supply rate is given by:
\begin{equation}
\xi_i(t)= \frac{M_T}{t_i} \exp \left( -\frac{t}{t_i} \right),
\end{equation}
where $M_T$ and $t_i$ are the initial gas mass in
the reservoir surrounding the starburst
region and the time-scale of gas supply, respectively (Arimoto et al. 1992).
Physically, the time-scales $t_i$ and $t_*$ can be expressed
in terms of the dynamical time, sound-crossing time, and
cooling time, depending on what triggers the starburst.
It is however difficult to ascertain the
characteristic time-scale of each starburst from observations.
Thus, we hereafter analyse the simplest case, in which a
starburst is characterized by only one evolutionary time-scale $t_0$, i.e., 
we adopt $ t_0\equiv t_i=t_*$.

For all cases in this paper, we adopt the Salpeter initial mass function
(IMF) with the lower and upper mass limit of
0.1 M$_\odot$ and 60 M$_\odot$, respectively.
Equations (1)-(3) are numerically solved by using an evolutionary population
synthesis code of Kodama \& Arimoto (1997). The adopted stellar libraries 
and evolutionary tracks are the same as those in the original code. 
Details of nucleosynthesis prescription is given there.


The effect of dust on the SEDs is significant and predominant in starburst regions. 
In this paper, we adopt a simple model in which the dust-to-metal ratio 
$\delta_0$ is constant; i.e., $M_D(t)=\delta_0 Z_g(t) M_g(t)$. 
As described below (Section 2.3), we use three types of dust model for 
the Milky Way (MW), Large and Small Magellanic Clouds (LMC; SMC). 
The adopted values of $\delta_0$ for MW, LMC and SMC 
are 0.40, 0.55, 0.75, respectively (Takagi 2001). 
Starbursts could be the most ideal case for constant $\delta (t)$,
since 1) only type II supernovae contribute to the supernova
rate and 2) the gas fraction in molecular clouds is large
(see Dwek 1998).

\subsection{Model geometry of starburst region}

We consider a starburst region in which stars and dust are distributed within 
a radius $r_t$. We introduce a mass-radius relation for the starburst region;
\begin{equation}
\frac{r_t}{1 \mathrm{kpc}} =
\Theta \left( \frac{M_*}{10^9 \mathrm{M}_{\odot}} \right)^\gamma\;,
\end{equation}
where $\Theta$ is a compactness factor which expresses the matter 
concentration; the mean density becomes higher for smaller $\Theta$. 
Starburst galaxies are characterized by a large surface brightness, 
which are roughly constant 
(Armus, Heckman \& Miley 1990; Meurer et al. 1995, 1997).
In the adopted mass-radius relation, 
$\gamma =1/2$ gives the constant surface brightness for constant $\Theta$. 
Therefore, 
we adopt $\gamma =1/2$ throughout this paper. 
A variation in the surface brightness can be expressed by 
the different values of $\Theta$ in our model. 
Note that not only the surface brightness 
but also the SED feature is preserved for different values of $M_*$ 
if $\gamma =1/2$,
since the source function within the starburst is conserved. Therefore, 
multiple systems with the total stellar mass of $M_*$, in which each component has the same surface brightness, have 
the same SED as that of a unit system with a stellar mass of $M_*$.

We assume
that the stellar density distribution $\rho (r)$ is given by a generalized King profile;
\begin{equation}
\rho(r)=\rho_0 \left[ 1+\left( \frac{\displaystyle r}
{\displaystyle r_{c}}\right) ^{2} \right] ^{-\beta} , \label{King}
\end{equation}
where $\rho_{0}$ is the stellar density at the centre of starburst region
and $r_{c}$ is a core radius of stars. 
We adopt the stellar density distribution of 
typical elliptical galaxies; i.e., the concentration parameter 
$c\equiv \log (r_t/r_c)=2.2$, and $\beta =\frac{3}{2}$ (Combes et al. 1995). 
Recently, Scoville et al. (2000) and James et al. (1999) presented $K$-band 
images of starburst galaxies, and demonstrated that luminosity profiles 
of starbursts are well represented by the $r^{1/4}$ profile 
which is quite similar to those of elliptical galaxies. 
However, in order to derive the stellar density distribution in starbursts, 
it is important to eliminate the effects 
of dust extinction, since the effect of radiative transfer
can easily alter the luminosity profile from the original one. 
As we will show later, this effect is not negligible even in the $K$-band 
in ULIRGs (see Section 3.3). Therefore, the true density distribution 
of stars in starburst galaxies is difficult to determine directly, 
and we therefore adopt the typical density profile of elliptical galaxies 
as a first guess. This topic is further discussed in Section 3.3 and 4.4.  


Following Takagi et al. (1999), we assume that dust is distributed 
homogeneously within a radius $r_t$ of the starburst region. 
It is plausible that the dust distributes more diffusely than the stars, 
due to feedback from supernovae. 
When the light from centrally concentrated stars dominates the SED, 
i.e., optically thin case, this geometry results in a similar SED to
the case of shell geometry, which is suggested for 
UVSBGs by Gordon et al. (1997), Meurer, Heckman \& Calzetti (1999), 
Witt \& Gordon (2000). 
However, in the optically thick case, like 
ULIRGs, the shell geometry results in the spectral cut-off around NIR
(see the results by Rowan-Robinson \& Efstathiou 1993), 
and therefore UV-NIR SED of ULIRGs cannot be reproduced with this 
geometry without invoking the other components, like underlying 
stellar populations and/or AGNs. As we show in Section 4.2, our model 
can essentially reproduce UV-NIR SEDs of ULIRGs only with 
starburst stellar populations. These SED fitting results are 
confirmed with emission line measurements and the observed effective 
radii in Section 4.3 and 4.4, respectively. 
Therefore, we believe that the adopted geometry is suitable to 
approximate the real geometry of starburst regions with various 
optical depths.

\subsection {Dust model}

The dust model is adopted from Takagi (2001) who successfully 
reproduced the extinction curves observed in MW, LMC and SMC, 
as well as the spectrum of Galactic cirrus. The difference among the MW, LMC and
SMC extinction curves is attributed to the variation 
of the ratio of carbonaceous dust (graphite and PAHs) to silicate grains. 

The extinction curve is given by the cross-section per hydrogen:
\begin{equation}
\sigma^\mathrm{H}_{0\lambda} = \frac{1}{n_\mathrm{H}}
\sum_k \int \sigma^e_{\lambda ,k}(a)
\frac{dn_k}{da} da,
\end{equation}
where $n_\mathrm{H}$ is the number density of hydrogen,
$\sigma^e_{\lambda ,k}(a)$ is the extinction-cross
section of dust particle with the dust constituent $k$ and size $a$.
The size distribution of each dust constituent $dn_k/da$ is taken from
Takagi (2001). For a constant dust-to-metal ratio,
the extinction curve at starburst age $t$ is given by:
\begin{equation}
\sigma _{\lambda }^\mathrm{H}(t)=\frac{ Z^\mathrm{H}(t)}
{Z^\mathrm{H}_0}\sigma _{0\lambda}^\mathrm{H},
\end{equation}
where $Z^\mathrm{H}(t)$ is the gas metallicity with respect to
hydrogen mass (as opposed to the total gas mass) and $Z_0^\mathrm{H}$ is
the metallicity of the ISM under the same definition.
We adopt $Z_0^\mathrm{H}=0.024$, 0.011, 0.0034 for the MW, LMC and SMC
extinction curves, respectively (Pagel 1997; Lequeux et al. 1979).
The time variation of optical depth $\tau_V (t)$
is given by $n_{\mathrm{H}}(t) \sigma^{\mathrm{H}}_\lambda (t) r_t(t)$.
Note that $r_t(t)$ is defined by the time
dependent total stellar mass $M_*(t)$.
According to the adopted mass-radius relation, 
$\tau_V \propto n_{\mathrm{H}} r_t\propto M_H r_t^{-2} 
\propto f_{H} M_T r_t^{-2} \propto f_{H}
M_T (f_{star} M_T)^{-1} \propto f_{H} f_{star}^{-1}$ where $M_H$ and 
$f_H$ are the total mass of hydrogen 
and the mass fraction of hydrogen, respectively; 
therefore, $\tau_V$ does not depend on the value of $M_T$.

\begin{figure}
  \resizebox{8cm}{!}{\includegraphics{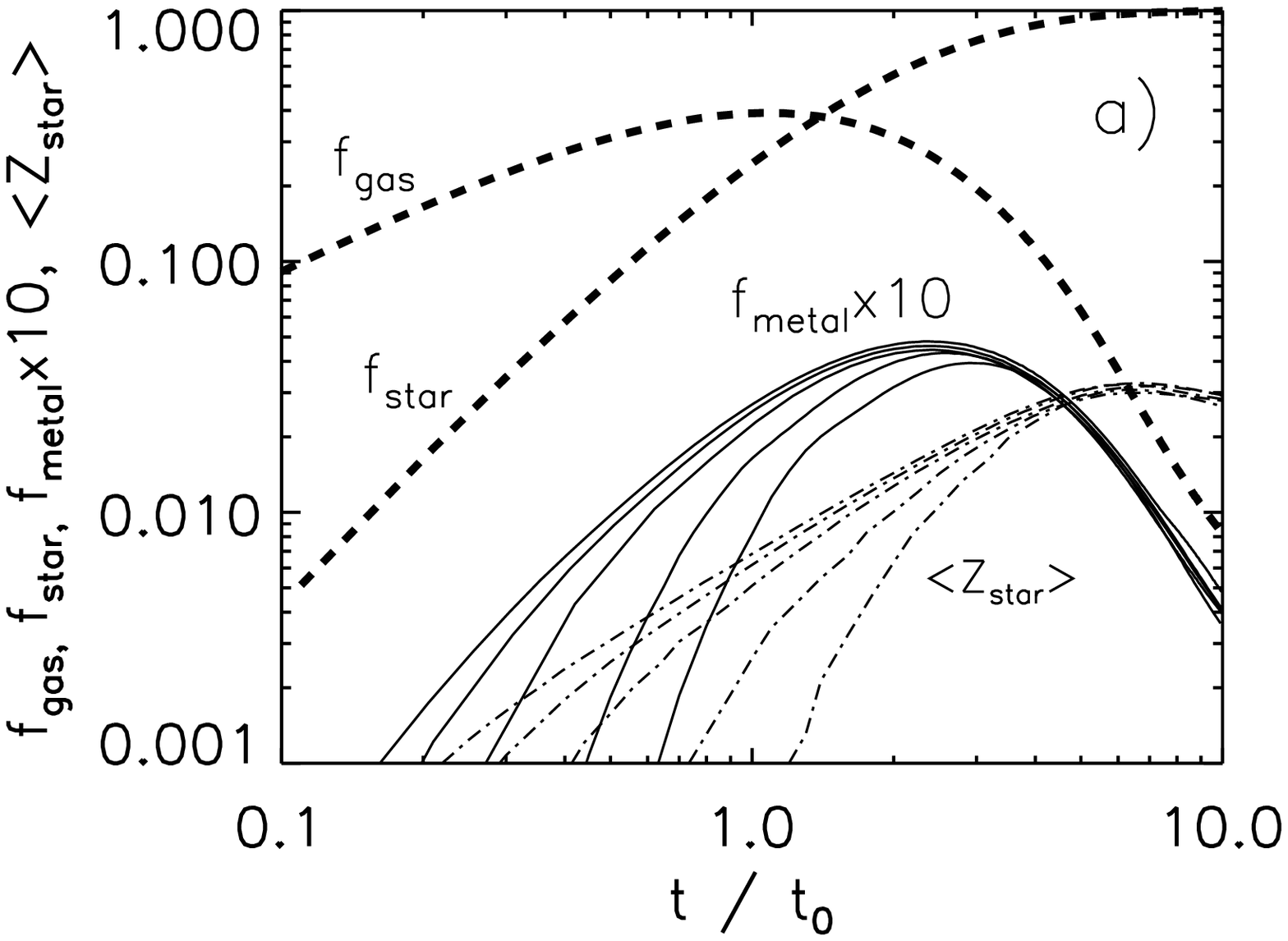}}
  \resizebox{8cm}{!}{\includegraphics{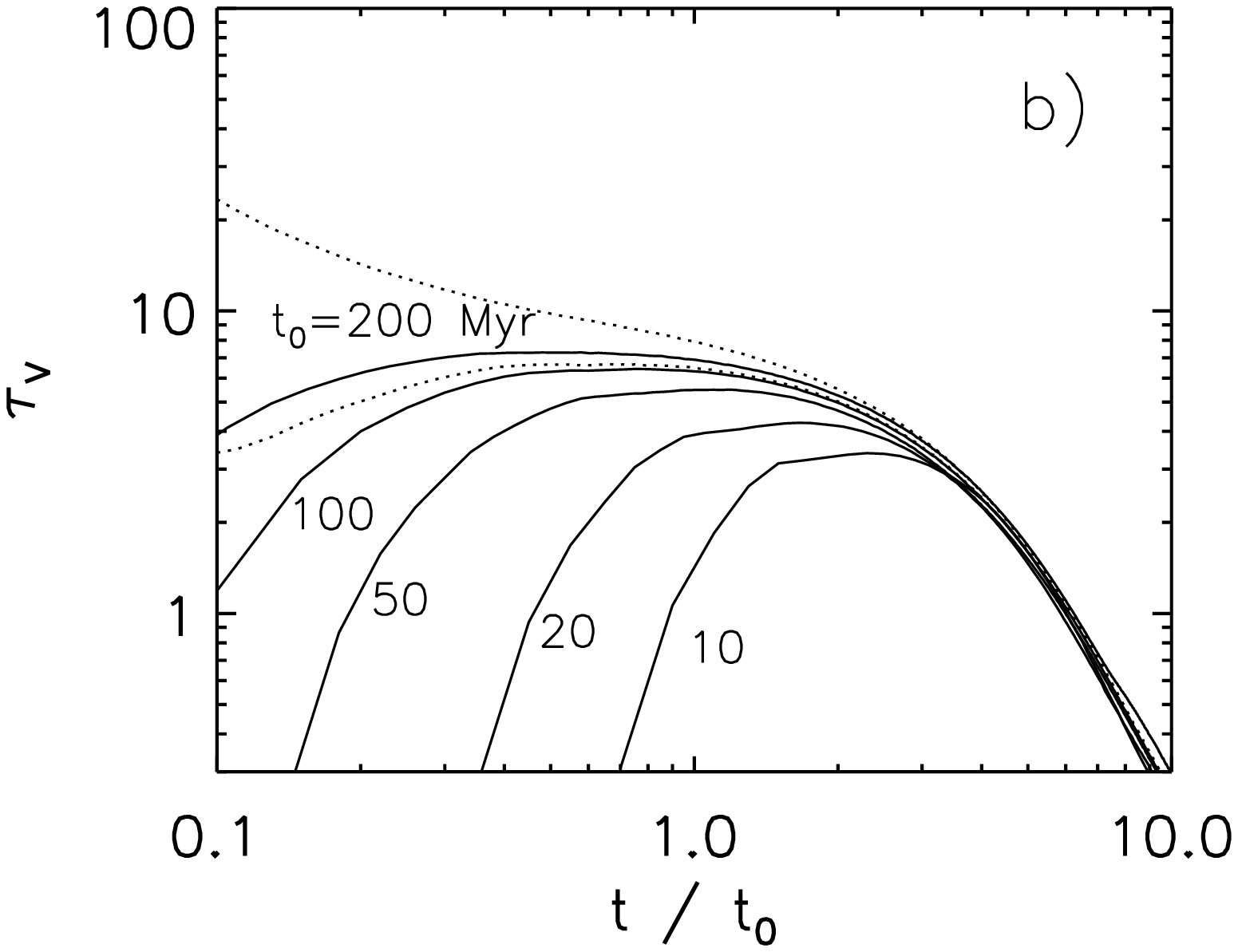}}
  \caption{Chemical evolution for
$t_0=$10, 20, 50, 100 and 200 Myr with $Z_i$=0. 
a) Evolution of gas mass fraction
$f_{gas}$, stellar mass fraction $f_{star}$, and metal mass fraction 
$f_{metal}$. 
Both $f_{gas}$ and $f_{star}$ are almost independent of 
$t_0$ so that the variation of both quantities is no more than 
the thickness of dashed curves. Luminosity-weighted mean 
stellar metallicities are indicated by dot-dashed lines. 
b) Evolution of optical depth at V-band. For two dotted lines, the 
different values of $Z_i$ (0.01$Z_\odot$ for lower and 
0.1$Z_\odot$ for upper) are adopted with $t_0=100$ Myr. 
}
  \label{chem_evo}
\end{figure}

\subsection{Intrinsic SED and radiative transfer}

We calculate unobscured stellar SEDs by using the 
population synthesis code of Kodama \& 
Arimoto (1997), in which the effect of 
stellar metallicity is fully taken into account.
We solve the equation of radiative transfer by using a computational code
developed by Takagi (2001).
Isotropic multiple scattering is assumed and the self-absorption of
re-emitted energy from dust is fully taken into account.
The temperature fluctuation of very small dust particles is calculated
consistently with the radiative transfer.

We assume no gradient of the stellar population along the radius of starburst region.
Although gas emission is not considered in our model,
a modification of the total SED due to gas emission is not significant unless
a starburst is considerably younger
than 10 Myr (Leitherer \& Heckman 1995; Fioc \& Rocca-Volmerange 1997).
The contribution from gas emission to the continuum light is especially
important in the NIR.
As we will show later, no systematic discrepancies in the NIR
flux are found between model results and observations.
Therefore, we believe that starburst galaxies are old enough to have
negligible contribution from gas emission to the continuum.

\section{Model Properties}
\subsection{Evolutionary properties}

Fig. \ref{chem_evo} shows how chemical properties and $\tau_V$
evolve as a function of $t/t_0$, for $t_0=10, 20, 50, 100$, and $200$ Myr.
The gas mass fraction $f_{gas}$, stellar mass fraction $f_{star}$, 
and metal mass fraction $f_{metal}$ are defined as
$M_g/M_T$, $M_*/M_T$, and $Z_gM_g/M_T$, respectively. 
The luminosity-weighted mean metallicity is expressed as
(Arimoto \& Yoshii 1987): 
\begin{equation}
<Z_{star}> = \frac{\Sigma Z_{star} L_{star}}{\Sigma L_{star}}, 
\end{equation}
where $Z_{star}$ and $L_{star}$ are the metallicity and luminosity of 
each star, respectively. 
In this section, we simply adopt $Z_i=0$, unless otherwise explicitly noted. 

The variation of $f_{gas}$ and $f_{star}$ as a function of $t/t_0$ is 
almost independent of $t_0$. 
After $t/t_0 \simeq 2$,
the evolution of $f_{metal} (\equiv f_{gas}Z_g)$ 
and $\tau_V$ depend little on $t_0$.
Dust mass evolves in proportion to $f_{metal}$, but $\tau_V$ evolves
somewhat differently, since $r_t$ increases gradually as stellar mass
accumulates. The effect of $Z_i$ on $\tau_V$ is negligible for $t/t_0>1$. 

\begin{figure*}
  \resizebox{7cm}{!}{\includegraphics{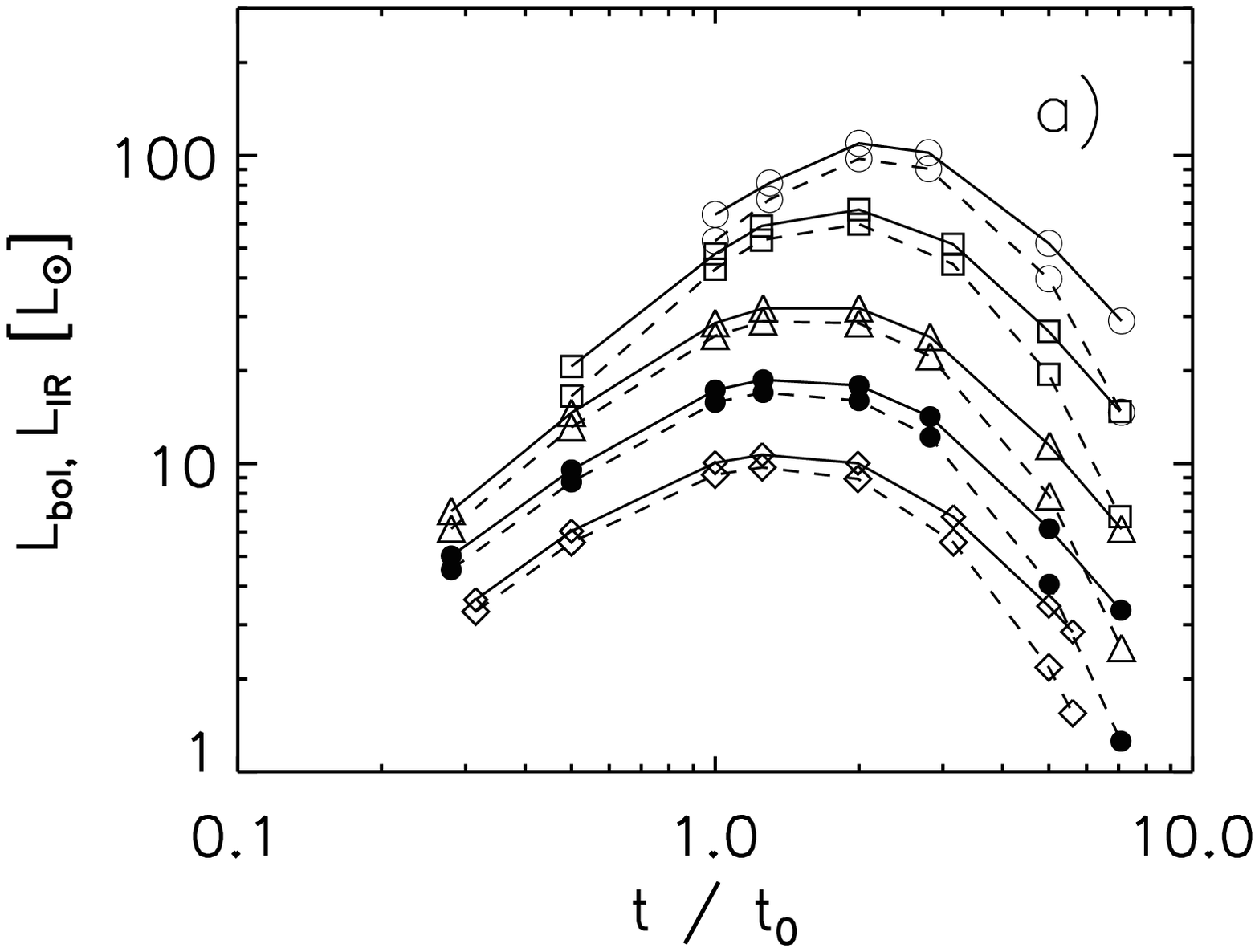}}
  \resizebox{7cm}{!}{\includegraphics{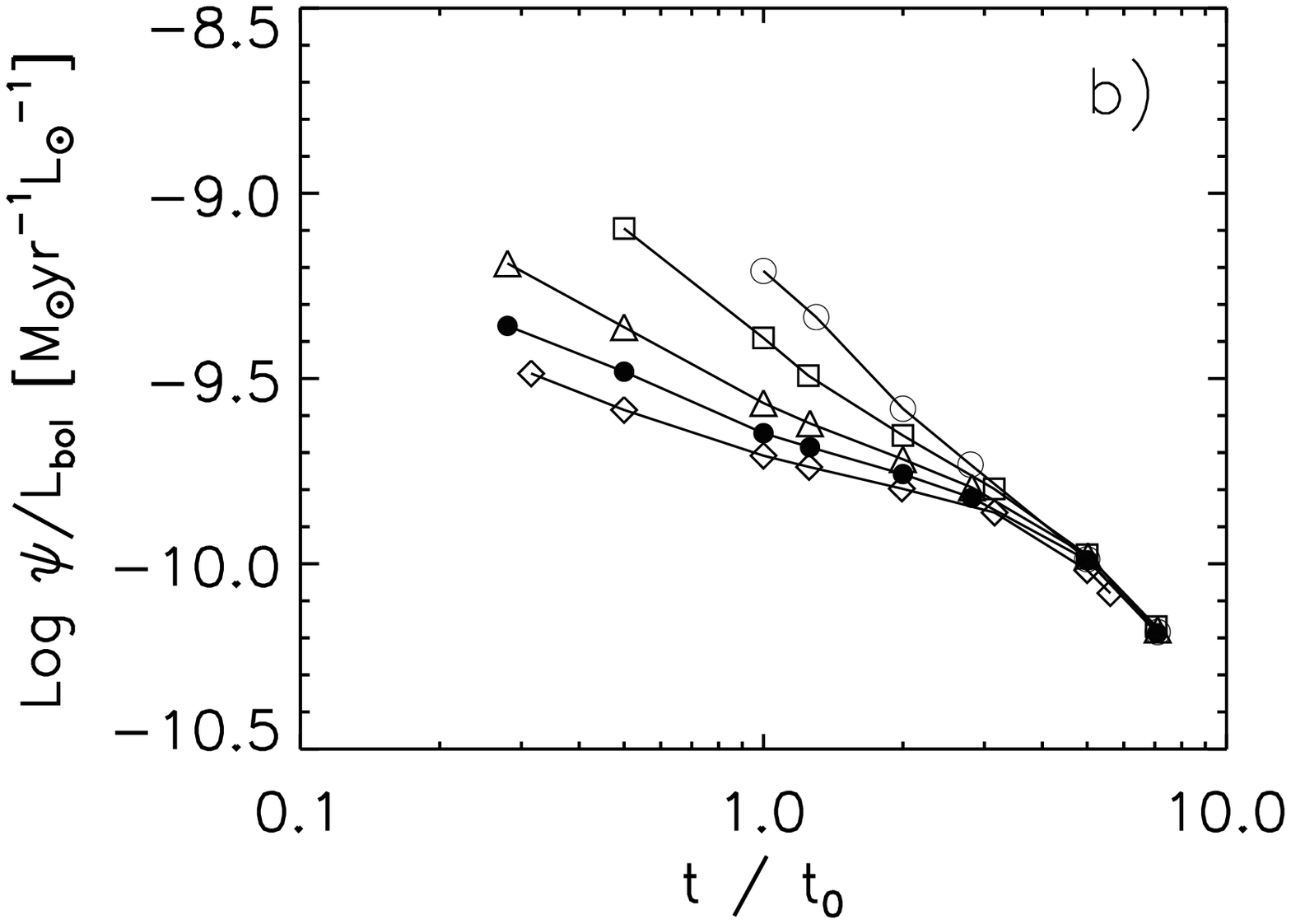}}
  \resizebox{7cm}{!}{\includegraphics{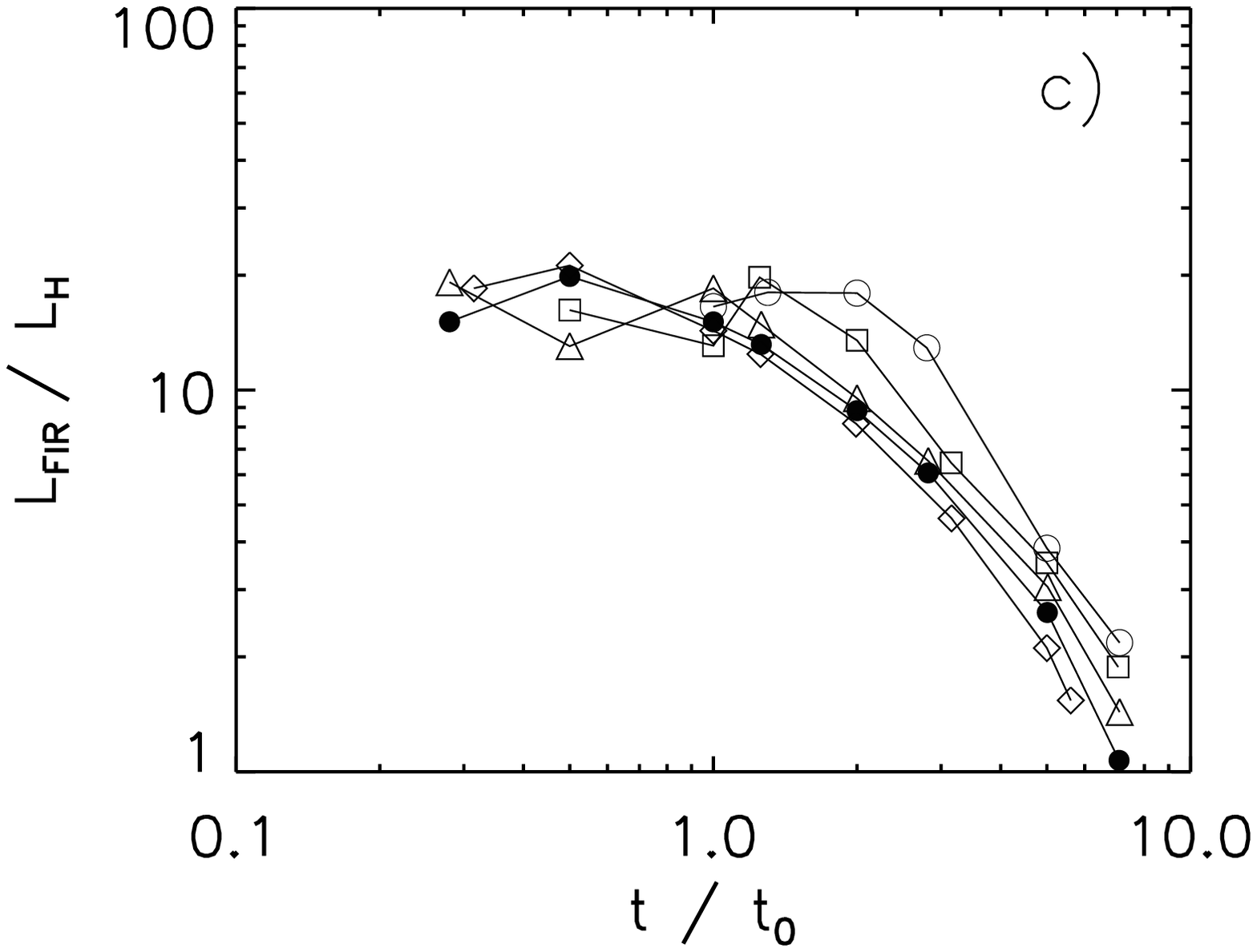}}
  \resizebox{7cm}{!}{\includegraphics{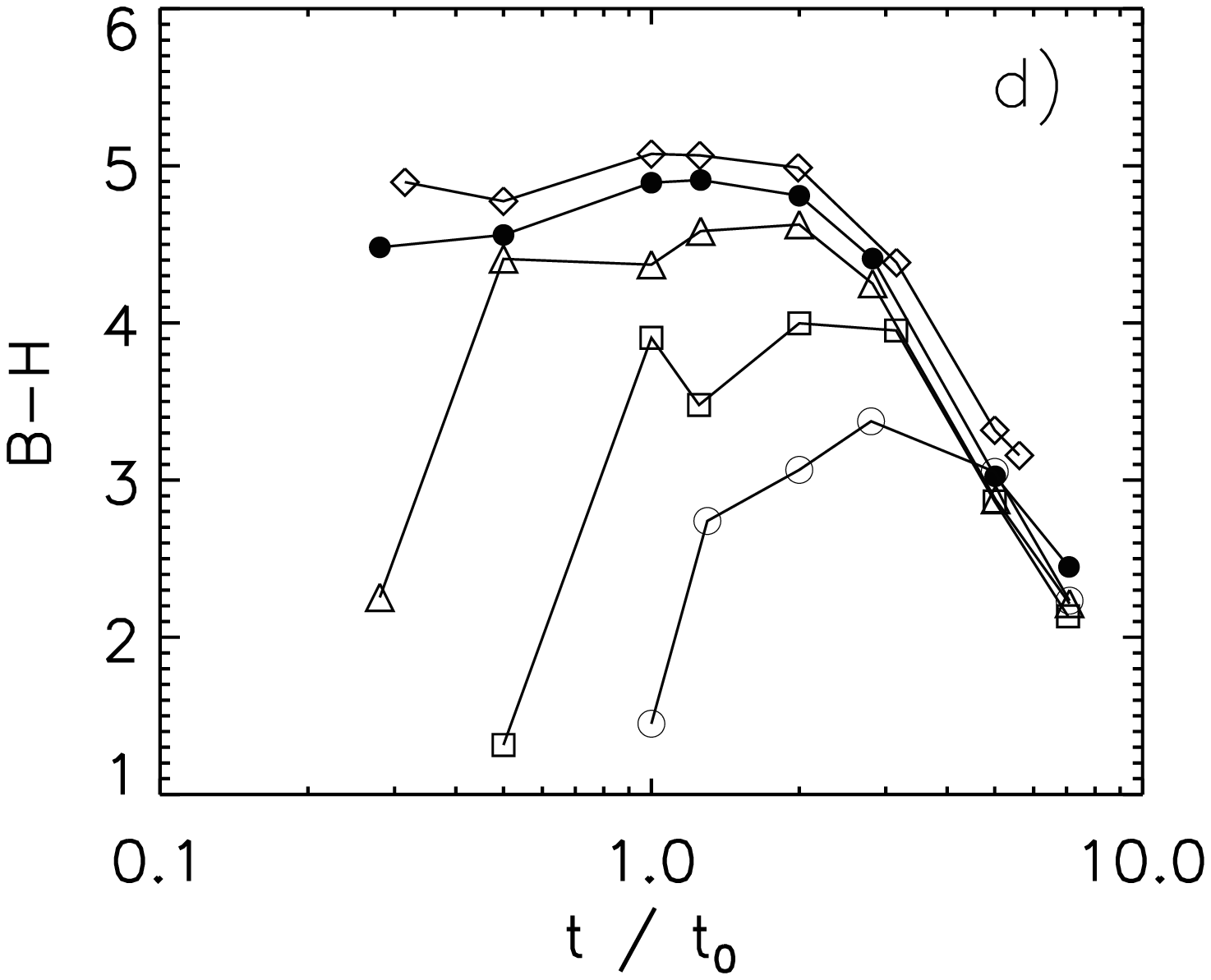}}
  \resizebox{7cm}{!}{\includegraphics{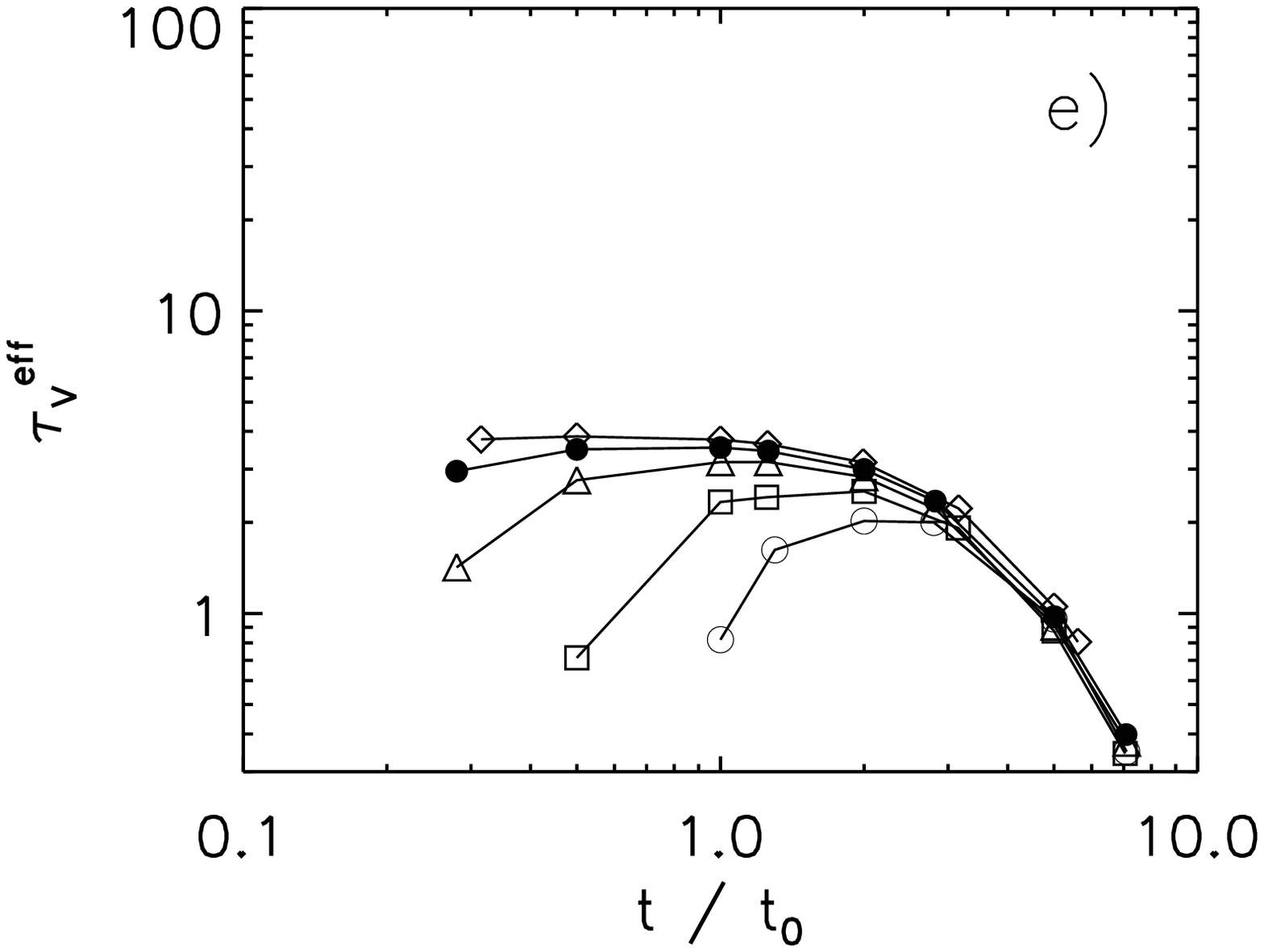}}
  \resizebox{7cm}{!}{\includegraphics{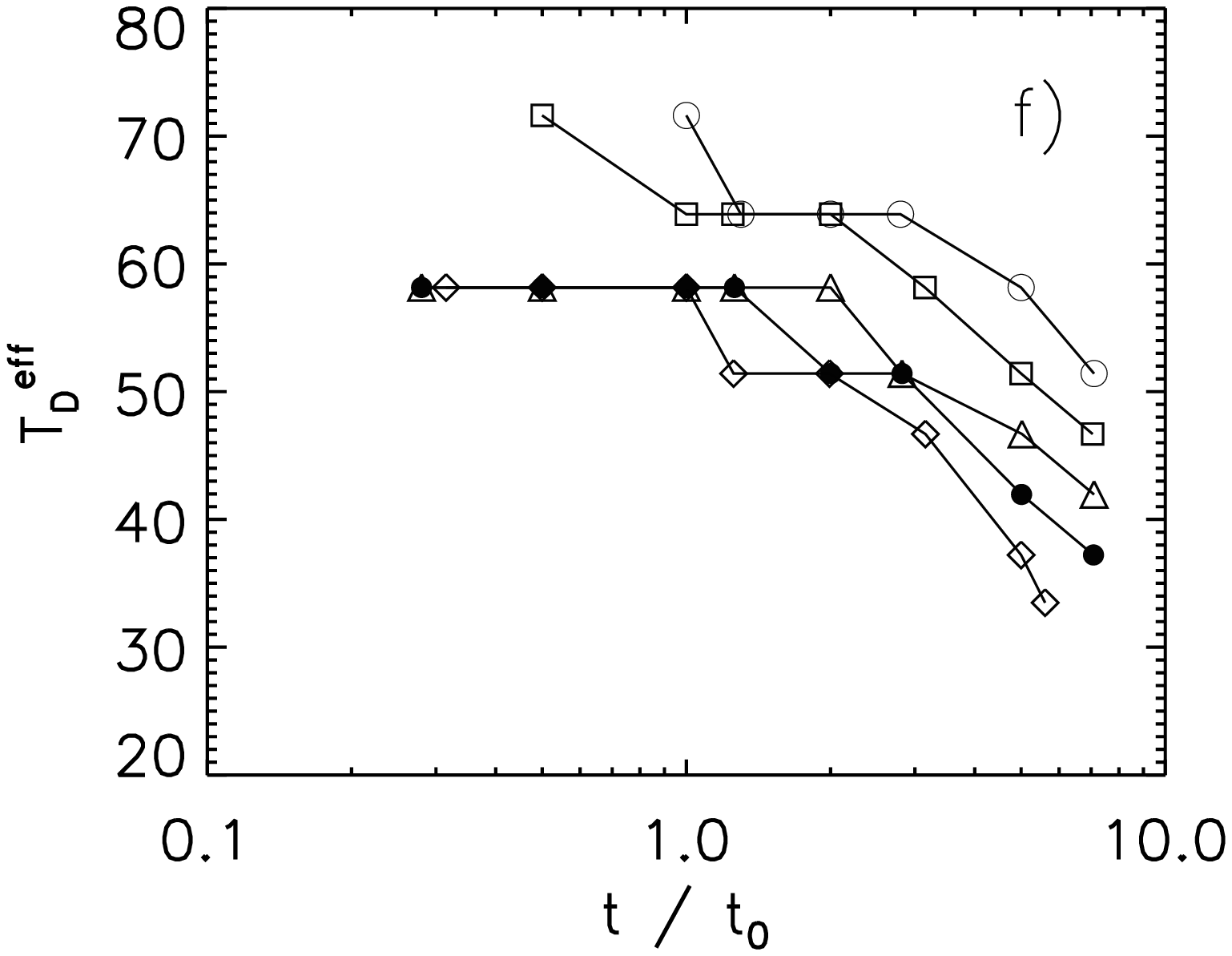}}
  \caption{The evolution of characteristic quantities of starburst
galaxies, a) $L_{bol}$ (solid line) and $L_{IR}$ (dashed line), 
b) the ratio of SFR to $L_{bol}$,
c) $L_{FIR}/L_H$, d) $B-H$, e) $\tau_V^{\mathrm{eff}}$, f) effective
temperature of dust $T_D^{\mathrm{eff}}$ for various $t_0$
(open circles for $t_0$= 10 Myr, squares for
20 Myr, triangles for 50 Myr, solid circles for 100 Myr, and
diamonds for 200 Myr).
$L_{IR}$ and $L_{FIR}$ are the luminosity in the wavelength range
of 8 -- 1000 $\mu$m and 42 -- 122 $\mu$m, respectively.
We adopt the MW extinction curve,
$\Theta=1$, $M_T=1$ M$_\odot$ and $Z_i=0.0$.
}
  \label{tau_evo}
\end{figure*}

Fig. \ref{tau_evo} shows the evolution of
$L_{bol}$, $L_{IR}$, $\psi(t)/L_{bol}$, $L_{FIR}/L_{H}$, $B-H$,
$\tau^{\mathrm{eff}}_V$, and $T_D^{\mathrm{eff}}$.
$\tau^{\mathrm{eff}}_V$ is the effective optical depth at $V$-band
defined as $L_V=L^0_V\exp (-\tau^{\mathrm{eff}}_V)$, where $L^0_V$ and
$L_V$ are the intrinsic and attenuated luminosity, 
respectively. $T_D^{\mathrm{eff}}$ is the effective dust temperature
which gives the same peak wavelength of dust emission as that of a 
model SED with an emissivity law $\propto \lambda^{-2}$.

We summarize the model properties as follows: 

1) The IR luminosity evolution suggests that starburst galaxies
should be found mostly in the late phase ($t/t_0 \ga 2$),
after the SFR and the dust mass become maximum.
The asymptotic relation between SFR $ \psi (t)$ and $L_{bol}$ for the 
late phase $t/t_0>2$ is
\begin{equation} \label{eq_bolsfr}
\frac{L_{bol}}{1.7 \times 10^9 \mathrm{L}_{\odot}} =  \frac{\psi(t)}{1
\mathrm{M}_{\odot} \mbox{yr}^{-1}} \frac{t}{t_0}  \; .
\end{equation}
We also derive an asymptotic relation for the mass-to-light ratio 
for the late phase $t/t_0 >2$;
\begin{equation} \label{eq_masslum}
 \frac{M_{\ast}}{L_{bol}}=  8.05 \times 10^{-3}
 \left ( \frac{t}{t_0} \right )^{1.8}
 \left ( \frac{t_0}{100 \mathrm{Myr}} \right )^{0.8}
\frac{\mathrm{M}_{\odot}}{\mathrm{L}_{\odot}} \; .
\end{equation}
These ratios will be used in the analysis of starbursts later (Section 5). 

2) The variation of $L_{FIR}/L_{H}$ as a function of $t/t_0$ 
is nearly independent of $t_0$. 
$L_{FIR}/L_{H}$ decreases monotonically with time.
Due to this weak dependence of SED quantities on $t_0$,
it is difficult to determine the absolute starburst age
from the SED fitting alone.

3) The dependence of $B-H$ on $t_0$ is similar to that of
$\tau_V$ and $\tau_V^{\mathrm{eff}}$.

\begin{figure*}
  \resizebox{13cm}{!}{\includegraphics{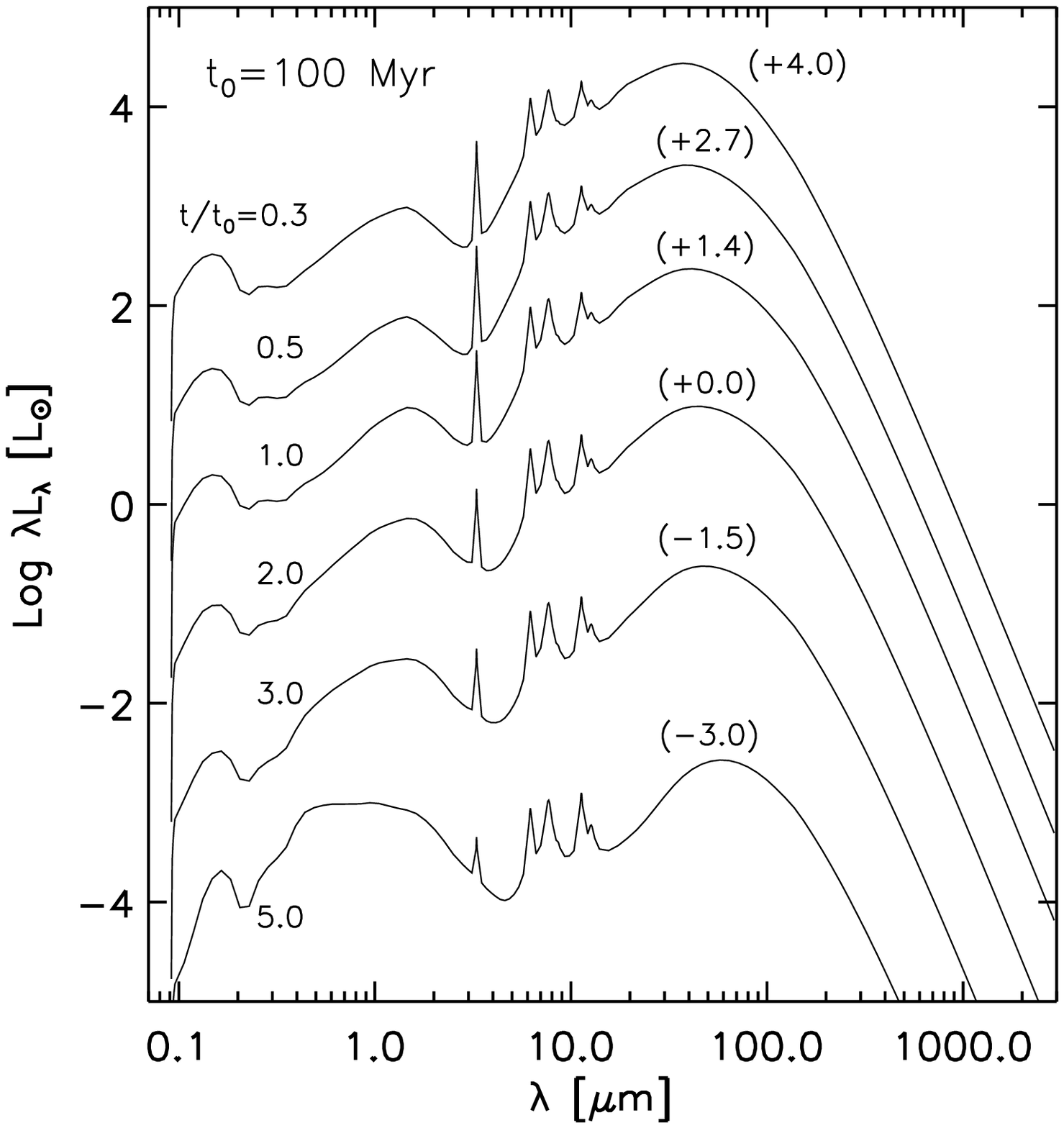}}
\caption{The spectral evolution of a starburst for $\Theta =1$ with
$M_T$=1 M$_\odot$, the MW extinction curve and $Z_i=0$. 
The evolutionary stage for each line is denoted near the lines. 
For clarity, the SEDs are shifted vertically with the denoted value in 
parenthesis at the emission peak.
}
\label{sed_evo}
\end{figure*}

Fig. \ref{sed_evo} shows the SED evolution 
for $t_0=100$ Myr, $\Theta=1$, the MW extinction curve, and $Z_i=0$.
The SEDs are quite similar to each other at $t/t_0 < 2$.
At $t/t_0 >2 $, the peak wavelength of dust emission
increases and the FIR excess ($L_{FIR}/L_H$) decreases.

\subsection{Dependence on compactness factor}

\begin{figure*} 
  \resizebox{7cm}{!}{\includegraphics{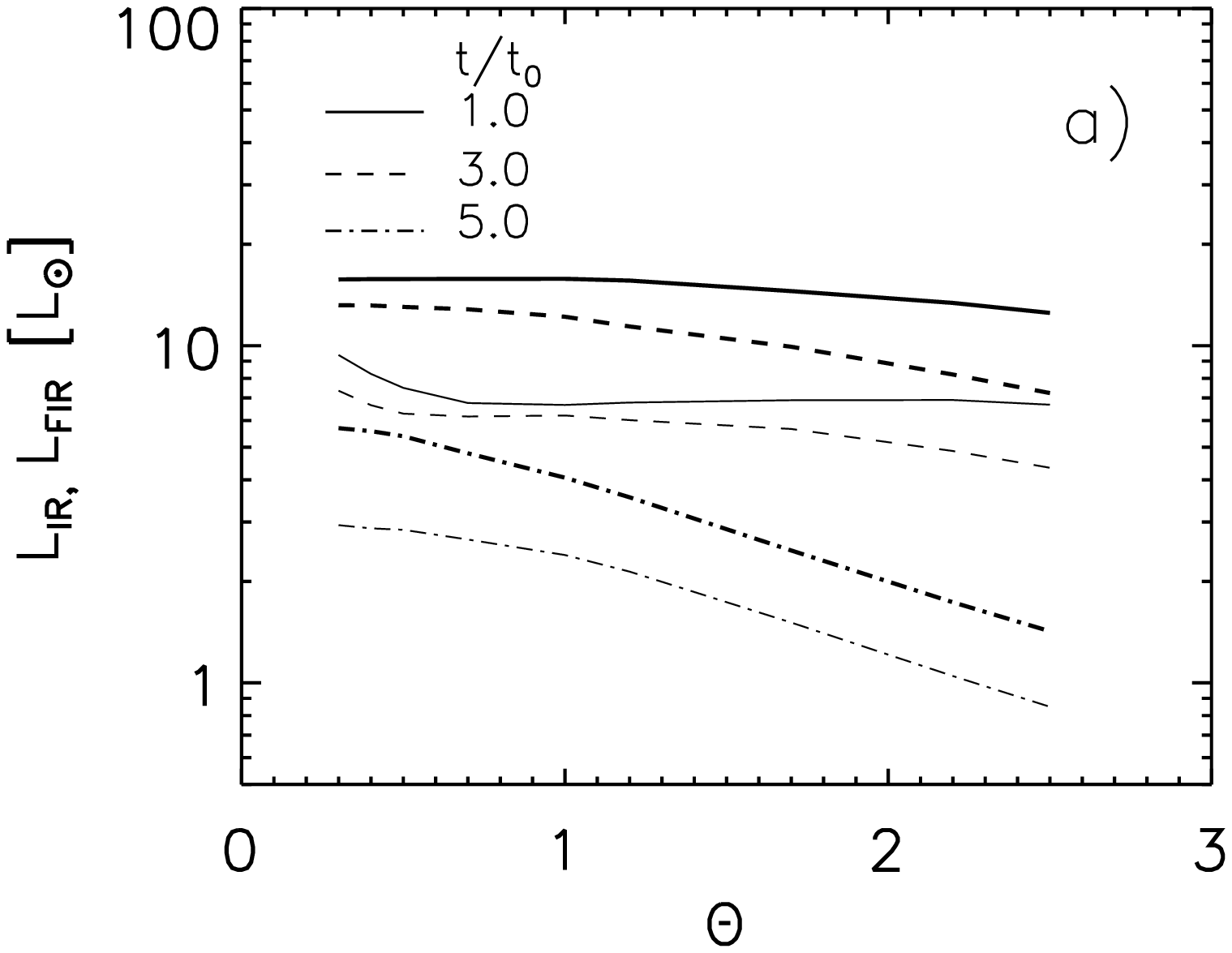}}
    \resizebox{7cm}{!}{\includegraphics{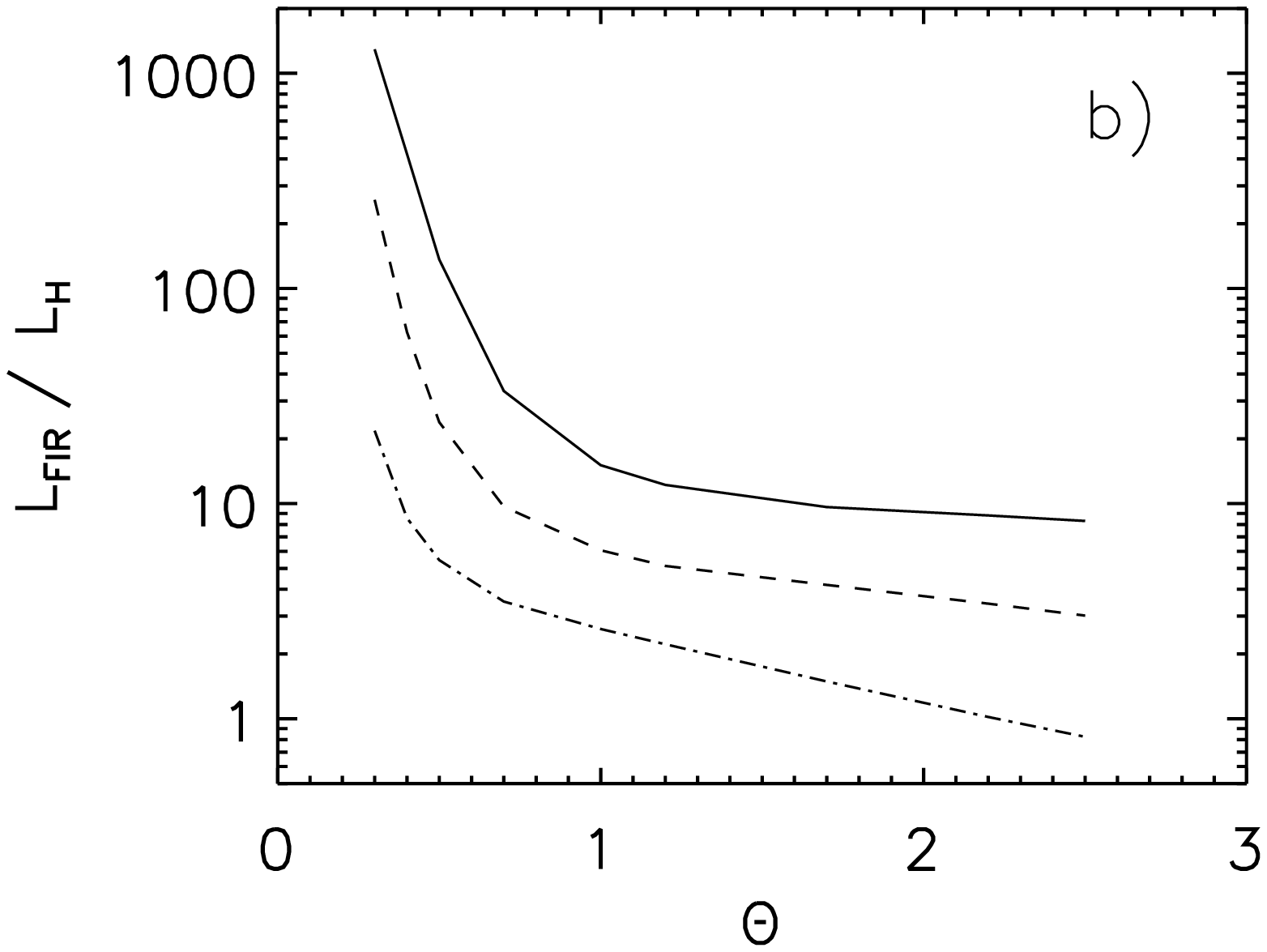}}
  \resizebox{7cm}{!}{\includegraphics{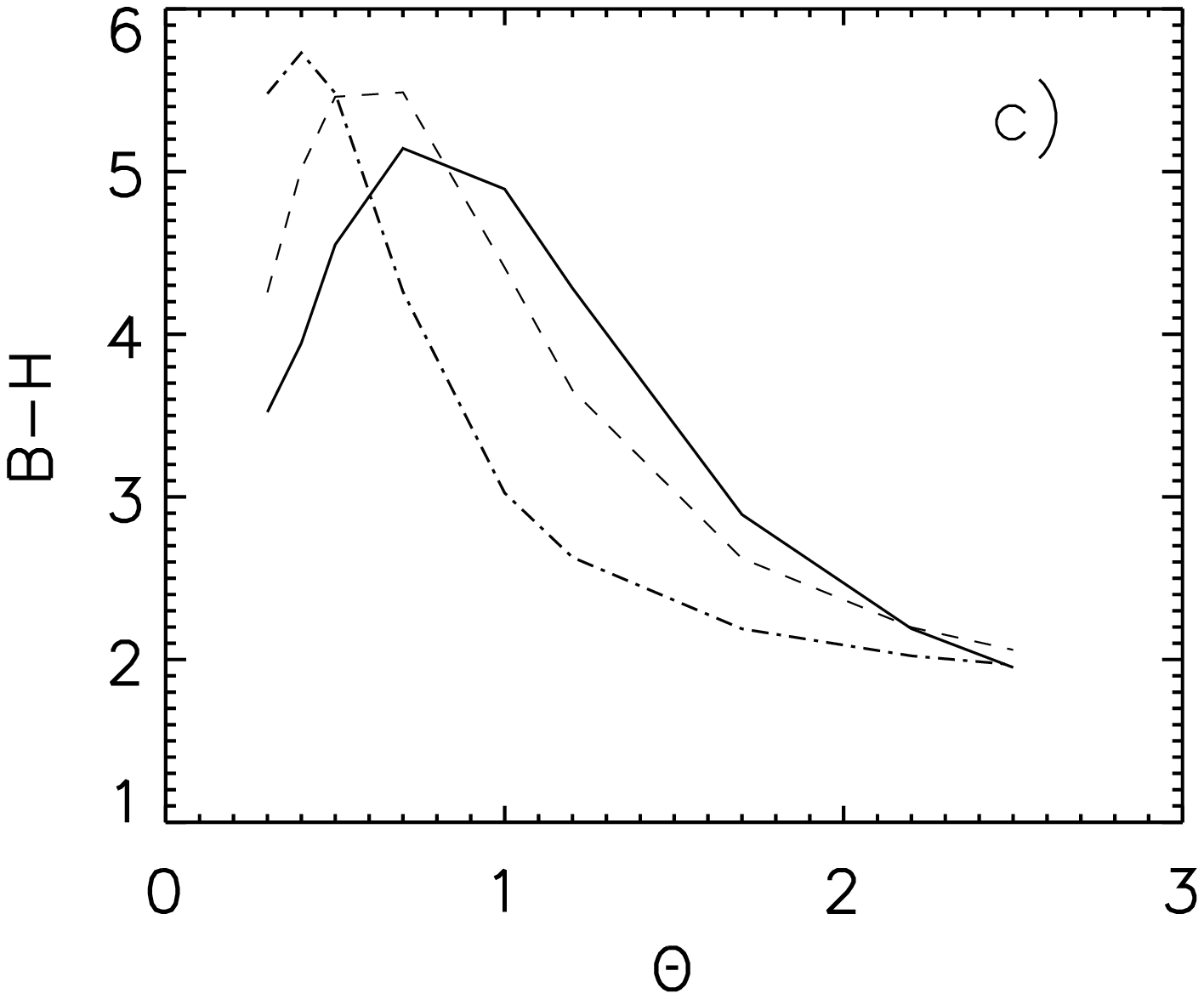}}
    \resizebox{7cm}{!}{\includegraphics{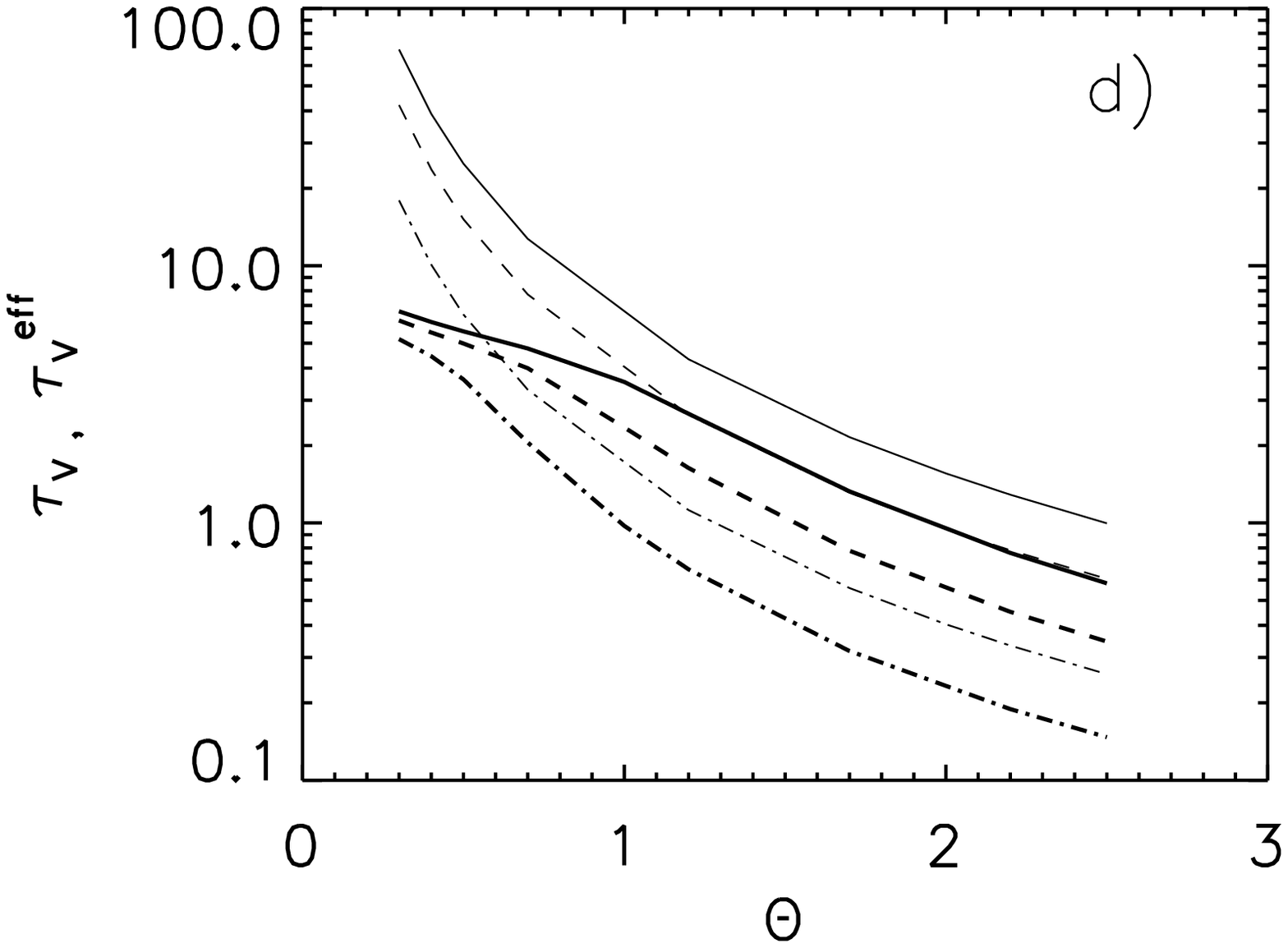}}
  \resizebox{7cm}{!}{\includegraphics{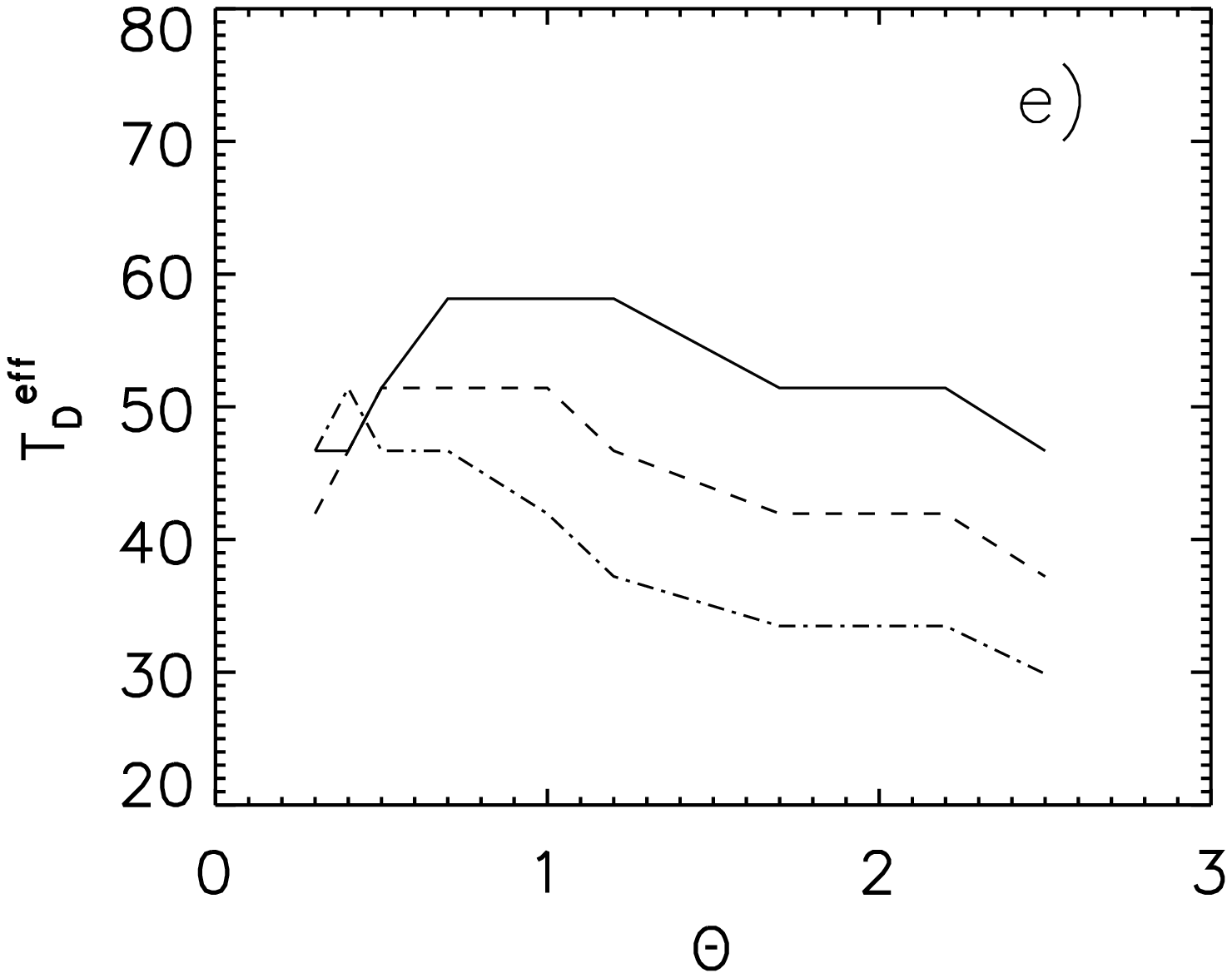}}
  \caption{The dependence of a) $L_{FIR}$ (thin lines) and $L_{IR}$ (thick lines), 
  b) $L_{FIR}/L_H$, c) $B-H$, d), $\tau_V$ (thin lines) and $\tau^{\mathrm{eff}}_V$ (thick lines), 
e) $T_D^{\mathrm{eff}}$ on $\Theta$. 
The case for $t_0=100$ Myr are shown in each panel.
Sold, dashed and dot-dashed lines indicate
the case of $t/t_0=1.0$, 3.0 and 5.0, respectively.
We adopt the MW extinction curve, $M_T=1$ M$_\odot$, and
$Z_i=0.0$.}
  \label{tau_evo2}
\end{figure*}

\begin{figure*}
  \resizebox{13cm}{!}{\includegraphics{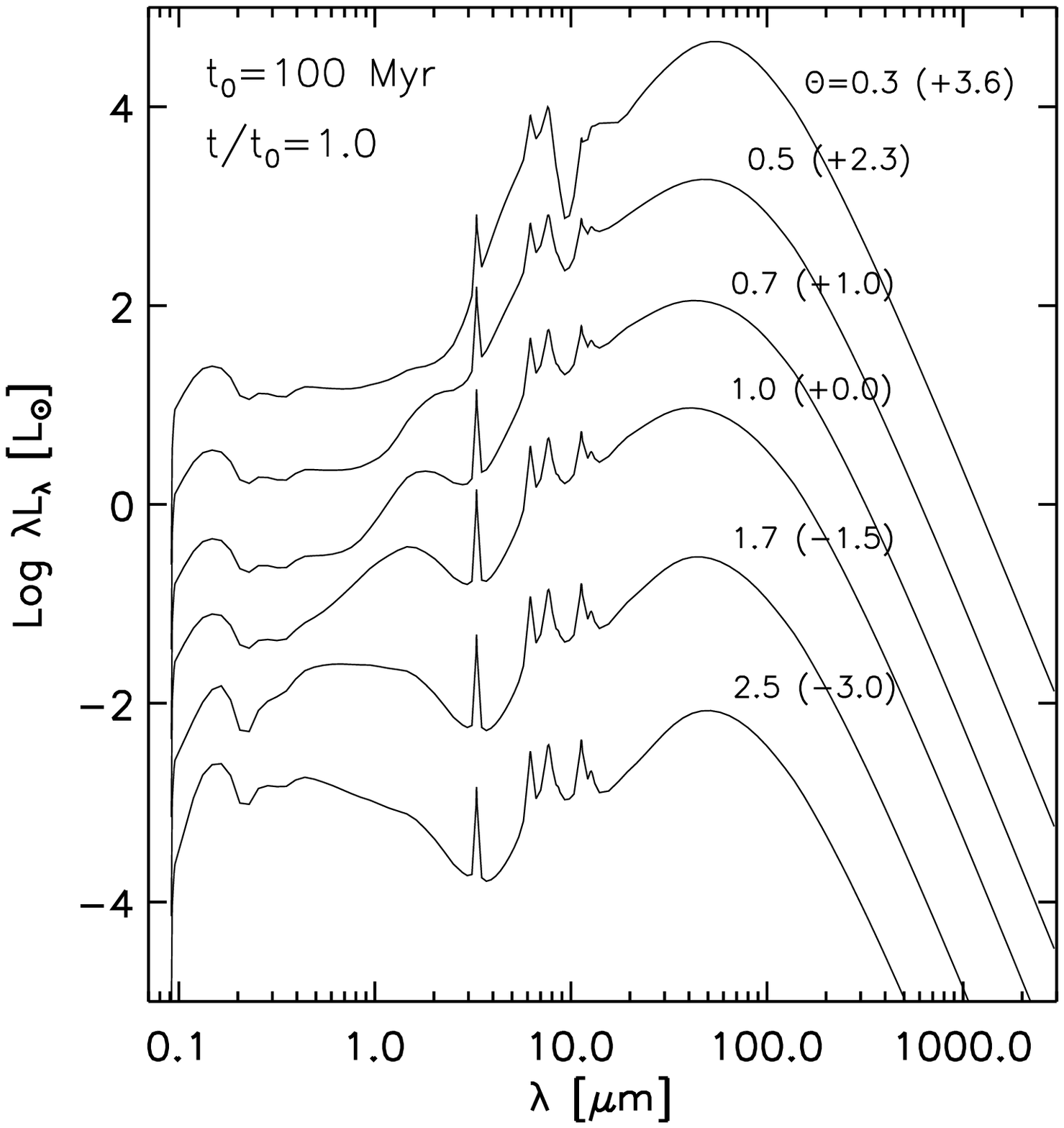}}
\caption{The SEDs of starburst galaxies for various $\Theta$ with 
$M_T$=1 M$_\odot$. 
We adopt the MW extinction curve and $Z_i=0$. 
For clarity, the SEDs are shifted vertically with the denoted value in 
parenthesis, following the values of $\theta$.}
\label{sed_theta}
\end{figure*}

Fig. \ref{tau_evo2} shows how the characteristic SED
quantities depend on the compactness factor $\Theta$. 
While $L_{FIR}$ is not sensitive to $\Theta$,
$L_{FIR}/L_H$ increases significantly for $\Theta <1$.
At a small $\Theta$ (i.e., optically thick), the SED at optical
wavelengths asymptotically reaches the source function of
the stellar light, since the light mainly comes from near the surface
for such a highly obscured region. This is the reason why a starburst
region becomes bluer for $\Theta \la 0.5$.

There are two possible reasons for large values of $\tau_V$; 
1) starbursts occur in metal rich clouds ($Z_i>0.1Z_\odot$) 
and are still very young ($t/t_0<0.3$) as shown in 
Fig.\ \ref{chem_evo}b, and 2) geometry of 
starbursts is compact ($\Theta < 1$) as shown in Fig.\ \ref{tau_evo2}d. 
If the former reason is the case, ULIRGs should be younger 
than UVSBGs, systematically. However, we find no such systematic difference 
as we show in Section 4. Instead, SED fitting results in smaller 
$\Theta$ for ULIRGs than those of UVSBGs. 
Thus, the variation of $\tau_V$ in starbursts is
mainly caused by the geometrical effect; i.e., $\Theta$. 
This result is reasonable, since the luminosity evolution 
suggests that starbursts are less luminous in 
the early phase of starbursts.
In later phases, the high luminosity and large fraction 
of life time during this phase as a starburst would increase the
chance of the starbursts being observed. 
The importance of $\Theta$ on the variation in starburst SED is
illustrated clearly in Fig.\ \ref{sed_theta}.

\subsection{Effective radius in various wavelengths}

\begin{figure*} 
  \resizebox{8cm}{!}{\includegraphics{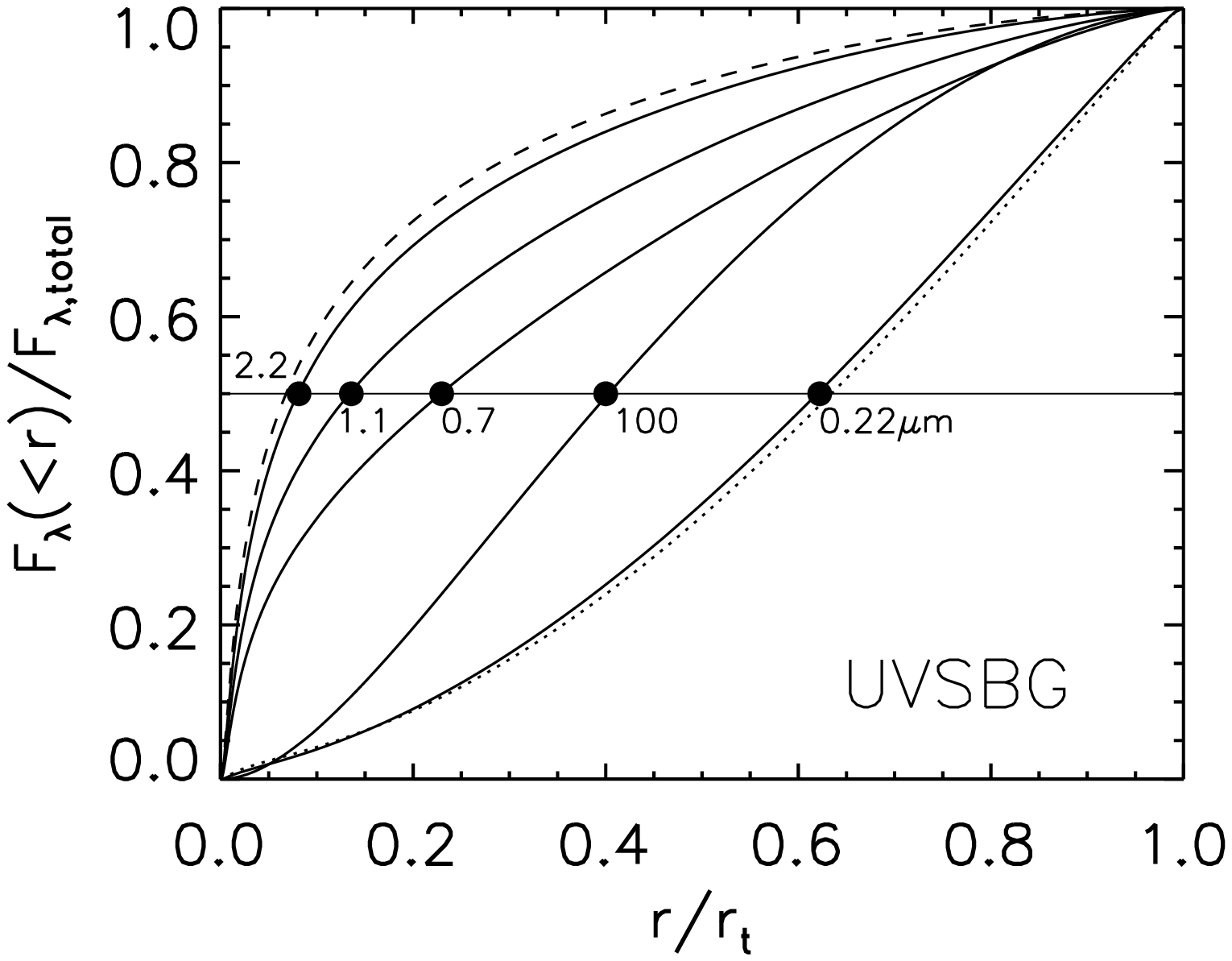}}
  \resizebox{8cm}{!}{\includegraphics{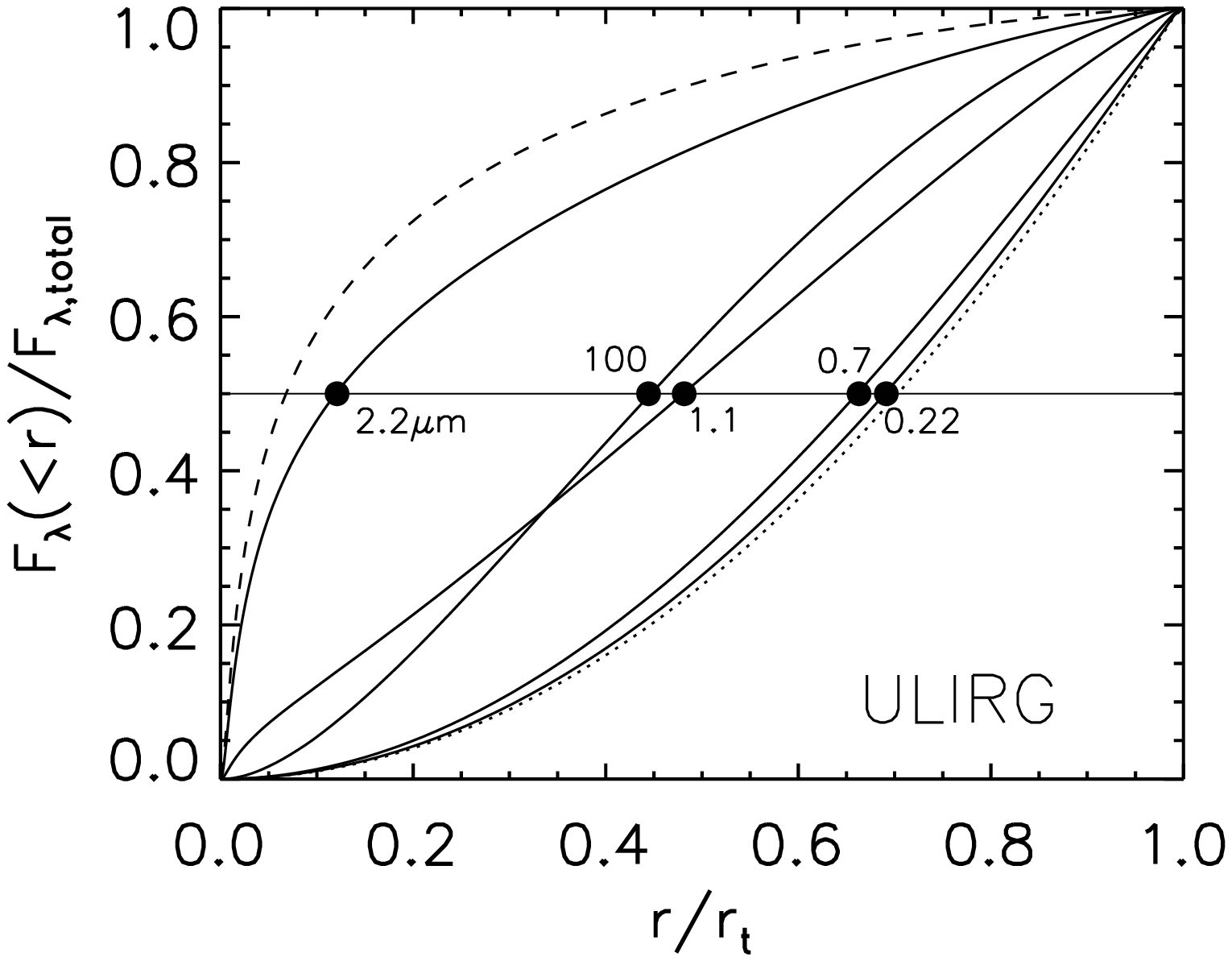}}
  \caption{Growth curves of starburst region at various wavelengths.
The vertical axis is the fraction of flux emitted within the projected
radius $r$. Dashed line is for the case without dust. Effective radii
for each wavelength is indicated by solid circles. 
Dotted line indicates the growth curve at 0.22 $\mu$m without a 
contribution from scattering light. 
Left panel: the case for UVSBGs ($t/t_0=2.0$, $\Theta =1.5$ and the MW
extinction curve), right panel: the case for ULIRGs
($t/t_0=2.0$, $\Theta =0.4$ and the SMC extinction curve).
}
  \label{profile}
\end{figure*}

Fig. \ref{profile} shows the growth curve of the
apparent surface brightness at 0.22, 0.7, 1.1,
2.2, and 100 $\mu$m for the models of UVSBGs
($t/t_0=2.0$, $\Theta =1.5$ and the MW extinction curve) and ULIRGs
($t/t_0=2.0$, $\Theta =0.4$ and the SMC extinction curve). 
Here we adopt the typical values of model parameters 
for each UVSBGs and ULIRGs, which are 
derived from the SED fitting described in the next section. 
At most of the observed wavelengths, as seen in Fig. \ref{profile},
the apparent effective radii are systematically larger than the intrinsic
effective radius which corresponds to 
the case without any extinction. In the FUV to
NIR wavelength region in which the dust emission is
negligible, the apparent effective radius
increases with decreasing wavelength. 
This is because the contribution from
stars in the central region decreases with increasing optical depth
toward the shorter wavelength region.
Therefore, only the growth curve in 2.2 $\mu$m follows
that of the input stellar distribution for the case of UVSBGs.
Meurer et al. (1995) presented the surface brightnesses of UVSBGs 
at FUV (0.22 $\mu$m) obtained with the {\it HST}/Faint Object Camera (FOC). 
These surface brightness is not well represented by the $r^{1/4}$ 
or exponential profile. 
This result is consistent with the model for UVSBGs 
in which the growth curve at 0.22 $\mu$m is significantly 
different from the intrinsic growth curve (dashed line). 
Note that both the $r^{1/4}$ and exponential profile have been 
used for normal galaxies in which the effect of dust is not significant, 
compared to starburst galaxies. 
On the other hand, in the case of ULIRGs,
the apparent effective radius even at 2.2 $\mu$m 
becomes twice the intrinsic effective radius, although the growth 
curve is somewhat similar to the intrinsic one.
In both UVSBGs and ULIRGs, 
the effect of scattering on the growth curve is negligible. 
The apparent effective radius at 100 $\mu$m is also different from
the intrinsic effective radius, since it is determined by the
distribution of dust, rather than that of stars.

\section {SED fitting to UVSBG{\sevensize s} and ULIRG{\sevensize s}}

We apply our evolutionary SED model to two samples of
local starburst galaxies; UVSBGs and ULIRGs.
All models are calculated with $t_0=100$ Myr, $Z_i=0.1\ Z_\odot$. 
A definitive value of $t_0$ is only important for the absolute 
time scale of starburst events which is not discussed here. 
The value of $Z_i$ should vary for each 
starburst galaxy, since $Z_i$ depends on the star formation 
history in precursors of a starburst galaxy. 
We expect that the gas fraction in precursors of starbursts 
should be large, and therefore 
assume that the level of chemical enrichment in precursors
is low and comparable to that in the SMC. 
Our conclusions are independent 
of the choice of $Z_i$. 

Fitting parameters are the starburst age $t$, 
and the compactness factor $\Theta$. We choose the best-fitting 
extinction curve from the MW, LMC and SMC type. The SEDs are 
normalized by the initial mass $M_T$ of a gas reservoir.
The best-fitting model for each
galaxy is sought by a standard least square method.
We adopt the cosmology of $H_0$ = 75 km sec$^{-1}$ Mpc$^{-1}$
and $q_0=0.5$ throughout this paper.

\subsection {UVSBGs}

\begin{figure*}
  \resizebox{7cm}{!}{\includegraphics{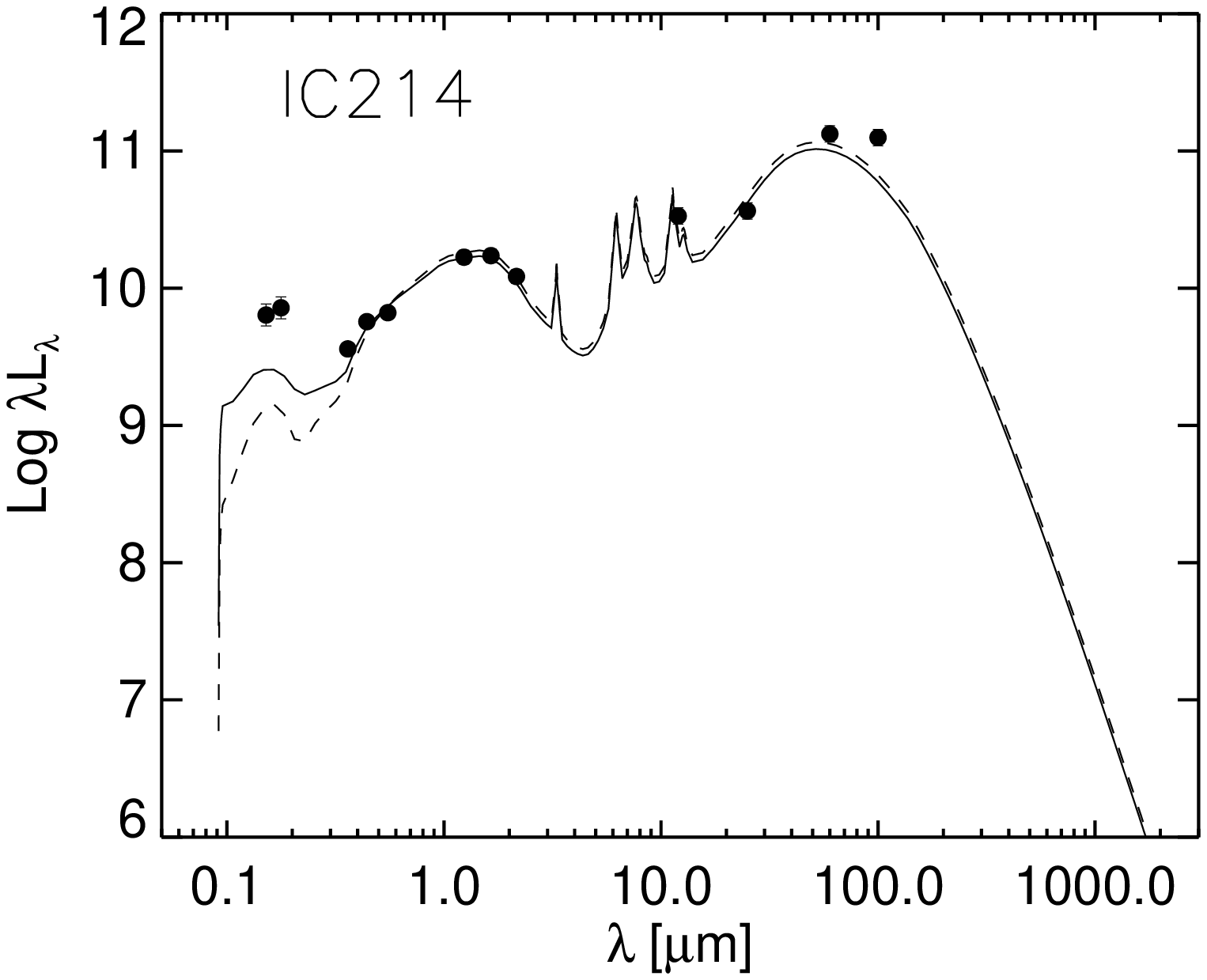}}
  \resizebox{7cm}{!}{\includegraphics{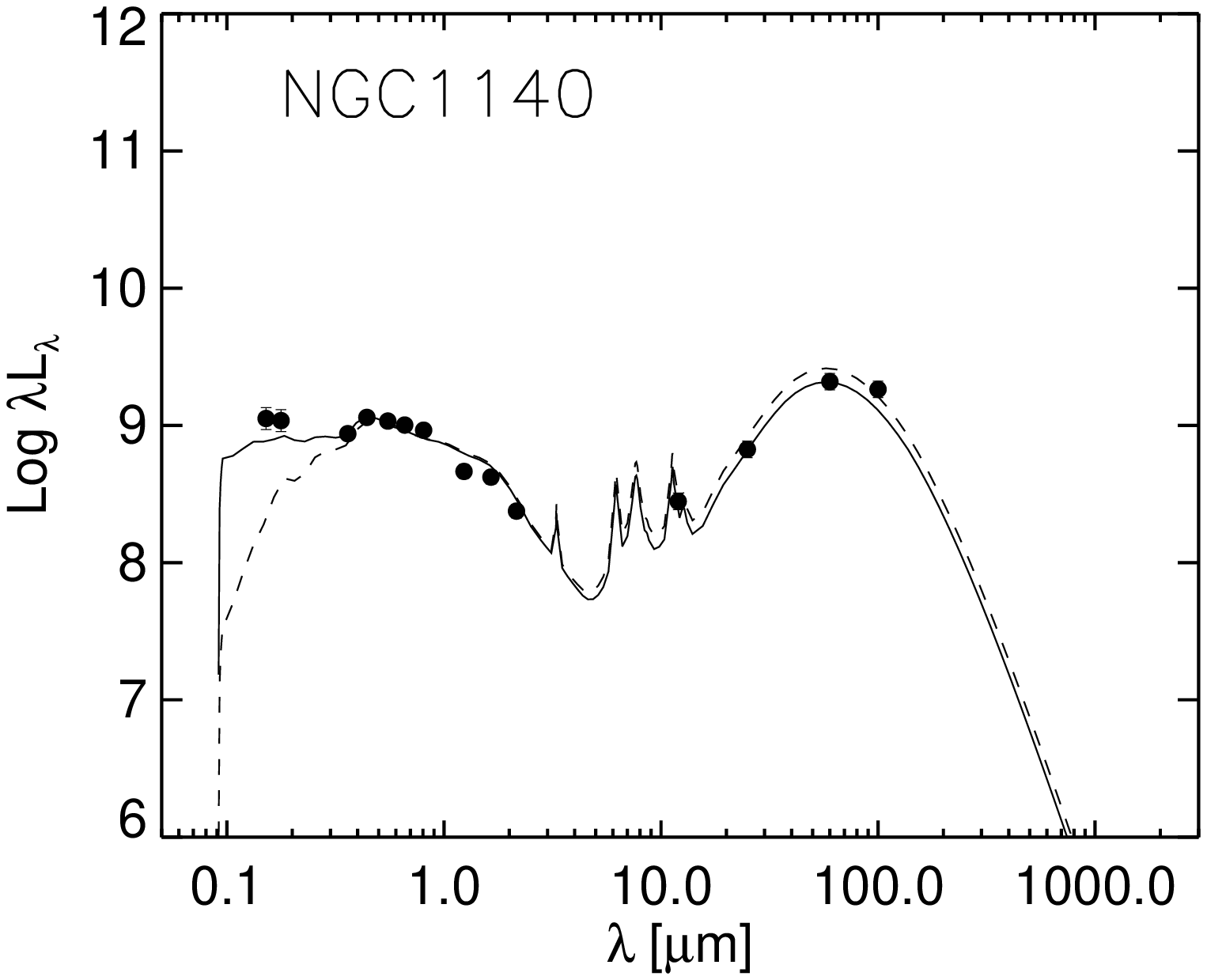}}
  \resizebox{7cm}{!}{\includegraphics{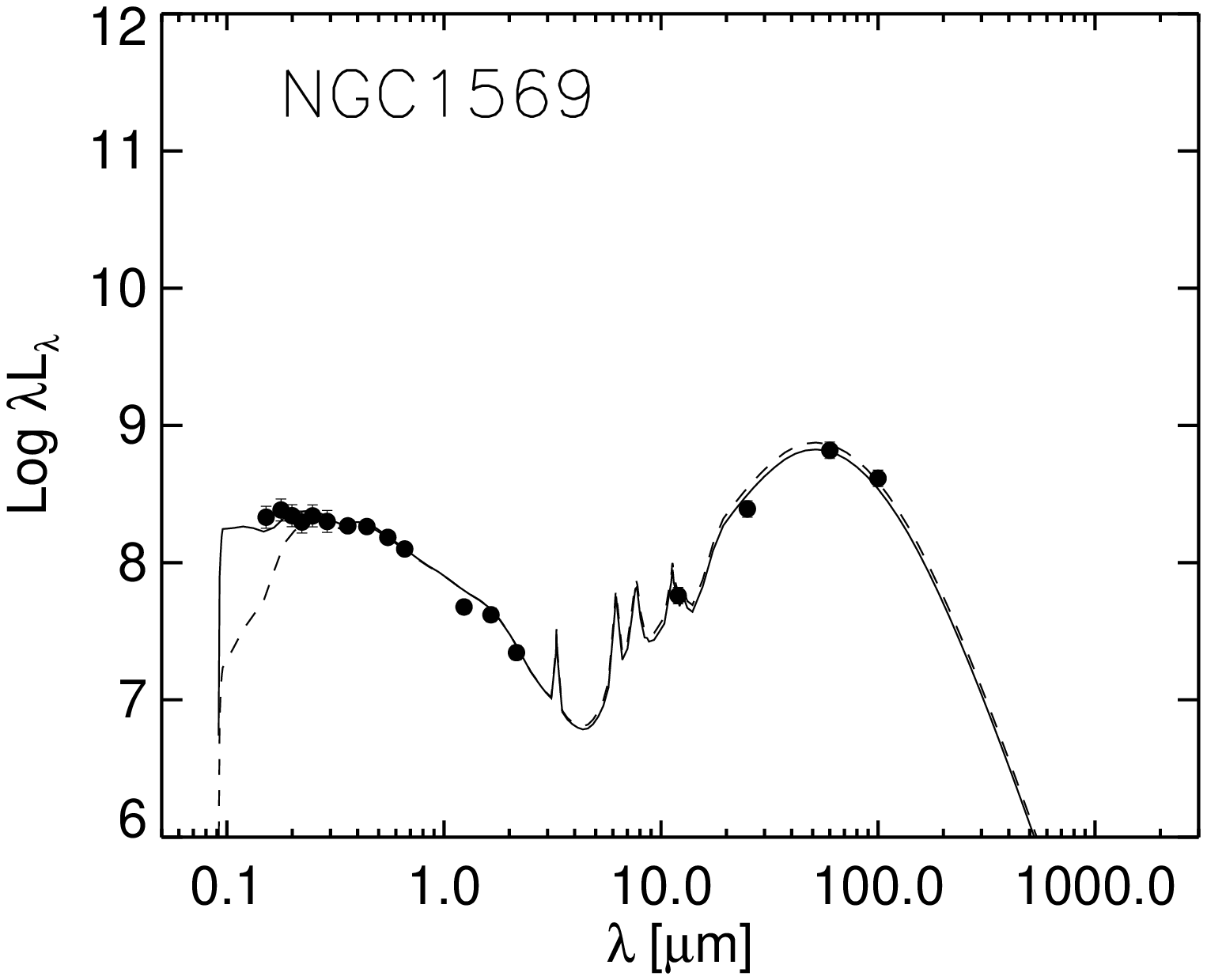}}
  \resizebox{7cm}{!}{\includegraphics{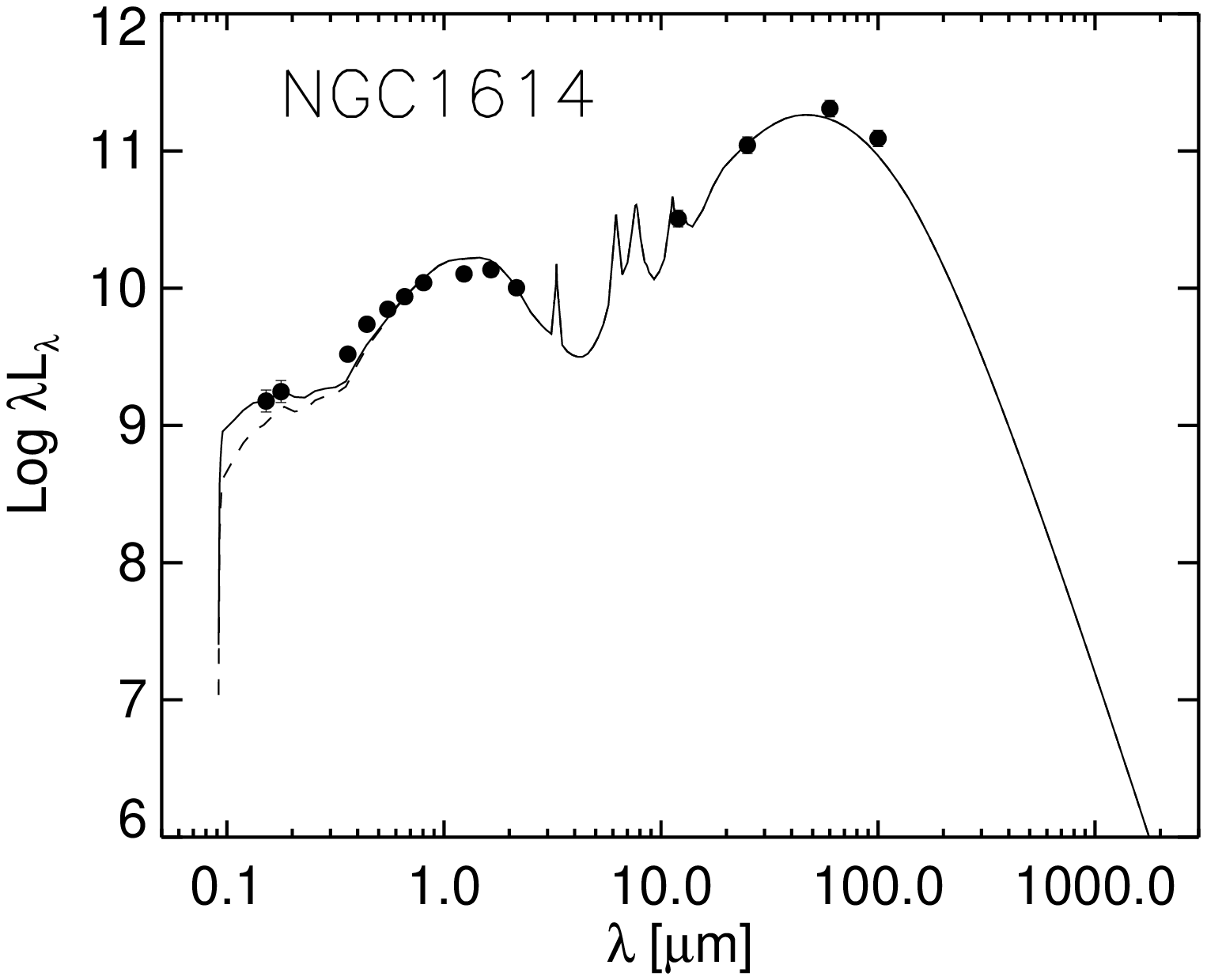}}
  \resizebox{7cm}{!}{\includegraphics{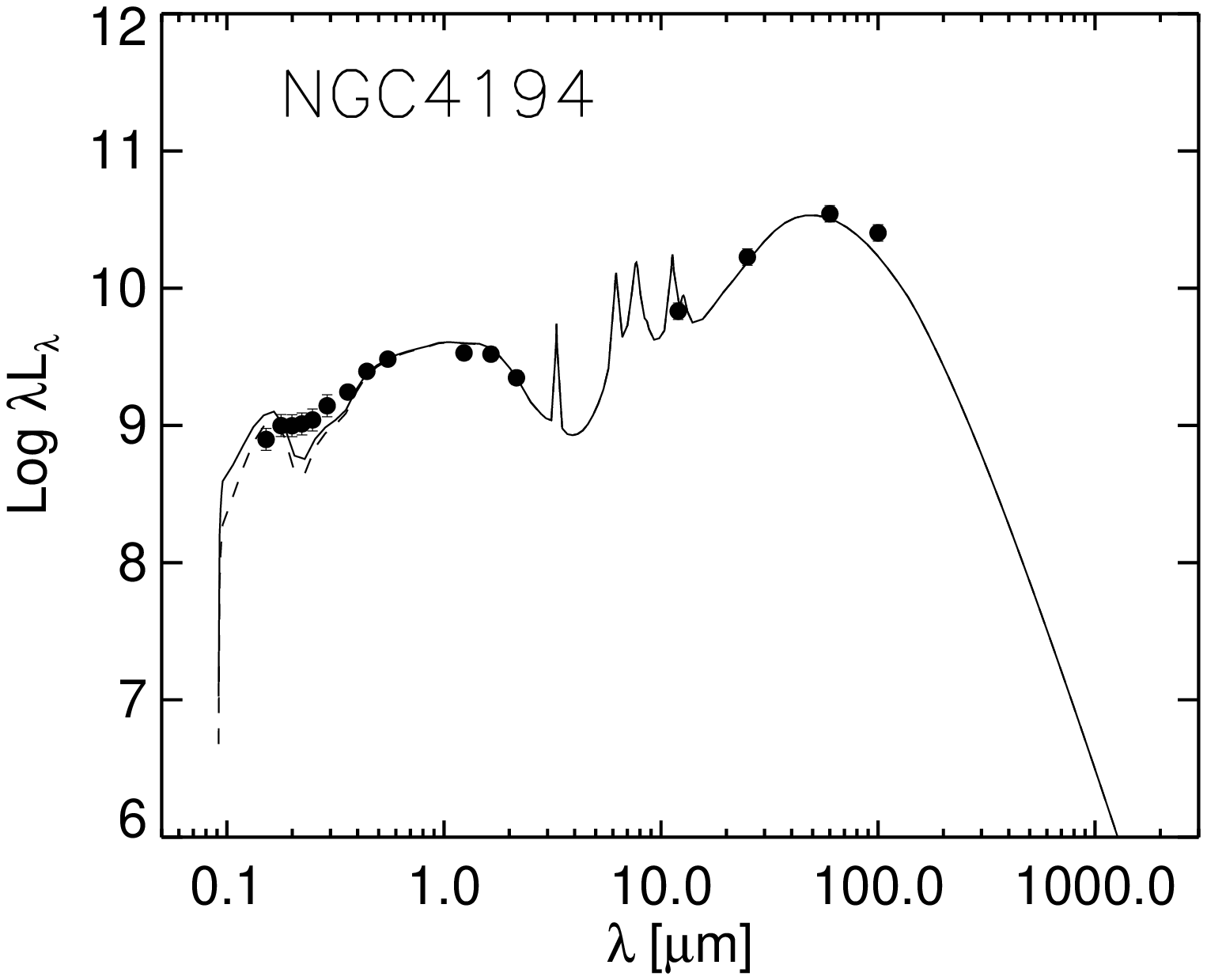}}
  \resizebox{7cm}{!}{\includegraphics{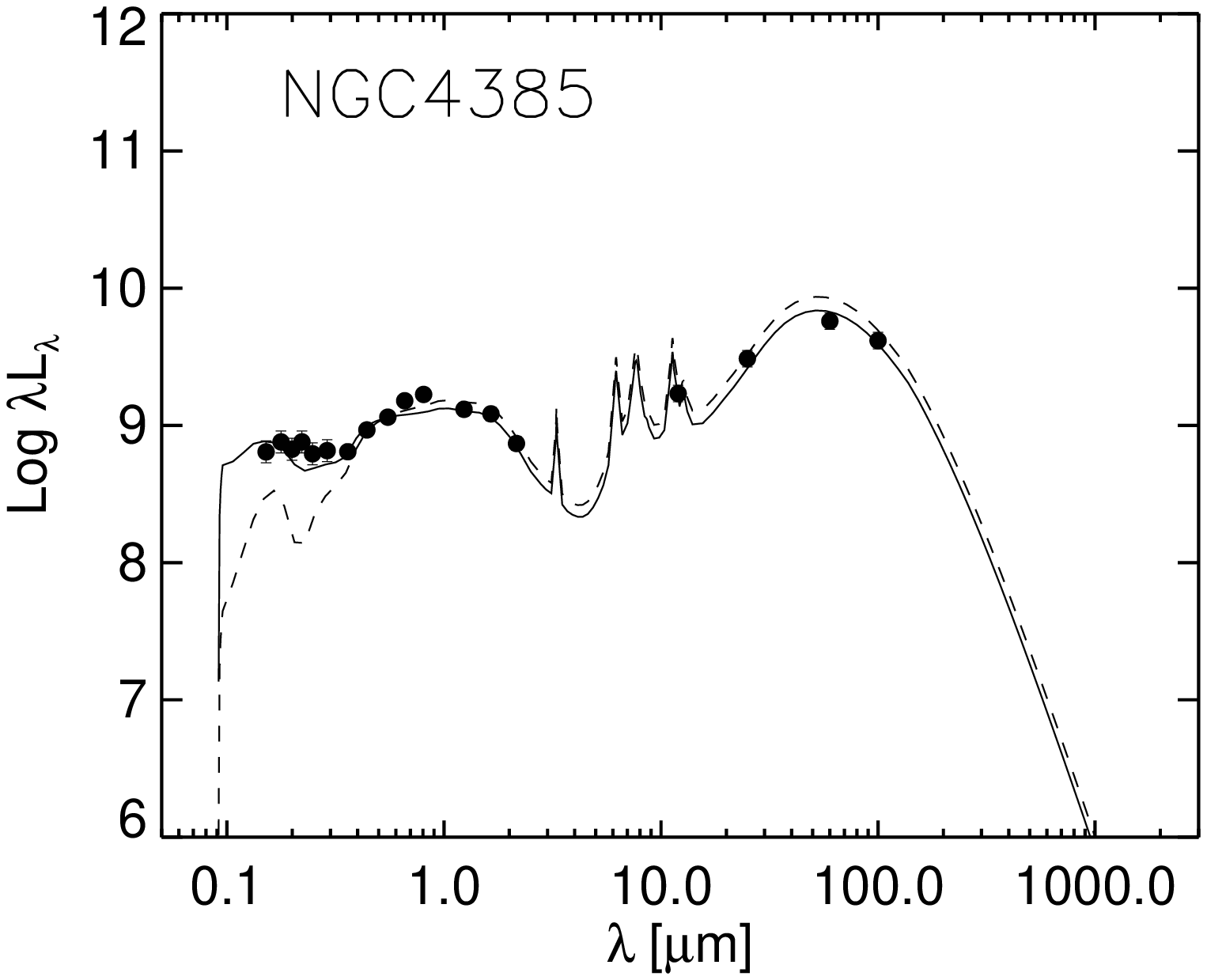}}
\caption{Fitting results for UV-selected starburst galaxies.
The dashed lines indicate the fitting results without
taking into account the photon leakage. In this case, we fit
the model SEDs to observations, except for FUV data.
The solid lines are the best-fitting models for observed SEDs including
the FUV data, considering the photon leakage. 
$\lambda L_{\lambda}$ is in solar unit.
}
\label{sb}
\end{figure*}

\addtocounter{figure}{-1}
\begin{figure*}
  \resizebox{7cm}{!}{\includegraphics{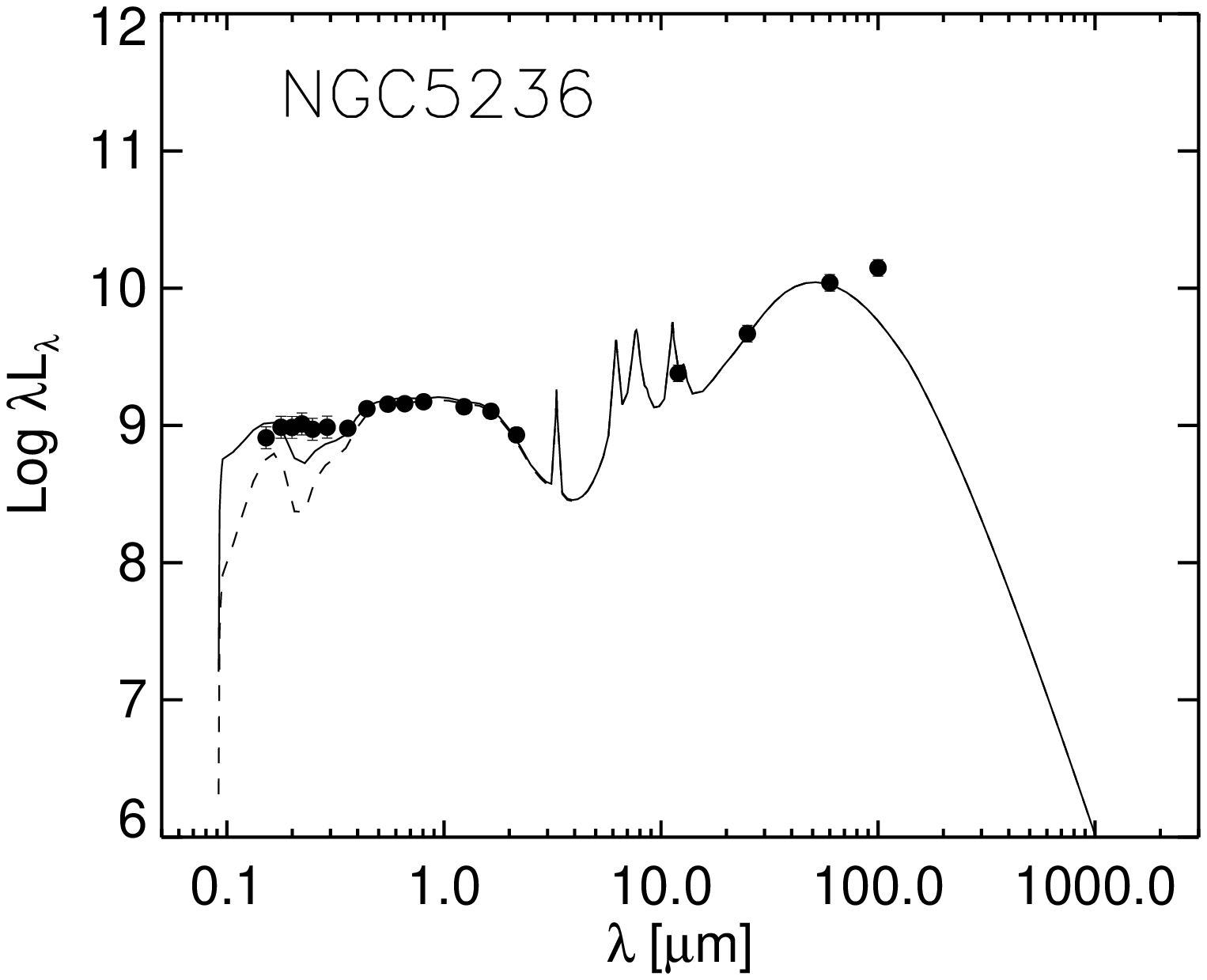}}
  \resizebox{7cm}{!}{\includegraphics{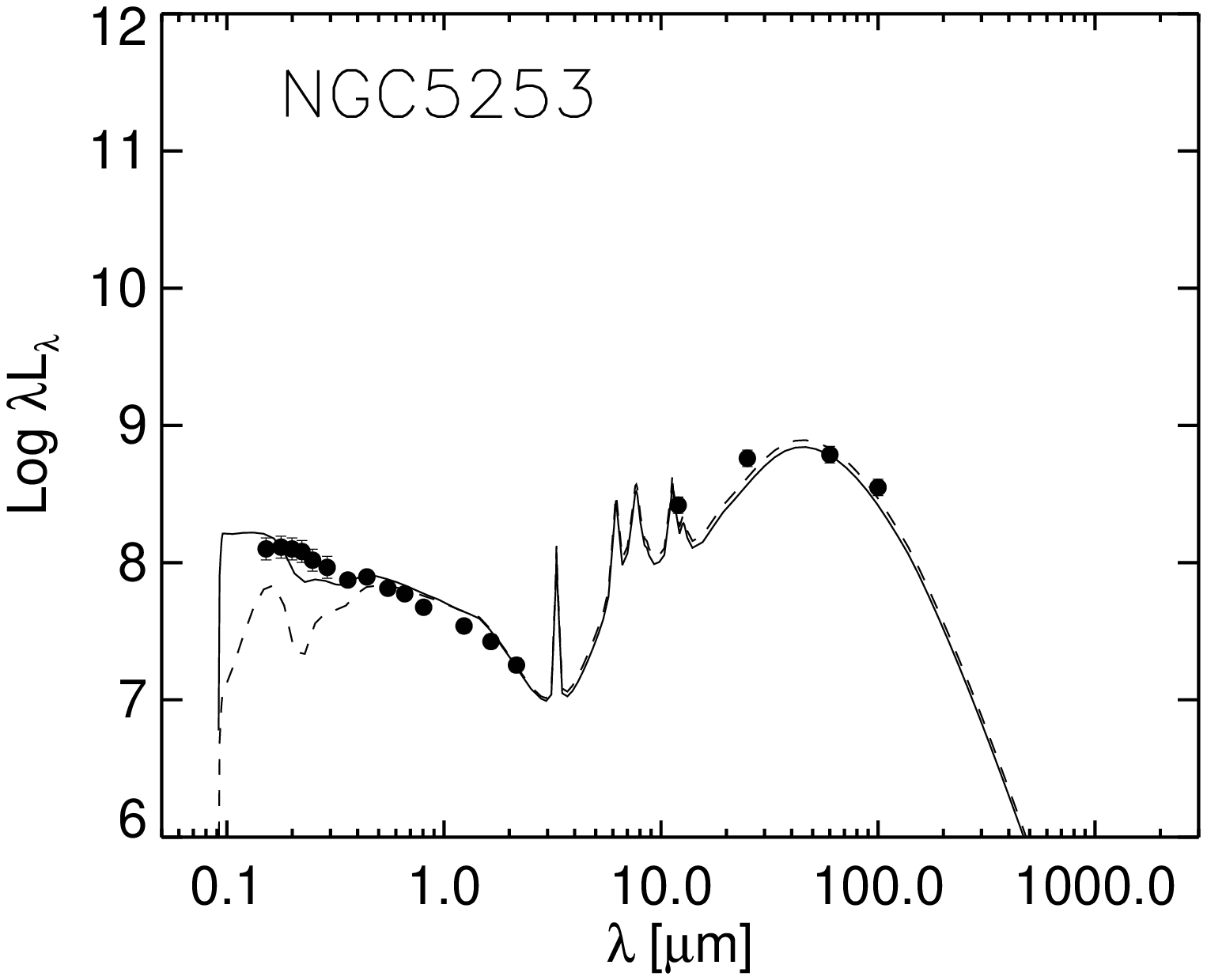}}
  \resizebox{7cm}{!}{\includegraphics{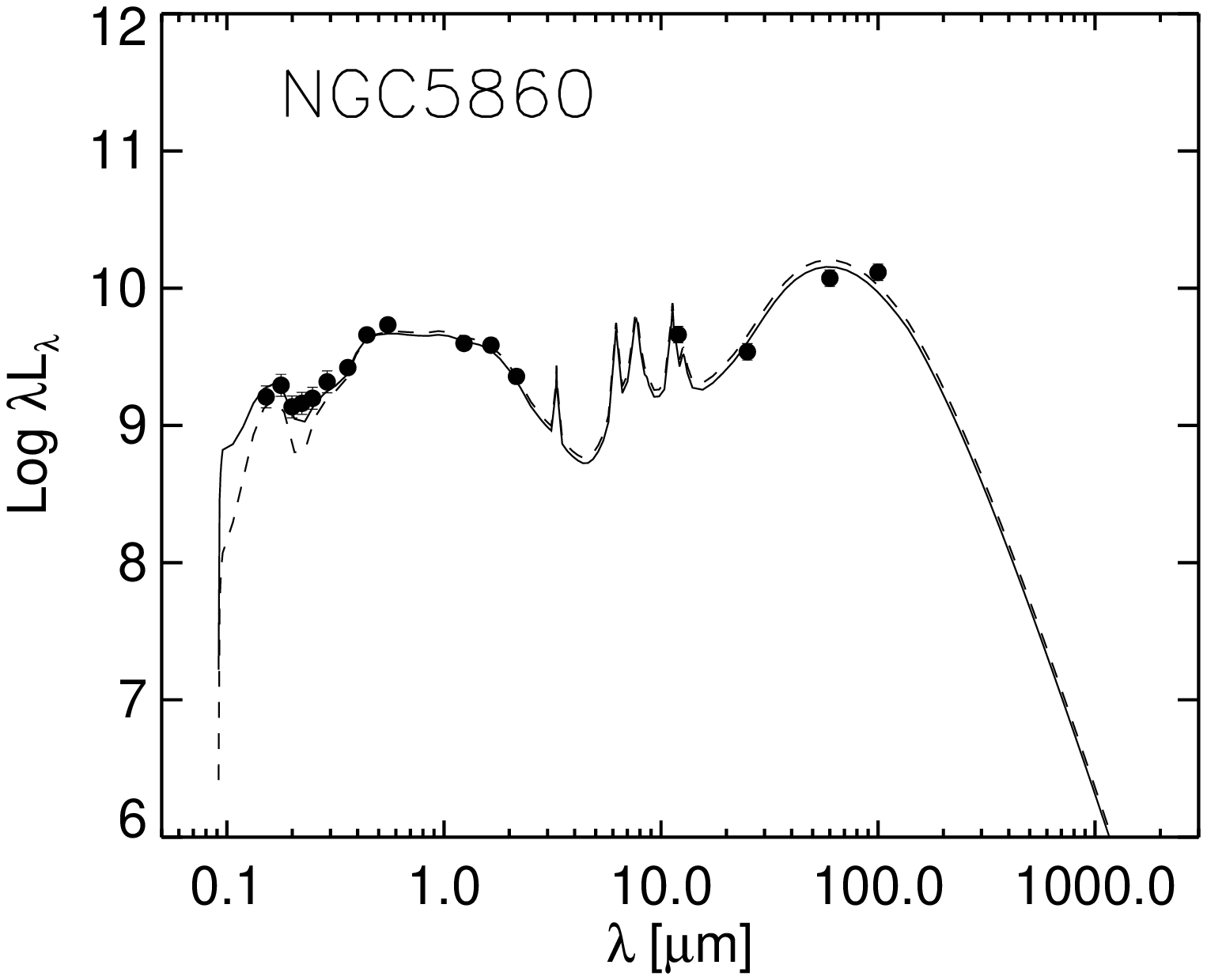}}
  \resizebox{7cm}{!}{\includegraphics{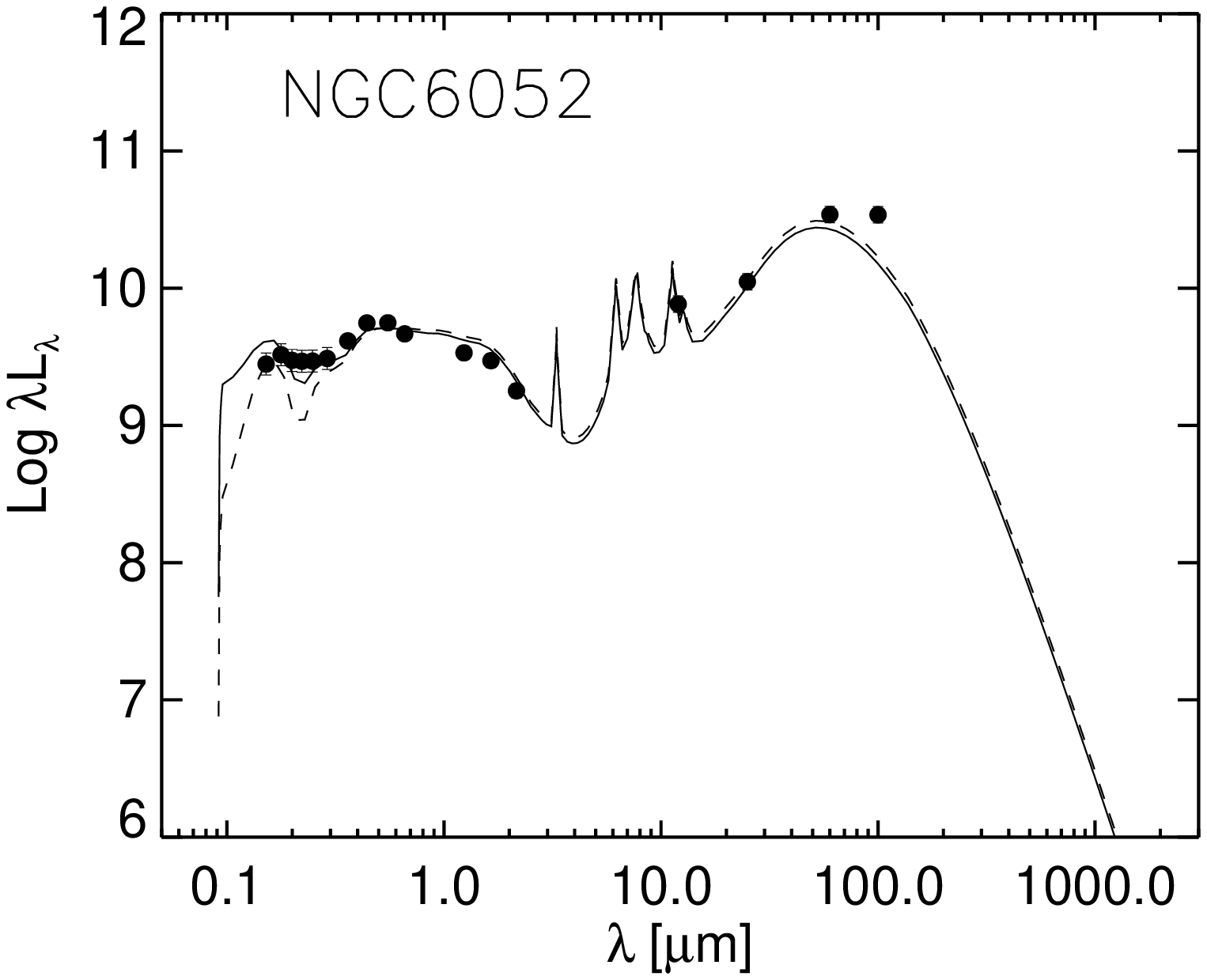}}
  \resizebox{7cm}{!}{\includegraphics{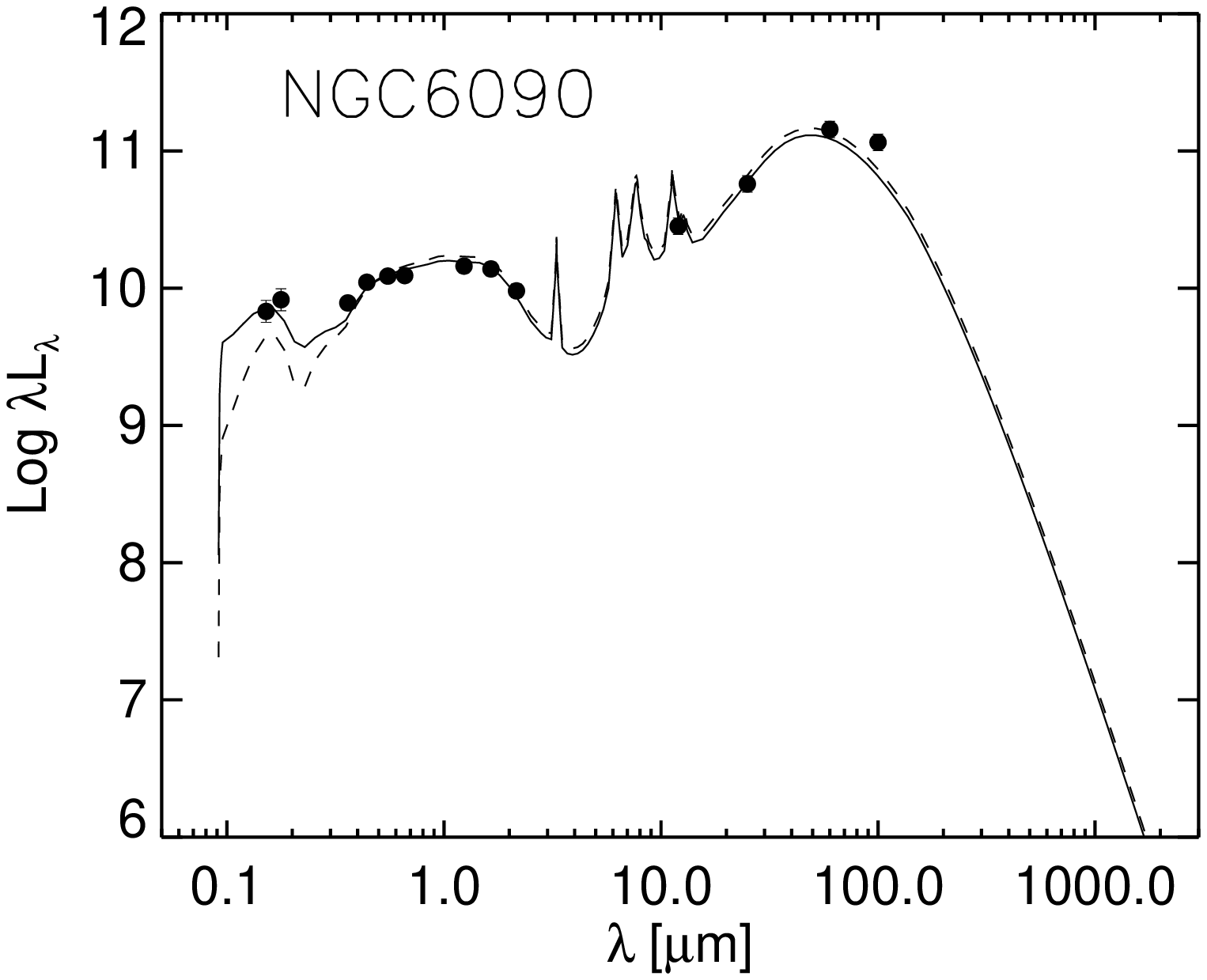}}
  \resizebox{7cm}{!}{\includegraphics{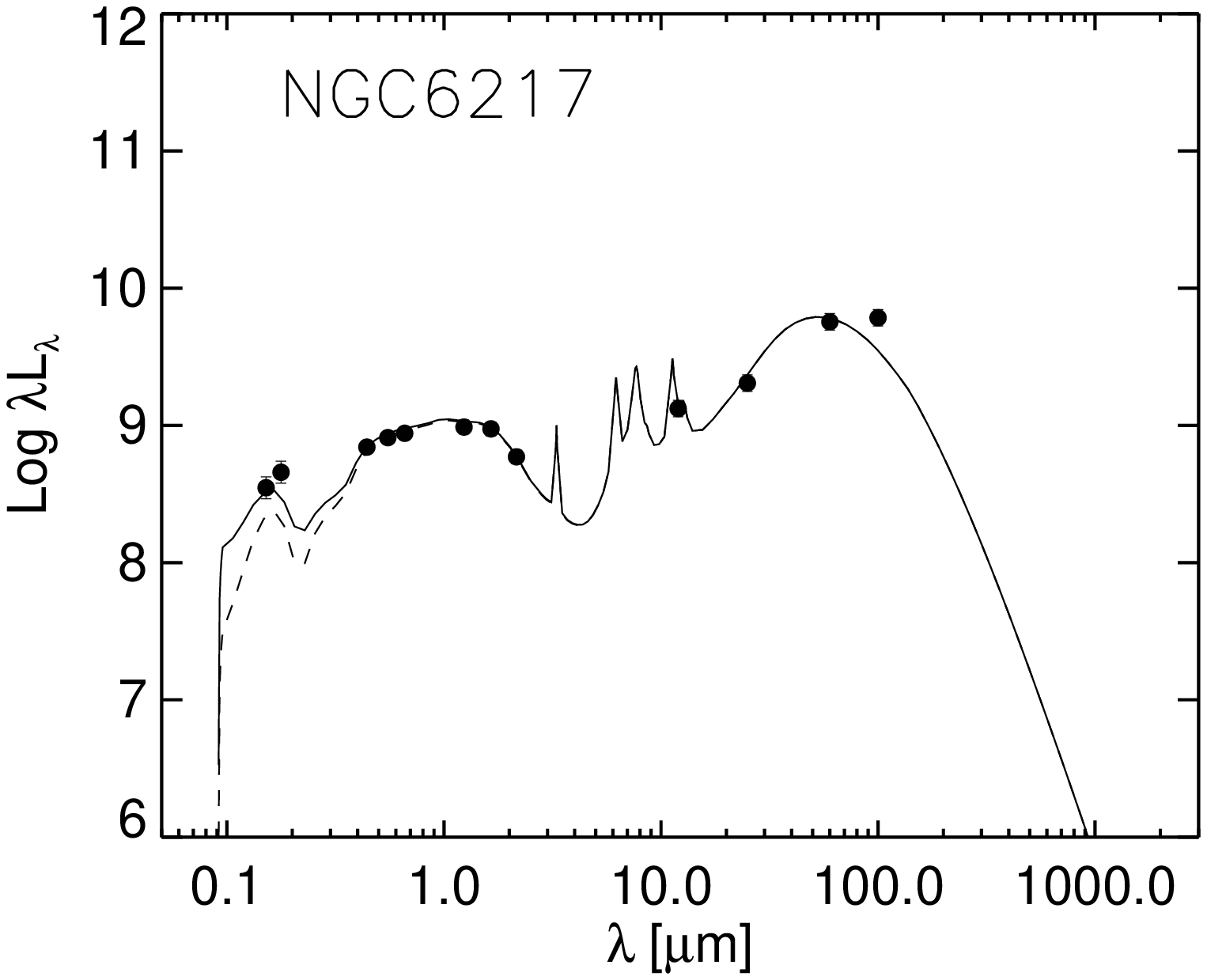}}
\caption{ { \it - continued.}
}
\end{figure*}

\addtocounter{figure}{-1}
\begin{figure*}
  \resizebox{7cm}{!}{\includegraphics{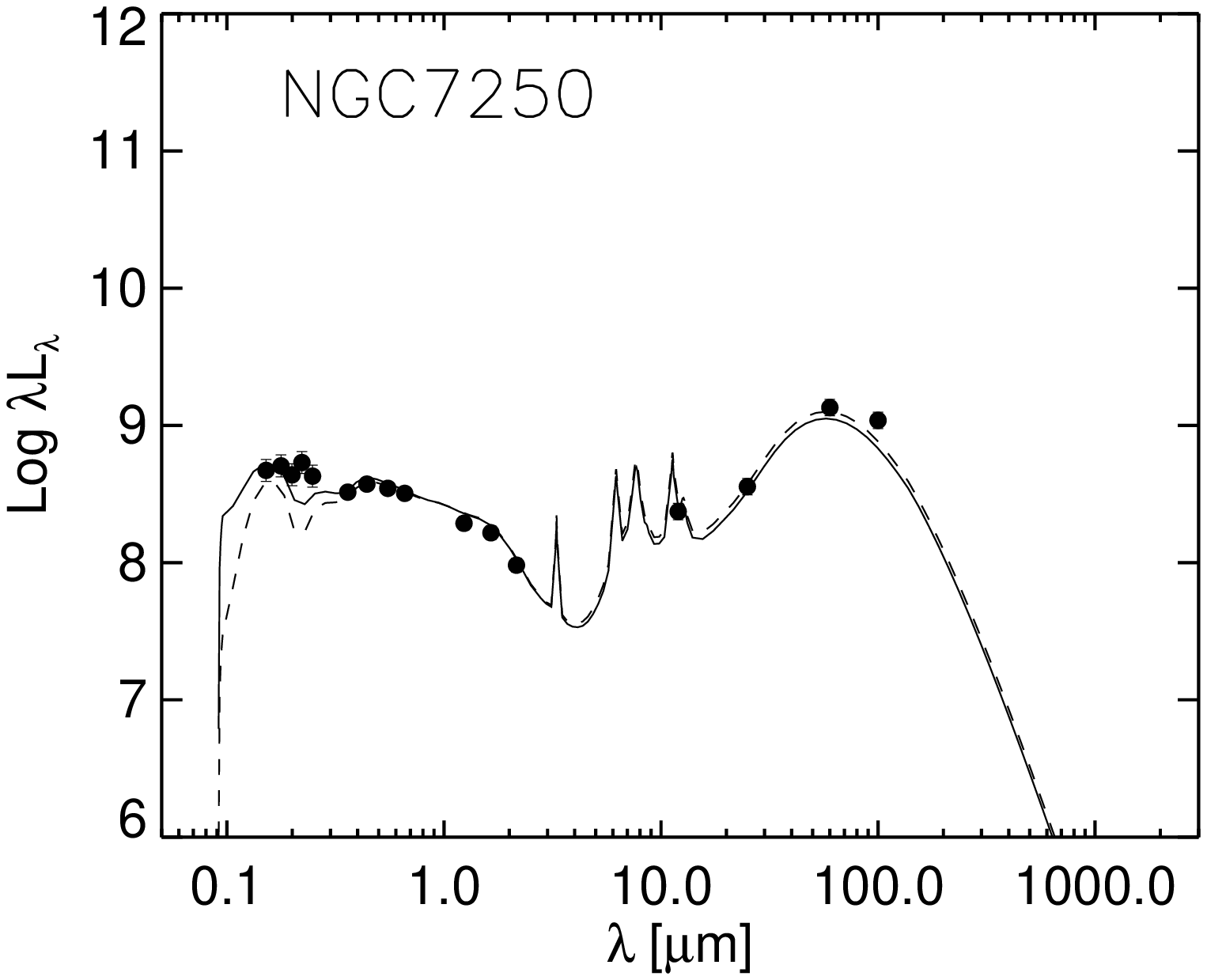}}
  \resizebox{7cm}{!}{\includegraphics{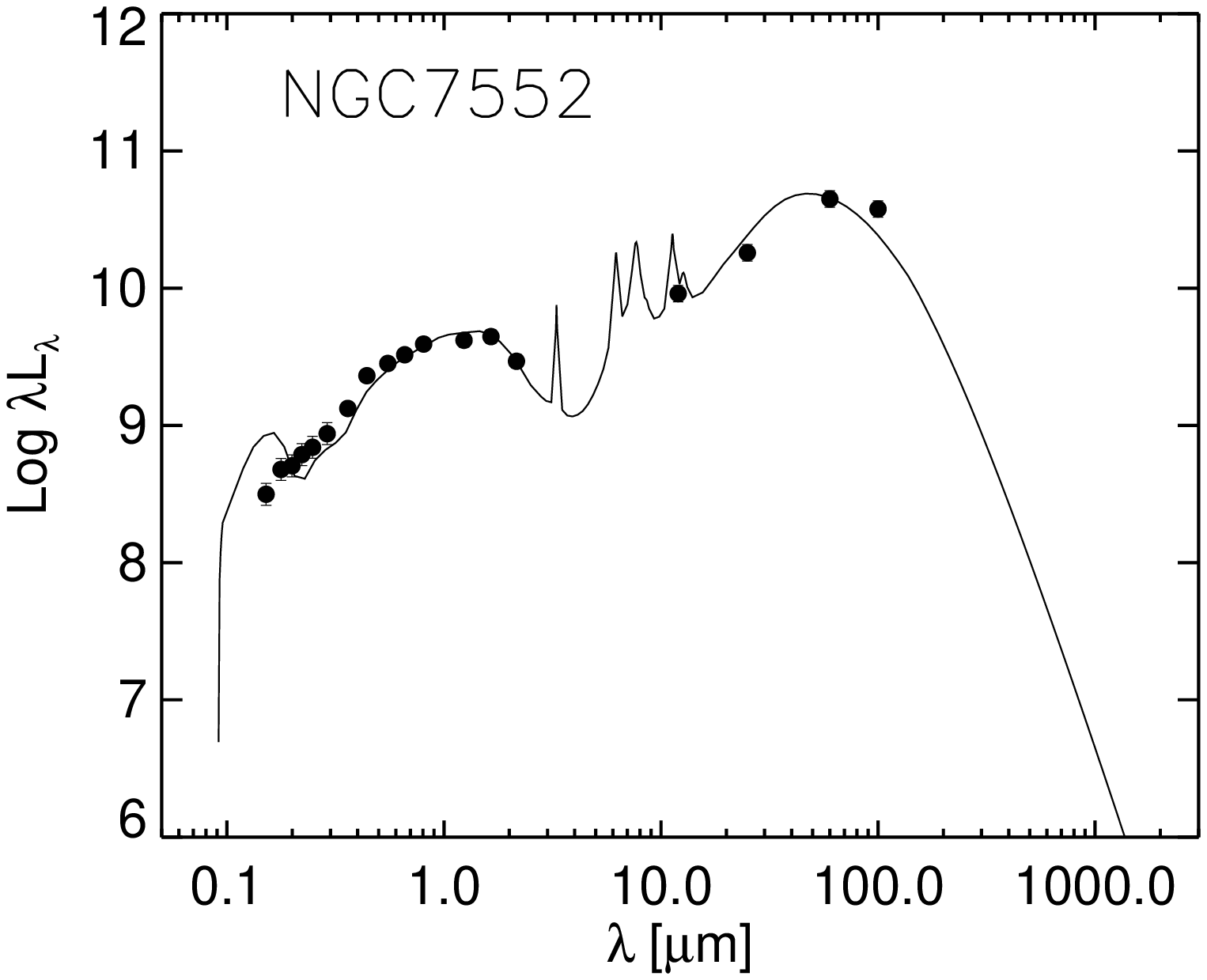}}
  \resizebox{7cm}{!}{\includegraphics{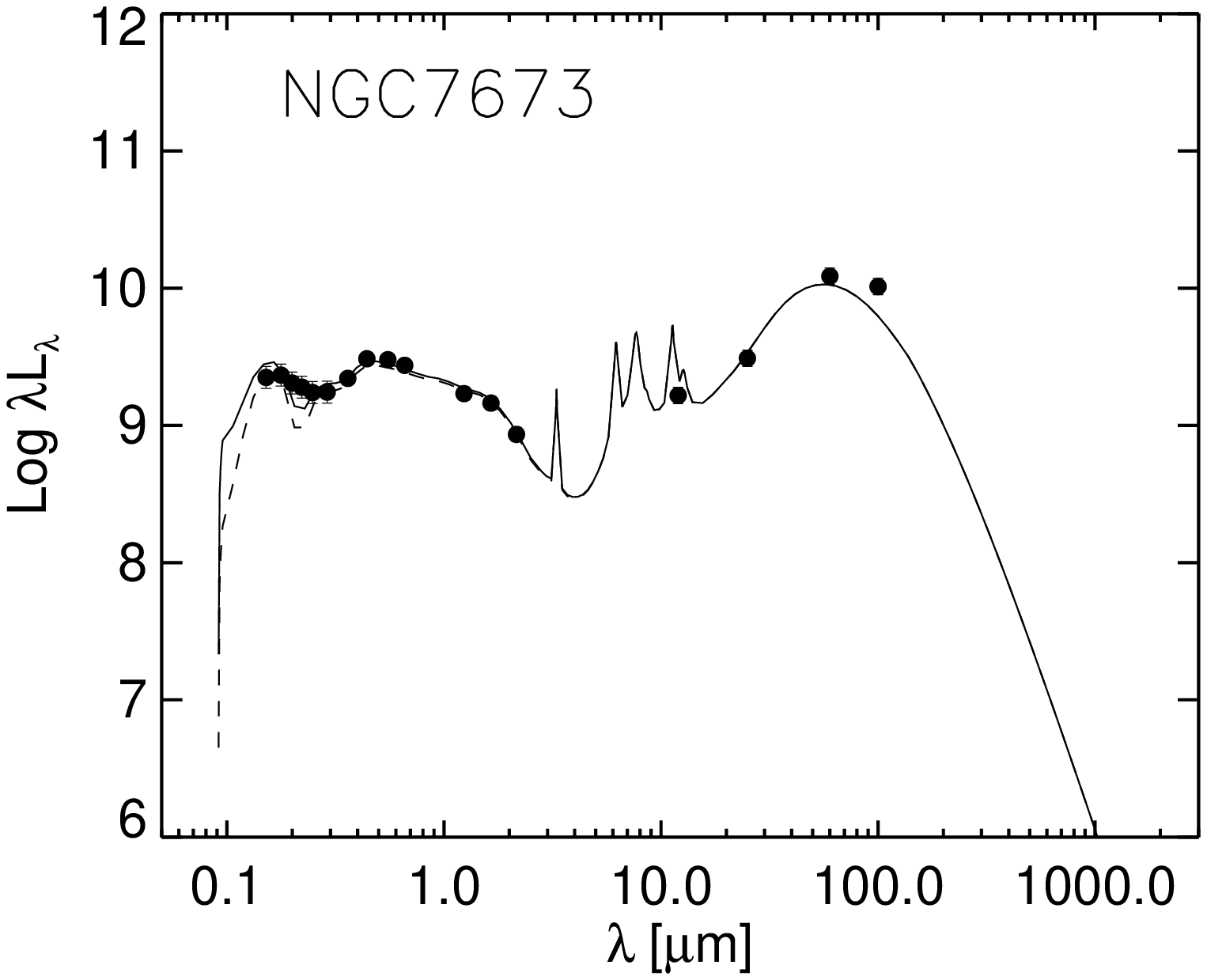}}
  \resizebox{7cm}{!}{\includegraphics{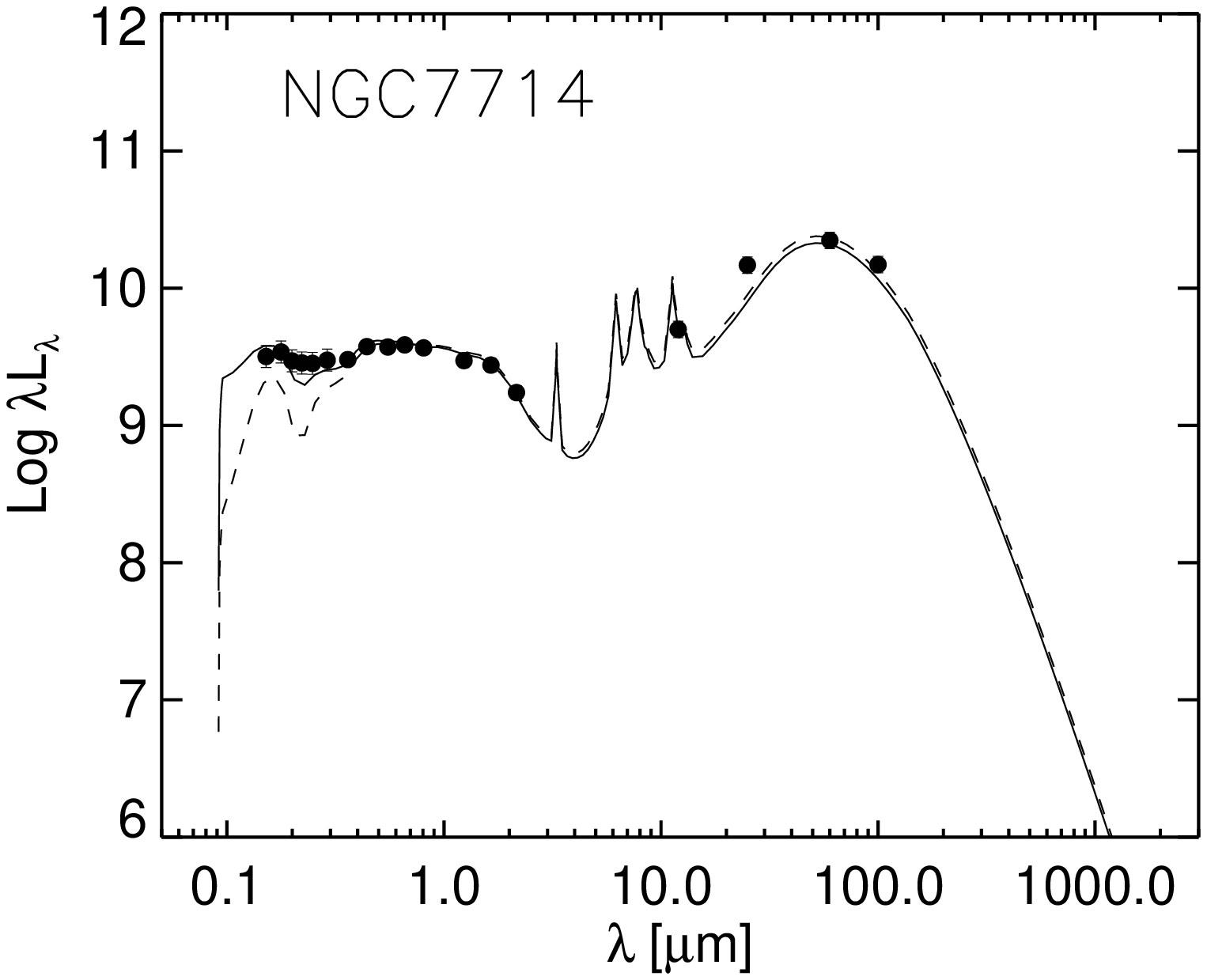}}
\caption{{\it - continued.}
}
\end{figure*}

We analyse the same sample of UVSBGs analysed by
Takagi et al. (1999). Photometric data from FUV to NIR are taken
from Gordon et al. (1997)
who presented a homogeneous sample of aperture-matched SEDs of $30$
starburst
galaxies. All of the sample was detected by the {\it IUE} satellite.
UV-selected starbursts are relatively bright in the UV 
and expected to be less attenuated by dust.
The aperture size of the FUV-NIR photometry is
$10^{\prime \prime} \times 20^{\prime \prime}$,
which is equal to $\sim 4.5$ kpc at their
median distance of $60$ Mpc.
The IRAS flux densities are taken from NED (NASA/IPAC Extragalactic
Database).
The IRAS fluxes are derived for the total galaxy,
while the UV-NIR photometric data are aperture limited in the
central region of starburst galaxies on average.
Nevertheless, the aperture correction does not change our main conclusions,
since the FIR emissions tend to concentrate at galactic centre
in the galaxies analysed here (Calzetti et al. 1995).
We fit the SEDs of 16 galaxies out of 30, which were detected in all
four IRAS bands (12$\mu$m, 25$\mu$m, 60$\mu$m and 100$\mu$m).

First, the best-fitting model is determined by the SED fitting from
the $U$-band to the FIR,
i.e., without FUV data obtained with the {\it IUE} satellite.
The fitting results from the $U$-band to the FIR are shown in Fig. \ref{sb}
with dashed lines. All the best-fitting models
result either in reasonable fits or in underestimates to the FUV data.

This discrepancy in the FUV flux can be reasonably explained by
the effect of photon leakage.
Although we consider a homogeneous distribution of dust, regions
with low gas density are naturally expected in the starbursts,
because of a cavity produced by successive supernova
explosions, for example. The effect of photon leakage is expected to be
more significant in the FUV compared with the longer wavelength region,
since the starburst population is intrinsically luminous in the FUV.
Therefore, the fitting results are consistent with our proposed picture
of the starburst region.

We estimate the effect of photon leakage as follows.
Although various levels of extinction for leaking photons are expected, 
depending on the degree of inhomogeneity in the dust distribution, 
the contribution of leaking photons from optically thiner 
regions could be much more significant. 
Therefore, for simplicity 
we assume that the extinction of leaking photons is negligible. 
We define the energy fraction of photons which leak out from
the stars directly through the starburst region, $f_{\mathrm{leak}}$, as:
\begin{equation}
F^{\mathrm{obs}}_\lambda =  (1-f_{\mathrm{leak}}) F_\lambda (\tau_\lambda)
+ f_{\mathrm{leak}} F_\lambda (\tau_\lambda=0),
\end{equation}
where $F^{\mathrm{obs}}_\lambda$ is the observed flux, and
$F_\lambda (\tau_\lambda)$ is the model flux for a given
optical depth without considering the photon leakage.
By using $F_\lambda (\tau_\lambda)$ and $F_\lambda (\tau_\lambda=0)$ for
each starburst galaxy derived from the $U$-band to FIR SED (i.e., without
FUV data), we determine the most plausible value of $f_{\mathrm{leak}}$
by the least square method by using all photometric data including the FUV;
i.e., in this fitting process, we use $f_{\mathrm{leak}}$ as a free parameter. 
Also, $M_T$ is re-adjusted to give the least square value for the SED with 
the leakage correction, which is necessary to derive a 
more accurate bolometric luminosity, and therefore SFR. 

The best-fitting models with the photon leakage are indicated
by solid lines in Fig. \ref{sb}. As a whole, these models show a good
agreement with the observed SEDs.
For most of the sample galaxies, the SEDs in the FUV 
can be reproduced with a $f_{\mathrm{leak}}$ less than 0.1.
Note that the correction of SED due to the photon leakage 
effects only the FUV.
In Table 1, the best-fitting parameters
are given together with the resulting $L_{bol}$, $M_*$, $M_D$, SFR and
$\tau^{\mathrm{eff}}_V$.
 From the SED fittings with our adopted evolutionary time-scale 
$t_0=100$ Myr, most of the derived starburst ages are within $200- 400$ Myr.
As we already noticed, starbursts are most luminous during
the ages of $t/t_0 \simeq 2-4 $. Therefore, the derived ages
with the SED fittings are reasonable for the starbursts.
For all the sample but three, we found that 
the MW extinction curve is most suitable for UVSBGs. 
Note that the absorption feature at 2175\AA\ is not used to determine 
the extinction curve. The MIR colours are important for the resulting 
extinction curve, since the difference in the relative fraction of 
dust constituents can effect the MIR SED in which very small 
graphite grains and PAHs play significant role. 
The mean value of $\Theta$ is 1.6, corresponding to 
$\tau_V^{\mathrm{eff}} = 1.2$ for $t/t_0=2.0$ in the case of 
the MW extinction curve. 

Some galaxies, such as NGC5236, NGC6052 and NGC6217, show a notable
excess of 100 $\mu$m flux compared to the model SEDs. As we noted,
the aperture size adopted for IRAS data is larger than those of the FUV --
NIR data. Therefore, it is likely that the FIR emission
from cold dust, heated by the interstellar radiation field generated by
the underlying stellar populations, contributes to the 100 $\mu$m flux.

For IC214, the model with the photon leakage still underestimates the 
FUV fluxes. Since this galaxy along with NGC1614 is the faintest 
in FUV among the UVSBG sample, the signal-to-noise ratio of the $IUE$ 
observation is low (3.2 for 1515\AA; Kinney et al. 1993). 
A more sensitive observation would be necessary for a detailed discussion 
on the origin of this discrepancy.

\begin{table*}
\begin{center}
Table 1.\hspace{4pt}
Fitting results for UVSBGs \\
\end{center}
\tabcolsep=3pt
\vspace{6pt}
\begin{tabular*}{\textwidth}{@{\hspace{\tabcolsep}
\extracolsep{\fill}}p{4pc}rcccllllcc}
\hline\hline\\ [-10pt]
\multicolumn{1}{c}{name} &
\multicolumn{1}{c}{$t/t_0$} &
\multicolumn{1}{c}{$\Theta$} &
\multicolumn{1}{c}{EC} &
\multicolumn{1}{c}{$f_{\mathrm{leak}}$} &
\multicolumn{1}{c}{$M_T$} &
\multicolumn{1}{c}{$L_{bol}$} &
\multicolumn{1}{c}{$M_*$} &
\multicolumn{1}{c}{$M_D$} &
\multicolumn{1}{c}{SFR} &
\multicolumn{1}{c}{$\tau_V^{\mathrm{eff}}$} \\
\multicolumn{1}{c}{} &
\multicolumn{1}{c}{} &
\multicolumn{1}{c}{} &
\multicolumn{1}{c}{} &
\multicolumn{1}{c}{} &
\multicolumn{1}{c}{[M$_\odot$]} &
\multicolumn{1}{c}{[L$_\odot$]} &
\multicolumn{1}{c}{[M$_\odot$]} &
\multicolumn{1}{c}{[M$_\odot$]} &
\multicolumn{1}{c}{[M$_\odot$ yr$^{-1}$]}&
\multicolumn{1}{c}{}  \\
\hline \\[-8pt]
IC214    & 4.0 & 0.8 & MW &0.012 & 2.2\ $10^{10}$ & 2.1\ $10^{11}$ & 2.0\
$10^{10}$  & 3.8\ $10^7$  &25.2 & 2.2\\
NGC1140  & 3.0  & 1.4  &  LMC  &0.165  & 3.5\ $10^8$  & 4.9\ $10^9$  &
2.7\ $10^8$  & 1.1\ $10^6$  &0.7 & 0.4\\
NGC1569  & 0.7  & 2.6 &  SMC  &0.101  & 8.9\ $10^7$  & 1.4\ $10^9$  & 1.3\
$10^7$  & 1.7\ $10^5$  &0.3 & 0.3\\
NGC1614  & 2.0 & 0.8 & LMC &0.002 &1.8\ $10^{10}$  & 3.5\ $10^{11}$ & 1.0\
$10^{10}$ & 6.1\ $10^7$  &57.7& 2.5\\
NGC4194  & 2.0  & 1.5  &  MW  &0.004  & 3.7\ $10^9$  & 7.1\ $10^{10}$ &
2.1\ $10^9$  & 9.1\ $10^6$  &11.7& 1.6\\
NGC4385  & 3.0  & 1.3  &  MW  &0.054  & 1.1\ $10^9$  & 1.5\ $10^{10}$ &
8.1\ $10^8$  & 2.4\ $10^6$  &2.1 & 1.3\\
NGC5236  & 2.0  & 1.6  &  MW  &0.028  & 1.2\ $10^9$  & 2.4\ $10^{10}$ &
6.9\ $10^8$  & 3.0\ $10^6$  &3.9 & 1.2\\
NGC5253  & 0.3  & 2.5  &  MW  &0.084  & 2.3\ $10^8$  & 1.4\ $10^9$  & 7.9\
$10^6$  & 7.3\ $10^4$  &0.5 & 0.9\\
NGC5860  & 4.0  & 1.3  &  MW  &0.035  & 3.6\ $10^9$  & 3.4\ $10^{10}$ &
3.2\ $10^9$  & 6.2\ $10^6$  &4.1 & 0.9\\
NGC6052  & 2.0  & 1.8  &  MW  &0.038  & 3.2\ $10^9$  & 6.0\ $10^{10}$ &
1.8\ $10^9$  & 7.7\ $10^6$  &10.0 & 0.9\\
NGC6090  & 2.0  & 1.5  &  MW  &0.017 & 1.4\ $10^{10}$ & 2.7\ $10^{11}$&
7.9\ $10^9$  & 3.5\ $10^7$  &45.0 & 1.5\\
NGC6217  & 3.0  & 1.3  &  MW  &0.013  & 9.6\ $10^8$ & 1.3\ $10^{10}$  &
7.3\ $10^8$  & 2.2\ $10^6$  &1.9 & 1.4\\
NGC7250  & 2.0  & 2.4  &  MW  &0.088  & 1.5\ $10^8$  & 2.8\ $10^9$  & 8.1\
$10^7$  & 3.5\ $10^5$  &0.4& 0.5\\
NGC7552  & 2.0  & 1.3  &  MW  &$<0.001$  & 4.6\ $10^9$ & 8.8\ $10^{10}$  &
2.6\ $10^9$  & 1.1\ $10^7$  &14.5 & 2.1\\
NGC7673  & 2.0  & 2.2  &  MW  &0.032  & 1.3\ $10^9$ & 2.5\ $10^{10}$  &
7.3\ $10^8$  & 3.2\ $10^6$  &4.1 & 0.6\\
NGC7714  & 2.0  & 1.8  &  MW  &0.057  & 2.5\ $10^9$ & 4.7\ $10^{10}$  &
1.4\ $10^9$  & 5.9\ $10^6$  &7.7 & 0.9\\
\hline \\
\end{tabular*}
\vspace{6pt}
\par\noindent
  \label{sb-table}
\end{table*}

\subsection {ULIRGs}

\begin{figure*}
  \resizebox{7cm}{!}{\includegraphics{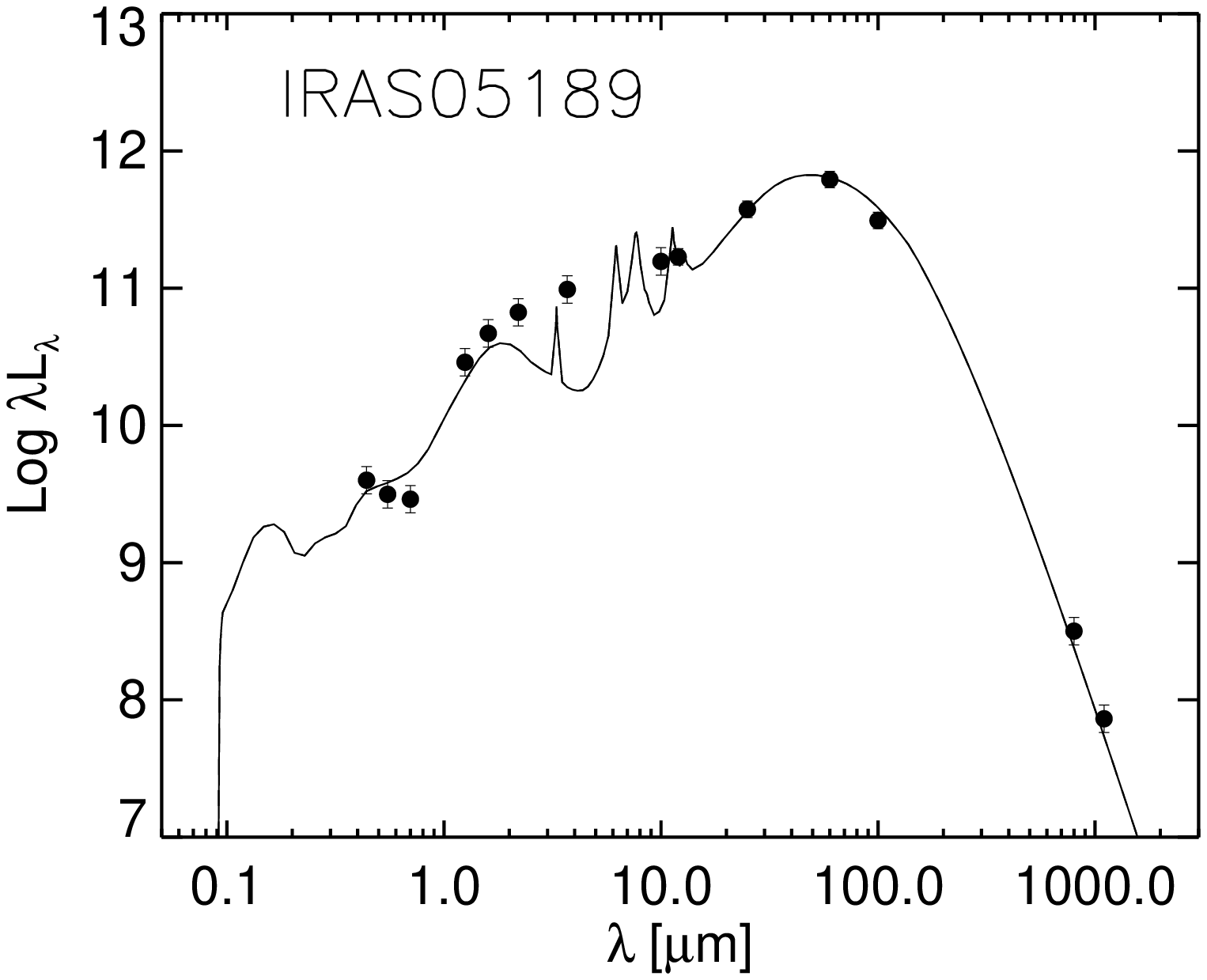}}
  \resizebox{7cm}{!}{\includegraphics{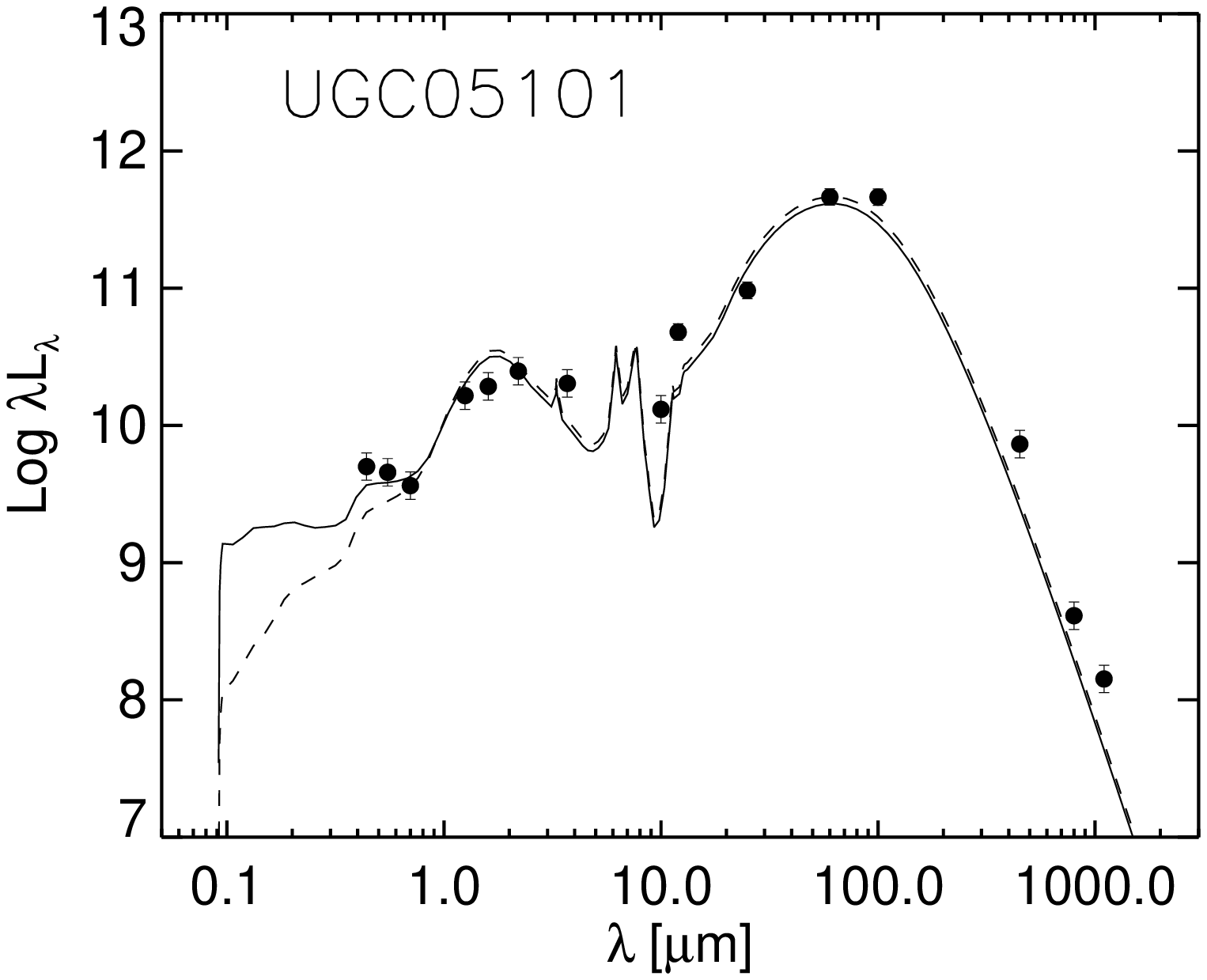}}
  \resizebox{7cm}{!}{\includegraphics{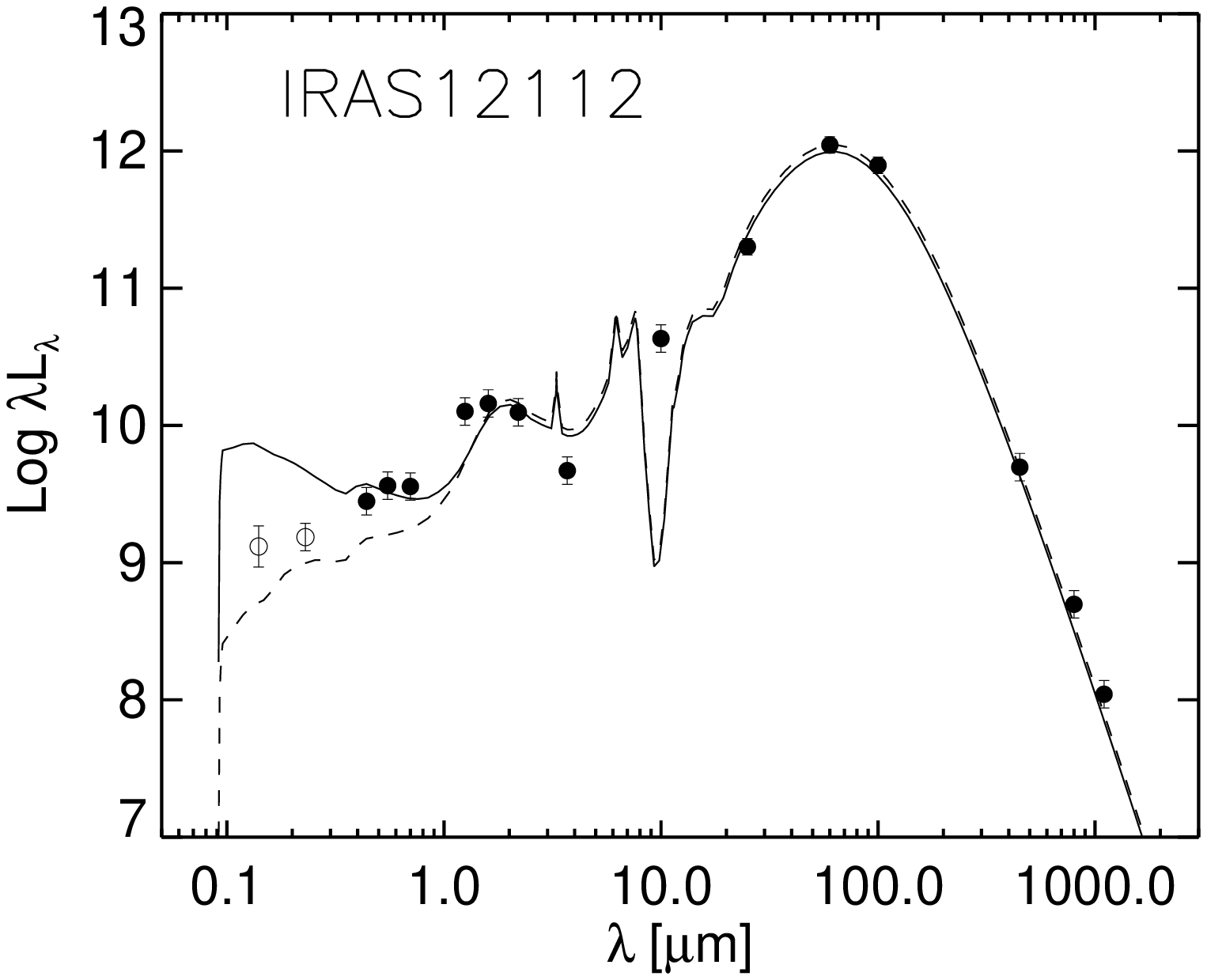}}
  \resizebox{7cm}{!}{\includegraphics{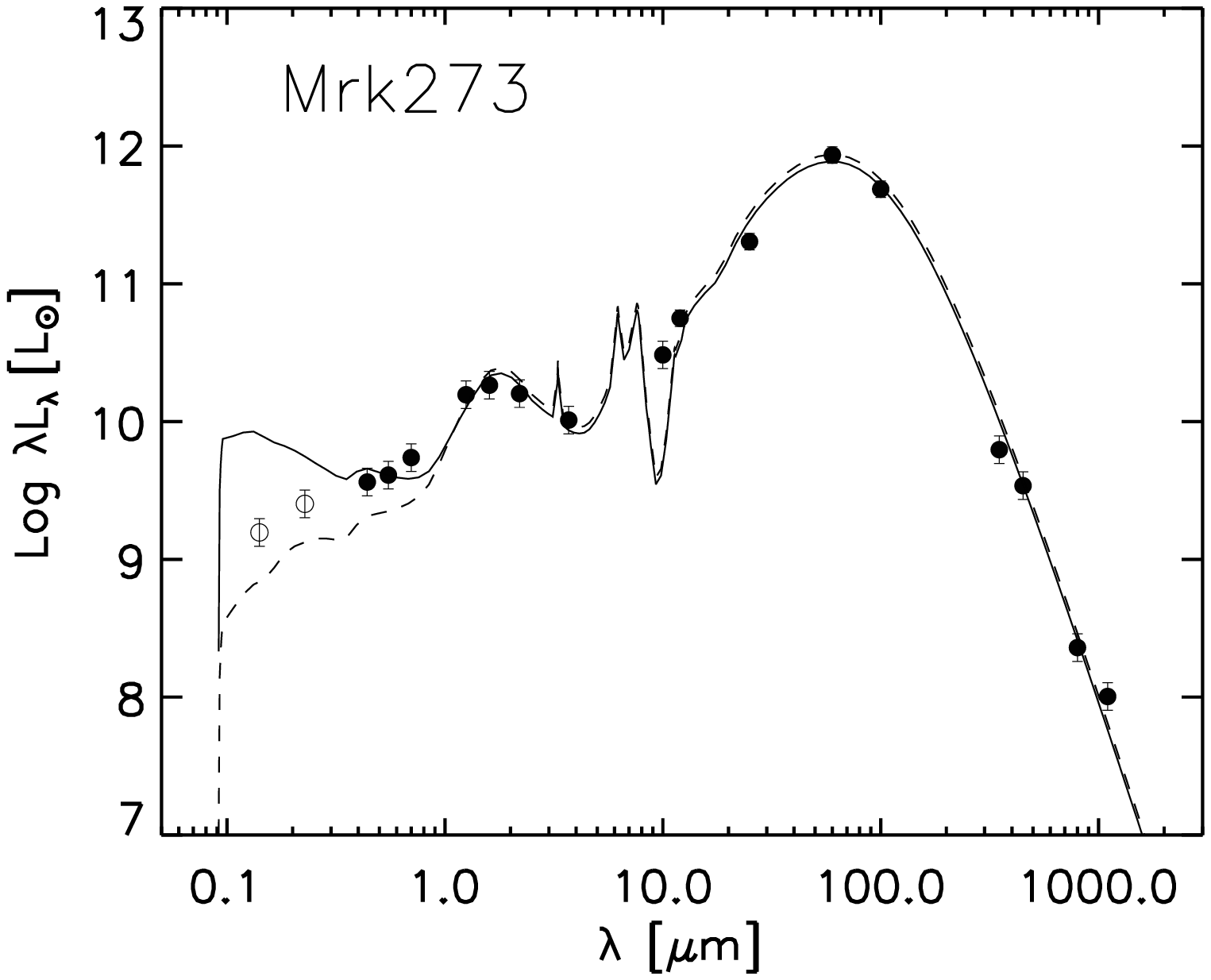}}
  \resizebox{7cm}{!}{\includegraphics{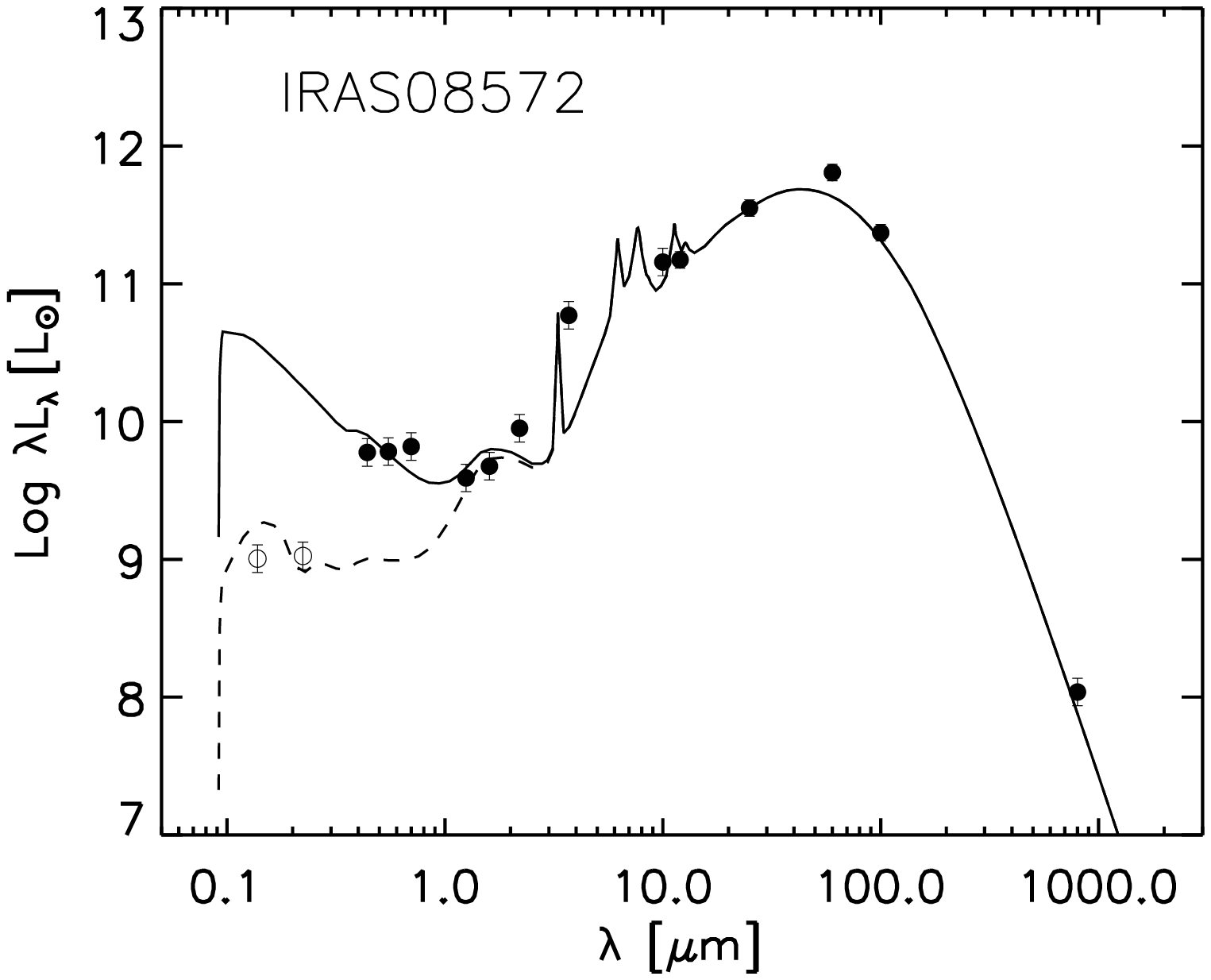}}
  \resizebox{7cm}{!}{\includegraphics{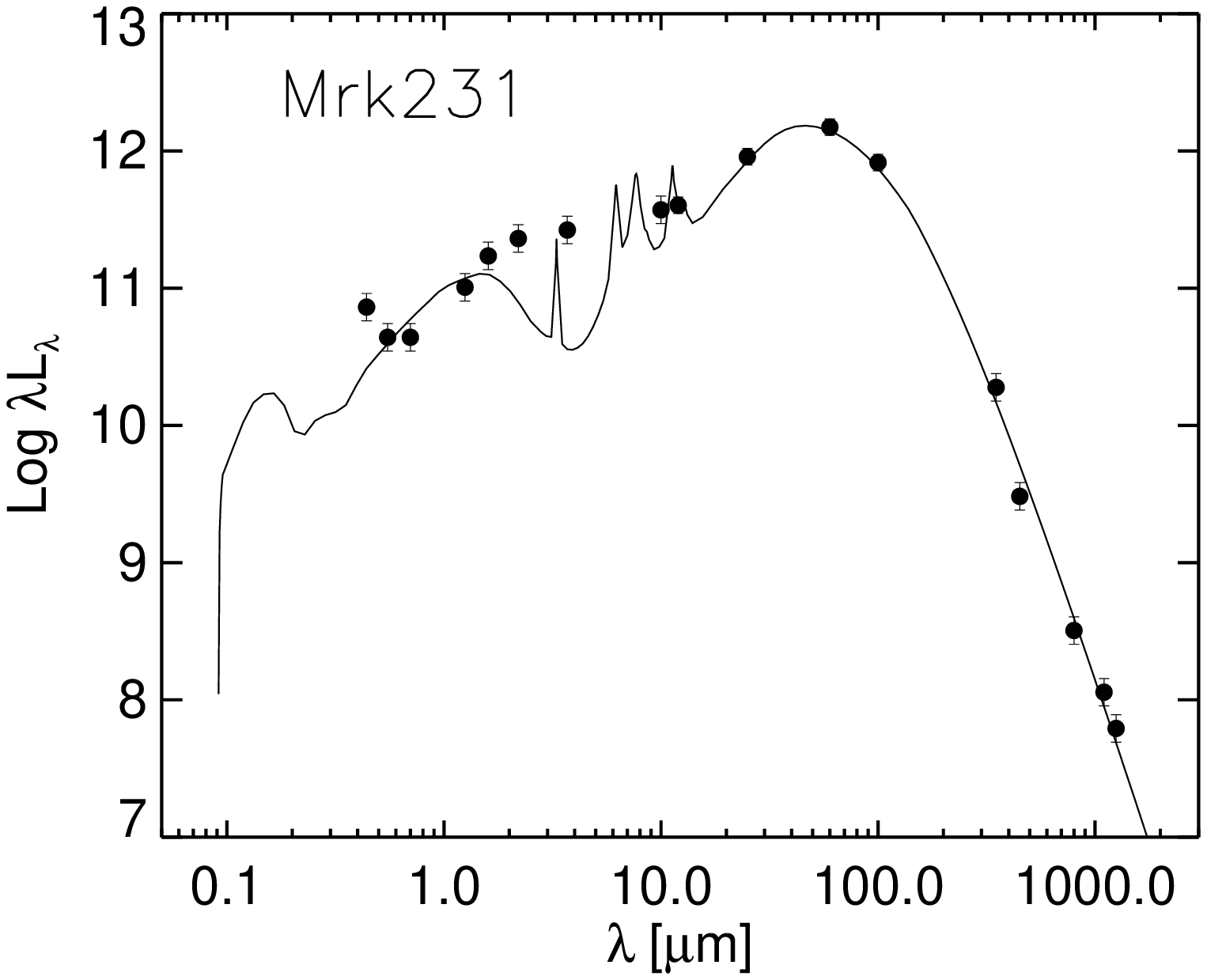}}
\caption{The same as Fig. \ref{sb}, but for ultraluminous
infrared galaxies. Open circles for IRAS12112, IRAS08572, 
IRAS15250, Arp220 and IRAS22491
indicate the photometric data, obtained by the Hubble Space
Telescope (Trentham, Kormendy \& Sanders 1999; Goldader et al. 2002).
}
  \label{ulirg}
\end{figure*}

\addtocounter{figure}{-1}
\begin{figure*}
  \resizebox{7cm}{!}{\includegraphics{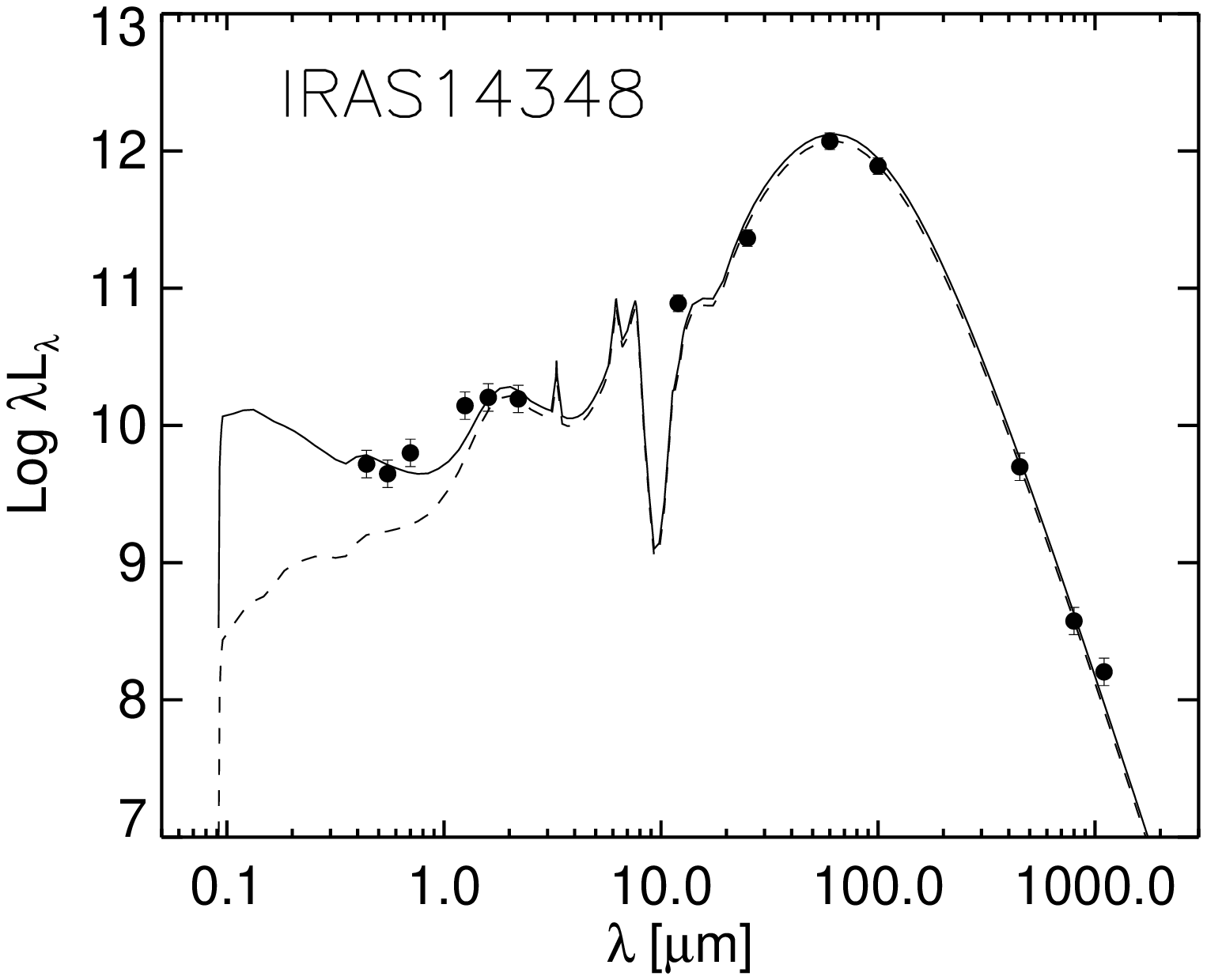}}
  \resizebox{7cm}{!}{\includegraphics{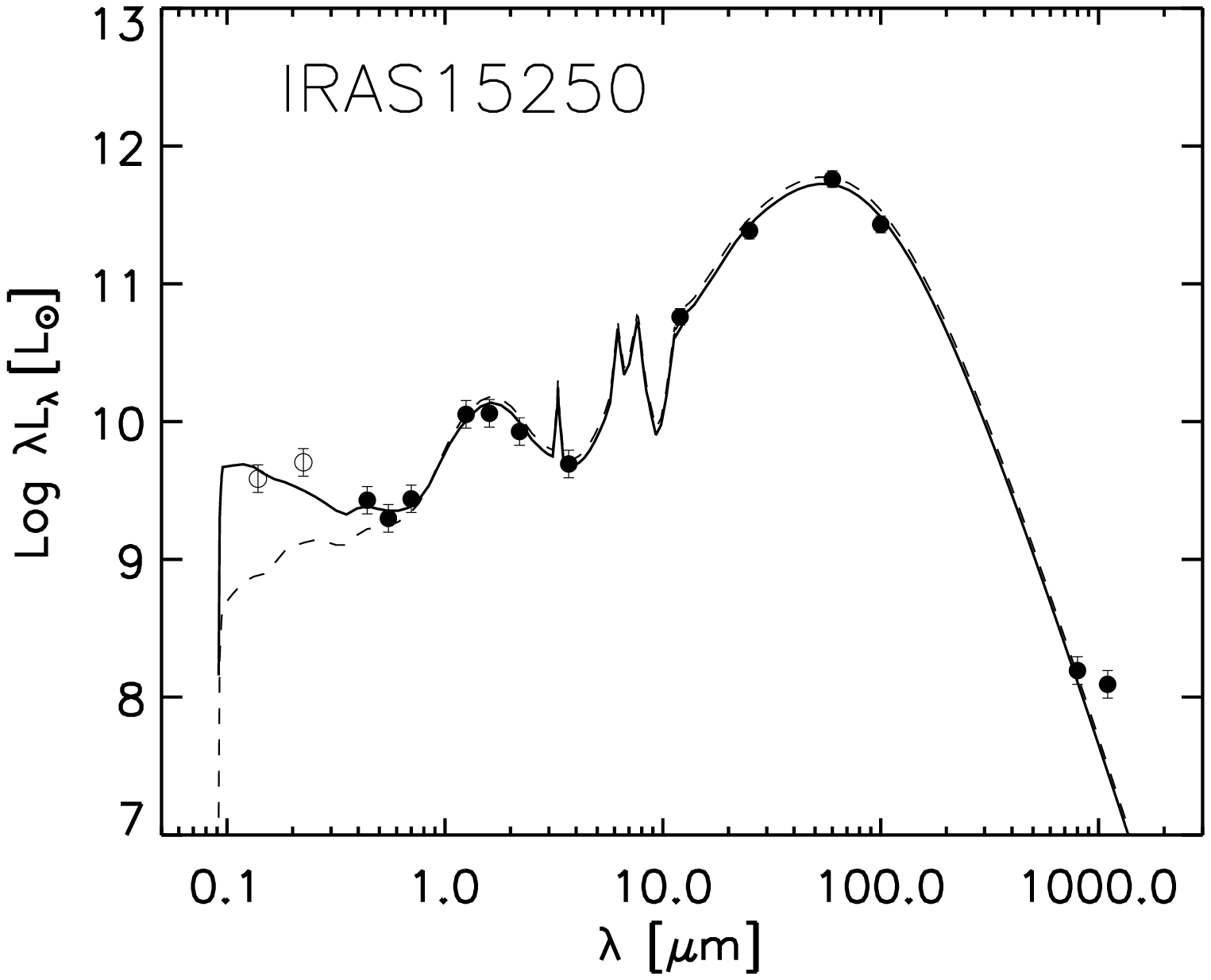}}
  \resizebox{7cm}{!}{\includegraphics{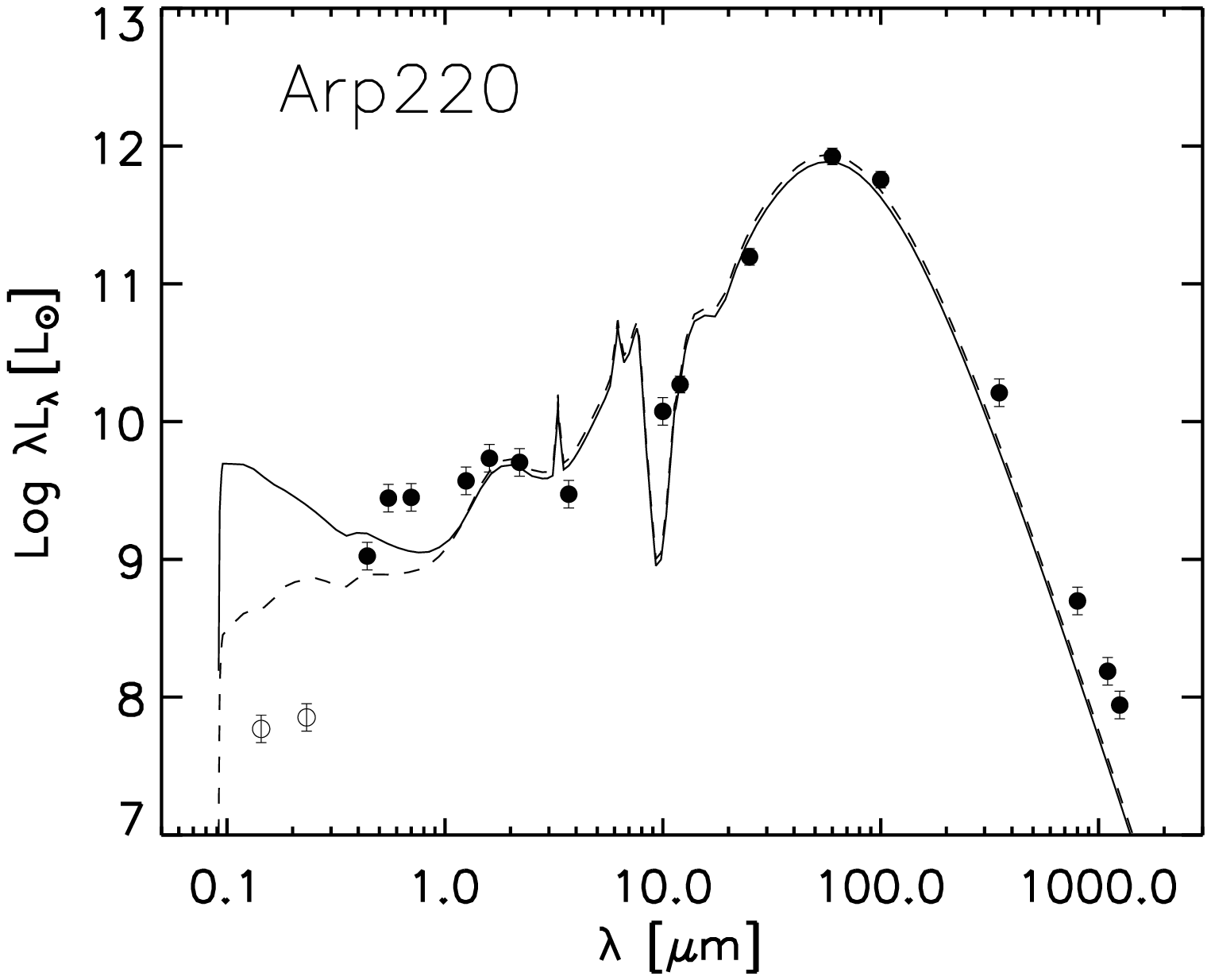}}
  \resizebox{7cm}{!}{\includegraphics{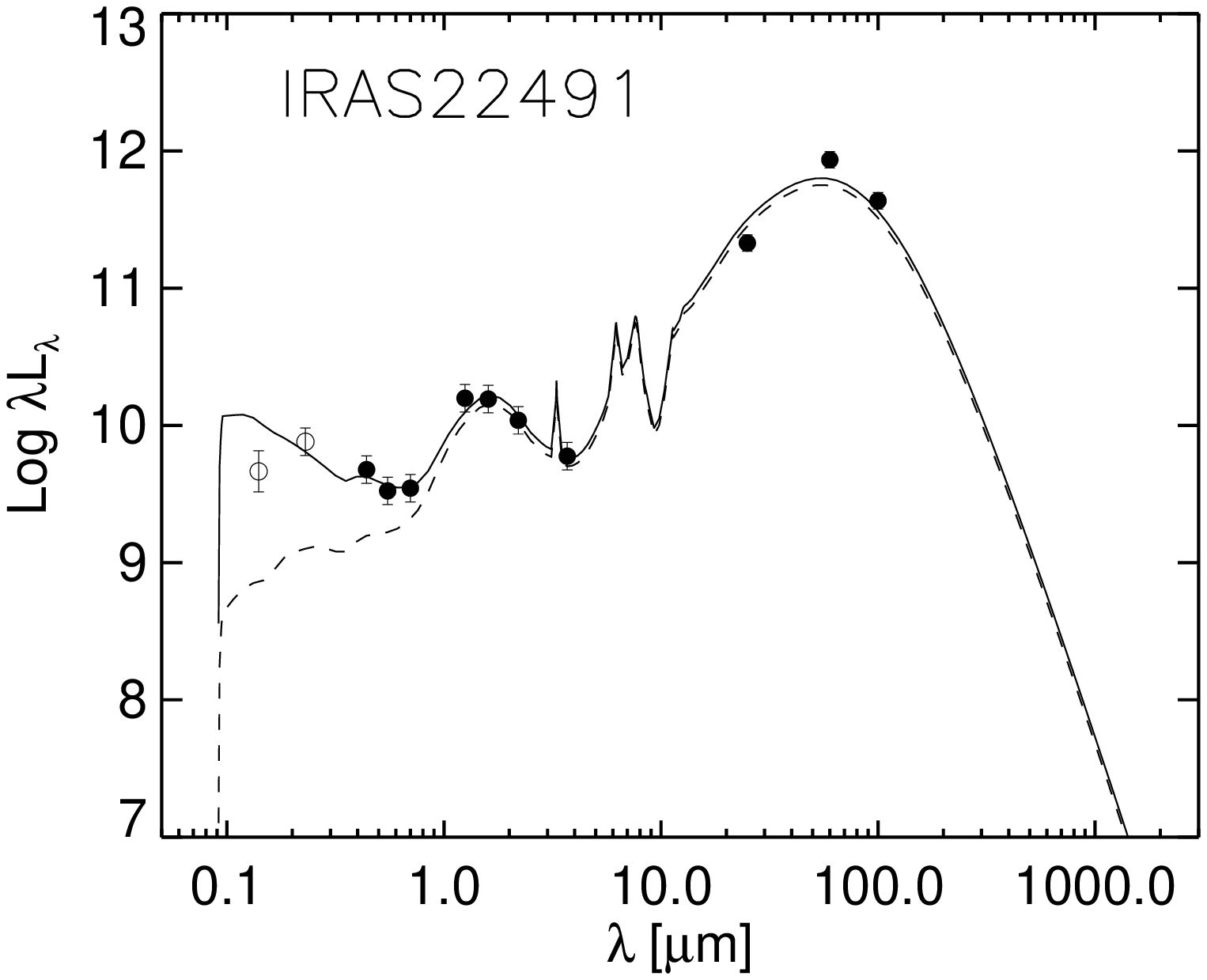}}
\caption{ {\it - continued.}
}
\end{figure*}

The sample of galaxies with $L_{\mathrm{IR}} > 10^{12}$ L$_\odot$, 
i.e., ULIRGs, 
analysed here were originally studied by Sanders et al. (1988).
Although a starburst is the only energy source in our model,
we also apply it to Seyfert galaxies in which
the AGN could considerably contribute to the dust heating.
According to Veilleux et al. (1999a),
IRAS05189, Mrk231 and Mrk273 are Seyfert galaxies,
although the line-to-continuum ratio of the 7.7 $\mu$m emission feature
indicates that Mrk273 is a starburst-dominanted 
ULIRG (Rigopoulou et al. 1999).

All photometric data from the $B$-band to the submm wavelengths are taken
from
Rigopoulou, Lawrence \& Rowan-Robinson (1996).
The adopted aperture size is 5$^{\prime \prime}$ for all
ULIRGs, corresponding to $\sim 5$ kpc for the mean redshift.
The adopted aperture is the smallest among those tabulated in
Sanders et al. (1988), where the original photometric data are
taken from. Although IRAS and submm data are obtained with larger
apertures,
the compactness of emission source ensures negligible corrections
for the aperture mismatch (Sanders et al. 1988;
Wynn-Williams \& Becklin 1993; Soifer et al. 2000).

For ULIRGs, galaxies with secure FUV photometry are relatively rare.
Thus, we simply determine the best-fitting model for all the photometric data
available. We find that our models again give a flux lower than that observed
in the shorter wavelength region.
We estimate $f_{\mathrm{leak}}$ for ULIRGs by attributing this
deficiency to the photon leakage.

The best-fitting models without considering the leakage are shown in Fig.
\ref{ulirg} with dashed lines.
Our model can reproduce the observed SEDs especially well 
at wavelengths longer than NIR, except for IRAS05189 and Mrk231,
which are candidates for AGN powered ULIRGs.
For both galaxies, it is found that the observed flux from NIR to MIR
show clear excess comparing with the best-fitting model, which could be
accounted for by the emission from hot dust around AGN.
Thus, our fitting results are consistent with the spectroscopic
classification of ULIRGs.
All of starburst powered ULIRGs are best-fit with the SMC type extinction 
curve except for IRAS08572. This shows a clear contrast to the results for
UVSBGs, in which the MW extinction curve is found to be better.
All the best-fitting models are characterized by the
small compactness factor $\Theta \sim 0.5$, corresponding to $\tau_V \sim
20$;
as a result, the strong silicate absorption feature at 9.7 $\mu$m
is a common factor in ULIRGs. This result suggests that
the starburst regions of ULIRGs are much more compact in size
compared with the UVSBGs.

As already mentioned above, we can see that the best-fitting models
tend to give a flux in the optical region lower than the observed one.
Following our treatment of UVSBGs, we determine $f_{\mathrm{leak}}$ with
the least square method,
assuming negligible extinction for leaking photons.
The best-fitting models with the leakage are
shown in Fig. \ref{ulirg} with solid lines. The resulting fitting
parameters are given in Table 2. Due to the simple assumption that the
extinction is negligible for the leaking photon, the resulting SEDs in
the UV are significantly bluer. Although the values of
$f_{\mathrm{leak}}$ are smaller than those for UVSBGs, the effect of the
leakage on the SED is prominent because of heavy obscuration of the
stellar continuum. 
In order to verify the fitting results,
we plot the FUV photometric data (Trentham et al. 1999; Goldader et al. 2002) 
for some of the ULIRG sample 
with open circles in Fig. \ref{ulirg}.
For Mrk273, IRAS08572 and IRAS12112, the FUV data taken by the {\it HST}
are consistent with the best-fitting model without considering the leakage
(dashed line), while for IRAS15250 and IRAS22491 
the FUV data can be explained by the best-fitting
model with the leakage (solid line). 
Only for Arp220, the FUV data cannot be explained by both cases. 
According to Goldader et al. (2002), these data are problematic due 
to uncertainty of sky level, since the angular size of Arp220 
is large enough to cover entire field of view of $HST$/STIS. 
Thus, fitting results are consistent with FUV observations, except for 
the problematic case of Arp220, when we consider 
the derived value of $f_{\mathrm{leak}}$ as the upper limit for some 
galaxies. 

As mentioned earlier, starburst ages are difficult to assign, 
but our resulting ages are in the range of $50-500$ Myr, 
which shows no systematic deviation from those derived for UVSBGs.
For starburst dominated ULIRGs, 
the mean value of $\Theta$ and $\tau_V^{\mathrm{eff}}$ is 0.46 and 
4.9, respectively. Thus, ULIRGs are more compact than 
UVSBGs by a factor of 3.5. 

For most of the ULIRGs, the SMC extinction curve is found to provide the best fit. 
Note that neither the observations of silicate absorption features nor
the bump at 2175 \AA\ are used to
distinguish the type of extinction curve in the SED fitting.
The SED from MIR to submm wavelengths
can be better reproduced with the model in which
the self-absorption of MIR emission by dust is significant; i.e.,
the model with the SMC extinction curve.

\begin{table*}
\begin{center}
Table 2.\hspace{4pt} Fitting results for ULIRGs \\
\end{center}
\tabcolsep=3pt
\vspace{6pt}
\begin{tabular*}{\textwidth}{@{\hspace{\tabcolsep}
\extracolsep{\fill}}p{4pc}rcccllllcc}
\hline\hline\\ [-10pt]
\multicolumn{1}{c}{name} &
\multicolumn{1}{c}{$t/t_0$} &
\multicolumn{1}{c}{$\Theta$} &
\multicolumn{1}{c}{EC} &
\multicolumn{1}{c}{$f_{\mathrm{leak}}$} &
\multicolumn{1}{c}{$M_T$} &
\multicolumn{1}{c}{$L_{bol}$} &
\multicolumn{1}{c}{$M_*$} &
\multicolumn{1}{c}{$M_D$} &
\multicolumn{1}{c}{SFR} &
\multicolumn{1}{c}{$\tau_V^{\mathrm{eff}}$} \\
\multicolumn{1}{c}{} &
\multicolumn{1}{c}{} &
\multicolumn{1}{c}{} &
\multicolumn{1}{c}{} &
\multicolumn{1}{c}{} &
\multicolumn{1}{c}{[M$_\odot$]} &
\multicolumn{1}{c}{[L$_\odot$]} &
\multicolumn{1}{c}{[M$_\odot$]} &
\multicolumn{1}{c}{[M$_\odot$]} &
\multicolumn{1}{c}{[M$_\odot$/yr]} &
\multicolumn{1}{c}{}  \\
\hline \\[-8pt]
UGC05101  & 5.0 & 0.3& SMC  &0.005  & 1.1\ $10^{11}$  & 2.1\ $10^{12}$  &
1.0\ $10^{11}$  & 2.5\ $10^8$  &69.& 4.7\\
IRAS12112 & 2.0 & 0.3& SMC  &0.006  & 7.8\ $10^{10}$  & 1.5\ $10^{12}$  &
4.3\ $10^{10}$  & 3.5\ $10^8$  &244.&5.4  \\
Mrk273   & 2.0 &0.4 & SMC  &0.008  & 6.6\ $10^{10}$  & 1.3\ $10^{12}$  &
3.7\ $10^{10}$  & 3.0\ $10^8$  &207.& 4.9 \\
IRAS08572 & 0.5 & 0.8  &  MW  &0.036  & 9.2\ $10^{10}$  & 1.7\ $10^{12}$
& 7.7\ $10^9$  & 6.0\ $10^7$  &285.& 4.9 \\
IRAS14348 & 2.0 & 0.3 & SMC &0.008  & 1.0\ $10^{11}$  & 2.0\ $10^{12}$  &
5.8\ $10^{10}$  & 4.8\ $10^8$  &327.& 5.4 \\
IRAS15250 & 1.0 & 0.6 & SMC &0.005  & 4.8\ $10^{10}$  & 9.1\ $10^{11}$  &
1.2\ $10^{10}$  & 1.4\ $10^8$  &185.& 4.5 \\
Arp220 & 0.7 & 0.4 &  SMC  &0.004  & 7.2\ $10^{10}$  & 1.4\ $10^{12}$  &
1.0\ $10^{10}$  & 1.4\ $10^8$  &260.& 5.4 \\
IRAS22491 & 1.0 & 0.6 & SMC &0.011 & 5.7\ $10^{10}$  & 1.1\ $10^{12}$  &
1.4\ $10^{10}$  & 1.6\ $10^8$  &221.& 4.4 \\
IRAS05189$^a$ & 4.0  & 0.4 & MW  &$<0.001$  & 1.4\ $10^{11}$  & 2.7\
$10^{12}$  & 1.2\ $10^{11}$  & 2.4\ $10^8$&158.& 4.7 \\
Mrk231$^a$ & 2.0 & 1.0 & MW  &$<0.001$  & 1.6\ $10^{11}$  & 3.1\ $10^{12}$
  & 8.9\ $10^{10}$  & 3.9\ $10^8$  &506.& 2.8 \\
\hline \\
\end{tabular*}
\vspace{6pt}
\par\noindent
Note: $^a$ candidates for AGN powered ULIRGs \\
\label{ul-table}
\end{table*}

\subsection {Comparison with emission line measurements}

\subsubsection {Star formation rate}

\begin{figure}
  \resizebox{\hsize}{!}{\includegraphics{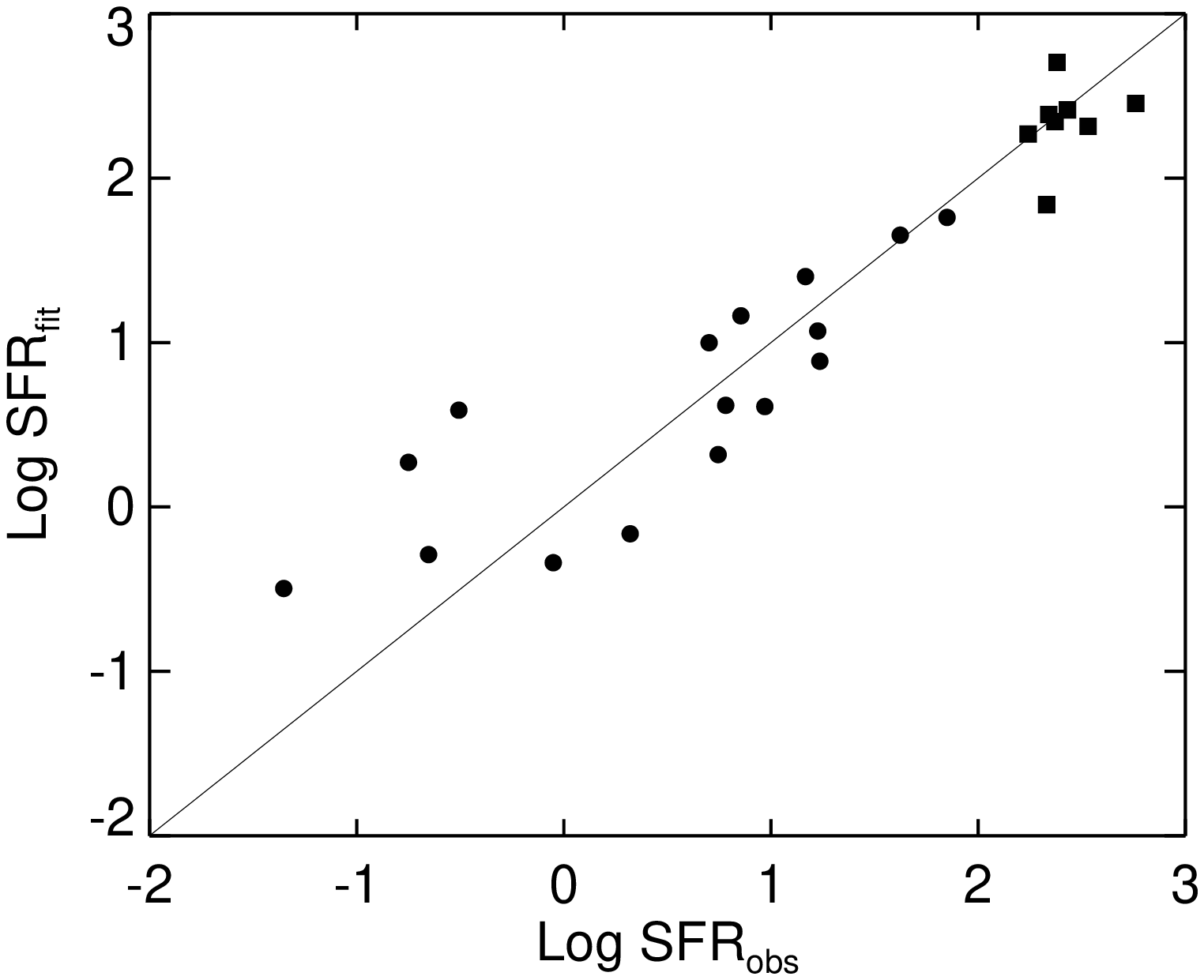}}
  \caption{Comparison of the SFR in M$_\odot$yr$^{-1}$.
The observed SFRs are derived
from the extinction corrected H$\alpha$ and FIR luminosity
for UVSBGs and ULIRGs, respectively.
AGN candidates, IRAS05189 and Mrk231, are not
plotted. Solid line indicates the linear relation.
}
  \label{sfr}
\end{figure}

In Fig. \ref{sfr}, we compare the resulting SFRs with
the observed values derived
from H$\alpha$ luminosities for UVSBGs (Storchi-Bergmann et al.
1994) and from the FIR luminosity
of ULIRGs by using the relation presented by Kennicutt (1998).
The SFRs in our model are basically determined from
the bolometric luminosity from UV to FIR. Therefore,
the observed SFRs for ULIRGs should be almost
identical to those from the SED fitting since the emission from the FIR 
dominates the SED of ULIRGs. Thus, we can see clearly this agreement
at the high SFRs. On the other hand,
our SFR derivation is independent of that
from H$\alpha$ line emission for UVSBGs.
Although the difference between the model results
and observation becomes large for galaxies with the SFR $\le$
1 M$_\odot$ yr$^{-1}$,
the resulting SFRs for UVSBGs agree well with those derived
from H$\alpha$ and IR luminosities over three orders of magnitude
without any systematic difference. 
Therefore, the SFR estimated by the SED fitting can be applied well
to any starburst galaxy regardless of their optical depth. 

Our SED model suggests that leaking photons contribute to the 
UV luminosities of starbursts. Therefore, if the UV luminosity 
is corrected for dust extinction without considering this 
contribution, the UV-derived SFR can be easily overestimated. 
This effect may explain the excess of UV luminosities 
for a given H$\alpha$ luminosity, reported by Sullivan et al. (2000). 
We emphasize that, in order to derive SFRs from the continuum emission, 
bolometric luminosity should be used to avoid systematic effects.

\subsubsection{Optical depth and extinction}

\begin{figure}
  \resizebox{\hsize}{!}{\includegraphics{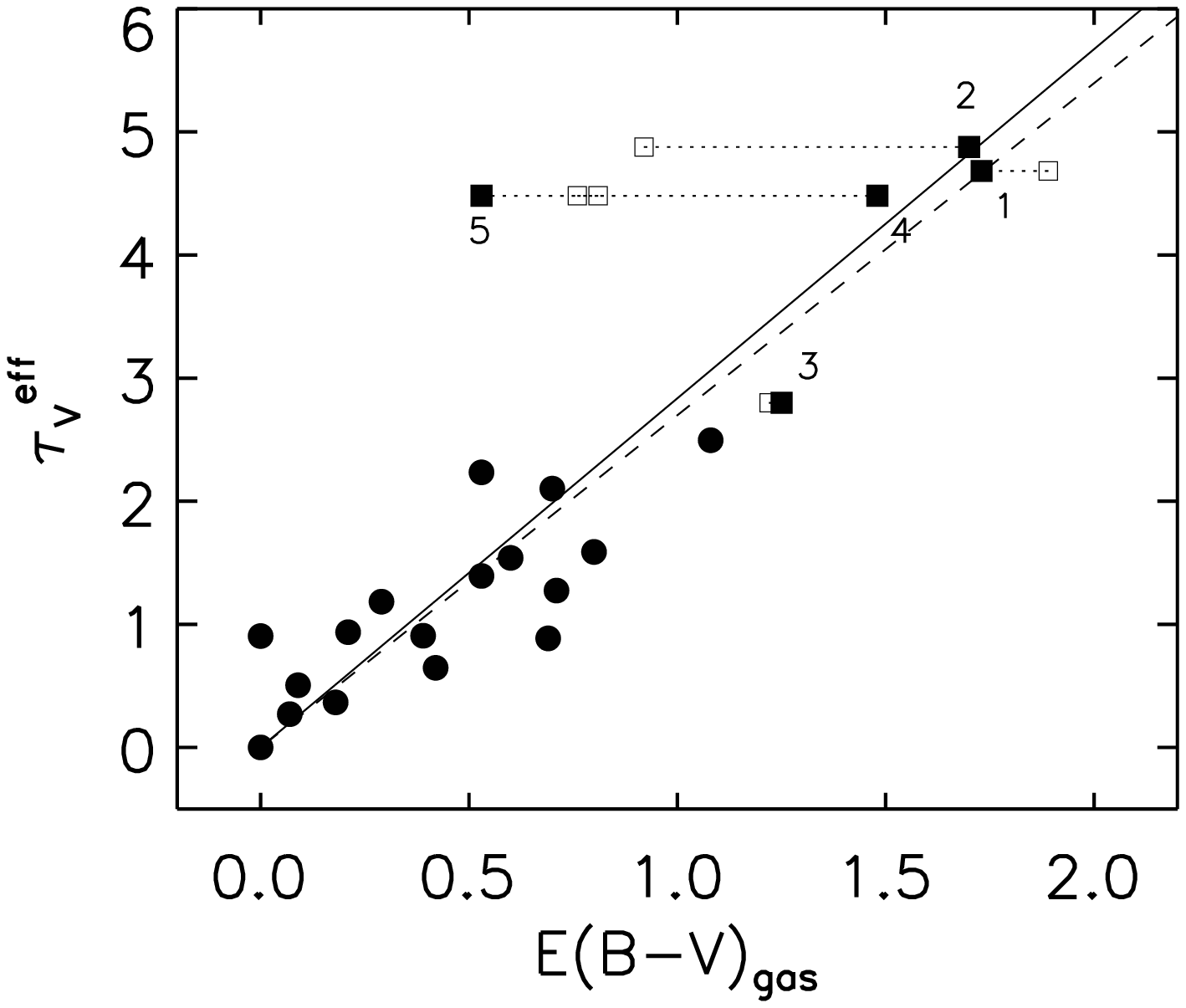}}
  \caption{Comparison of the extinction. The observed colour excesses
$E(B-V)_{\mathrm{gas}}$ are derived from the ratios of emission lines.
For UVSBGs, H$\alpha$/H$\beta$ is used, taken from
Storchi-Bergmann et al. (1994).
For ULIRGs, only galaxies which have observed value of Pa$\alpha$ are
plotted;
depicted numbers indicate galaxies; 1 - IRAS08572, 2 - IRAS12112,
3 - Mrk273, 4 - IRAS15250, 5 - IRAS22491. Filled and open
squares indicate $E(B-V)_{\mathrm{gas}}$ derived from Pa$\alpha$/H$\alpha$
and H$\alpha$/H$\beta$, respectively (Veilleux et al. 1999b).
Solid and dashed line indicate the relation for the ratio of 
total-to-selective extinction $A_V/E(B-V)=3.08$ for MW and 
2.93 for SMC (Pei 1992), respectively.
The derived relation between $\tau^{\mathrm{eff}}_V$
and $E(B-V)_{\mathrm{gas}}$ should be similar to 
this relation in the case that 
the distribution of ionized gas is the same as that of stars.
}
  \label{tau}
\end{figure}

The effective optical depth can be determined independently of the SFR.
In Fig. \ref{tau}, we compare the effective optical depth at $V$-band
$\tau_V^{\mathrm{eff}}$ with the colour excess $E(B-V)_{\mathrm{gas}}$
derived from the Balmer line ratios.
We see that $\tau^{\mathrm{eff}}_V$ is consistent with 
$E(B-V)_{\mathrm{gas}}$.
Calzetti et al. (1994) suggested that
the extinction inferred from the line ratios in starburst galaxies
is typically twice as high as that derived from the stellar continuum.
For optically thick UVSBGs in our sample,
the resulting $\tau_V^{\mathrm{eff}}$ tends to be slightly less
(within a factor of 2)
than the expected value for the MW extinction curve and 
therefore, consistent with the results of Calzetti et al. (1994).

The observed colour excesses $E(B-V)_{\mathrm{gas}}$ 
of ULIRGs should be derived
from the less attenuated lines, such as Paschen and Brackett lines,
since the Balmer line ratios could be 
saturated due to the large optical depth.
Veilleux et al. (1999b) presented $E(B-V)_{\mathrm{gas}}$ measurements 
derived from
both Pa$\alpha$/H$\alpha$ and H$\alpha$/H$\beta$. Both are plotted
in Fig. \ref{tau}, connected with dotted lines for each galaxy.
Half of our sample of galaxies have the observed Pa$\alpha$/H$\alpha$.
Except for IRAS22491, $\tau_V^{\mathrm{eff}}$ of the ULIRG sample are
consistent with the simply extrapolated relation from
the less attenuated UVSBGs.
For IRAS22491 (the only HII galaxy in our sample),
$E(B-V)_{\mathrm{gas}}$ derived from
Pa$\alpha$/H$\alpha$ and H$\alpha$/H$\beta$ differs in the opposite
sense. Galaxies showing such anomalous behaviour are predominantly
Seyfert 2 galaxies in the sample studied by
Veilleux et al (1999b),
and HII galaxies constitute only a small fraction.
The extinction of ULIRGs could be investigated more accurately
by spectroscopic studies at longer wavelengths with higher
sensitivity, by future FIR satellite missions, such as SIRTF and ASTRO-F.

In summary, the SED fitting model can reproduce well
the relation between $E(B-V)_{\mathrm{gas}}$ and $\tau_V^{\mathrm{eff}}$, 
and therefore, the dimming and reddening
effects of the light from starburst regions are 
consistently introduced in the model.

\subsection{Comparison of apparent effective radius}

\begin{figure}
  \resizebox{\hsize}{!}{\includegraphics{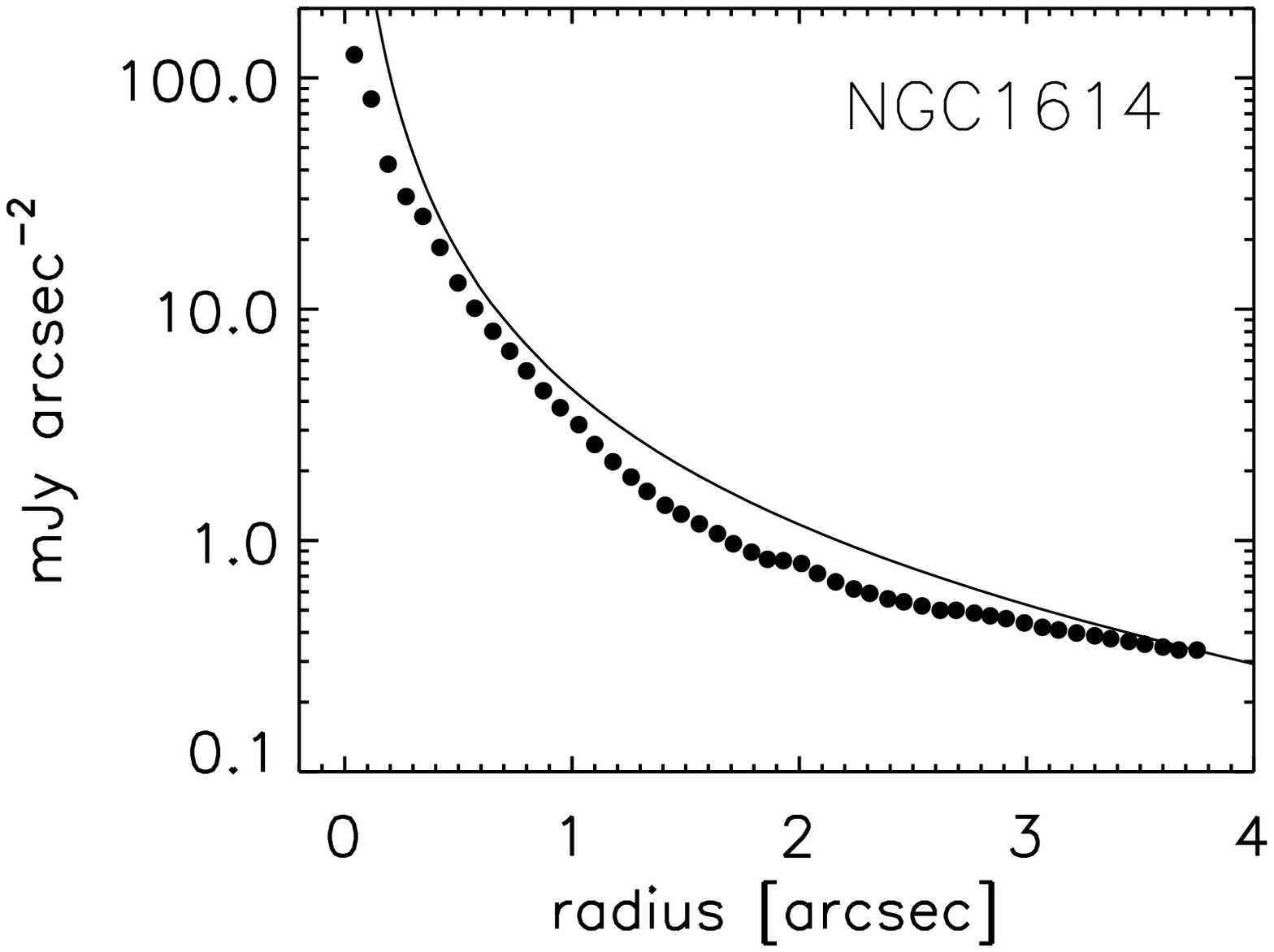}}
  \caption{Comparison of $K$-band surface brightness profile of NGC1614.
Data plotted are taken from Alonso-Herrero et al.\ (2001) obtained with
the {\it HST}/NICMOS. Solid line is the surface brightness of the
best-fitting SED model for NGC1614.
}
  \label{reK}
\end{figure}

\begin{figure}
  \resizebox{\hsize}{!}{\includegraphics{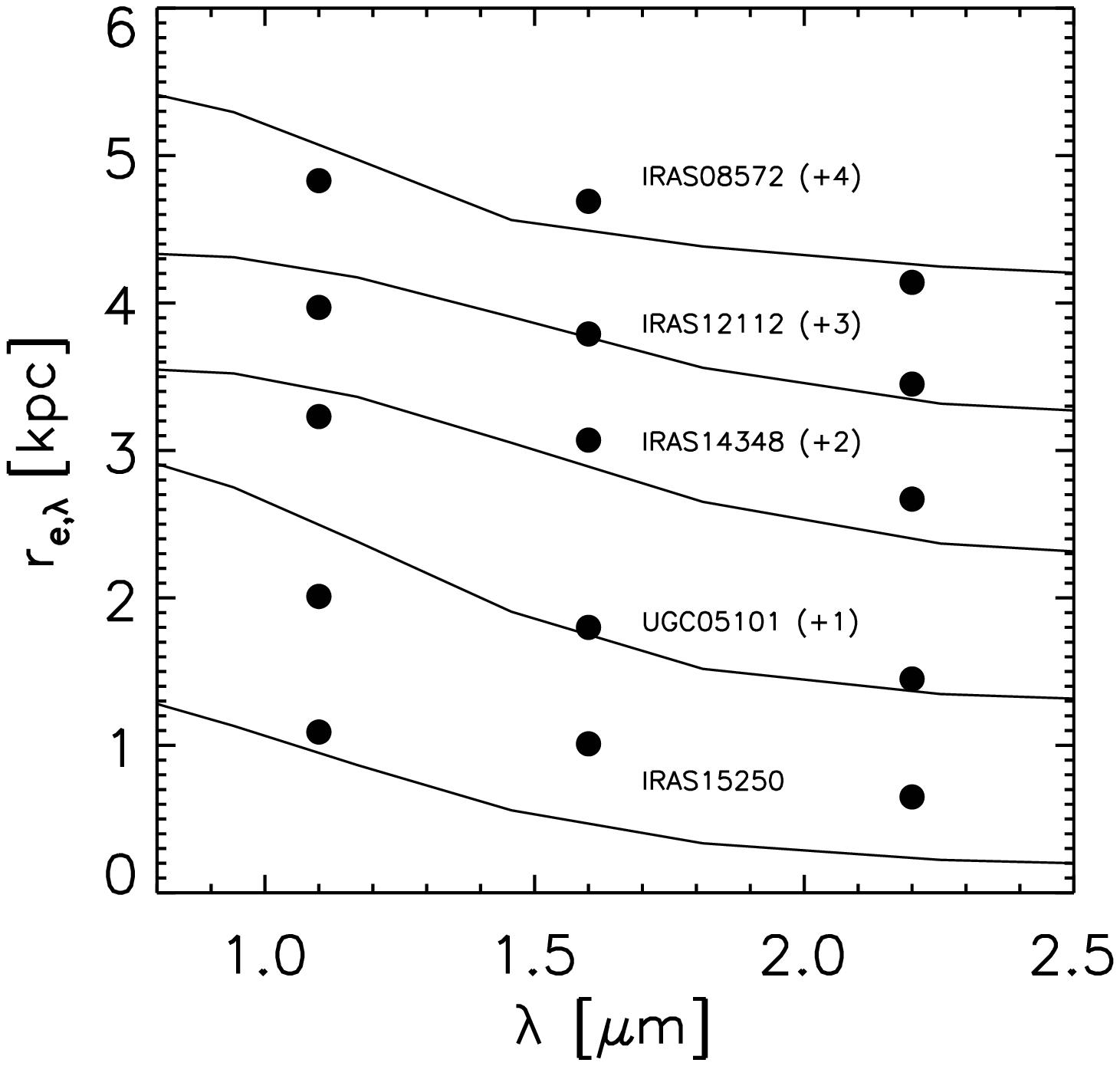}}
  \caption{Comparison of the apparent effective radii at $1.1 \mu$m,
$1.6 \mu$m and $2.2 \mu$m with the {\it HST} observations (Scoville et al.
2000; solid circles) for galaxies in which the effective radius is
determined for a single nucleus (see text in detail).
For clarity, we add the value noted in parenthesis to
the effective radii.
}
  \label{recheck}
\end{figure}


Since we applied the model not to a whole galaxy,
but to the central starburst region,
the effective radius derived from the
SED fitting should be compared with that of the central starburst region.
For the UVSBG sample, unfortunately no homogeneous presentation of
effective radii of starburst regions is found in the literature. 
Prugniel \& Heraudeau (1998) presented a catalogue of aperture
photometry of 7744 galaxies and derived $B$-band effective radii
for a whole galaxy, including all our sample UVSBGs, but three.
Note that these effective radii cannot be compared with the model
results directly, since the underlying population could enlarge the
$B$-band effective radii, systematically. 
It is found that $B$-band effective radii published by Prugniel \& Heraudeau
(1998) are systematically larger than those derived from the SED fitting
by a factor of three.
However, it is not yet clear 
whether the contribution from underlying population is
enough to explain this discrepancy or not.
This discrepancy may be partly due to the adopted
concentration parameter $c$ being too large for UVSBGs.
Alonso-Herrero et al. (2001) studied NGC1614 in detail
with {\it HST}/NICMOS observations, and presented a surface brightness
profile of the central starburst region in the $K$-band.
We compare the $K$-band surface
brightness profile of NGC1614 with that of the best-fitting SED model
in Fig. \ref{reK}. Clearly, the model result is in good agreement
with the observation. Since this galaxy is relatively luminous and obscured
for a typical UVSBG, we need a more comprehensive study using 
images with high spatial resolution to confirm the
stellar distribution of UVSBGs unambiguously.

Scoville et al. (2000) presented NIR images of ULIRGs obtained with
the {\it HST}/NICMOS, and derived the effective radii at
$1.1 \mu$m, $1.6 \mu$m and $2.2 \mu$m.
In most cases, the observed effective radii are less than
1$^{\prime \prime}$, and are therefore consistent with the adopted
aperture size as starburst regions (5$^{\prime \prime}$) in the SED fitting.
We focus on the ULIRGs for which the effective radii are
determined for isolated major starburst regions, since the multiple
structure could enlarge the effective radius, systematically.
In Fig. \ref{recheck}, 
the effective radii of the models for
ULIRGs are compared with the apparent effective radii
from the {\it HST}/NICMOS images.
As a result, we can see that the radii from the models and the
observations are consistent with each other within a factor of 2.
As shown in the Fig. 12,
the effective radius of isolated starburst region in the $K$-band is
$< 0.7$ kpc, which corresponds to an intrinsic effective radius
(i.e., for the case without dust) of $\sim$0.3 kpc.
Scoville et al. (2000) have shown that
the light profiles of these ULIRGs can be also fitted well with
the $ r^{1/4} $ law, which can be reproduced by the King profile.
Thus, the stellar distribution in a starburst region
can be roughly represented with our model geometry of
the King profile with $\log(r_t/r_c)=2.2$ for ULIRGs.

As shown by Scoville et al. (2000),
many ULIRGs have multiple starburst regions -- such as Arp 220.
As long as each starburst region has a
constant surface brightness, the SED feature is not affected by
this multiplicity as remarked in the model description (Section 2.2).
Thus, the model geometry can still work
as long as the stellar distribution
of individual starburst regions are considered.

\subsection {Correlations between starburst properties: signature of 
bimodality}

\begin{figure*}
  \resizebox{8cm}{!}{\includegraphics{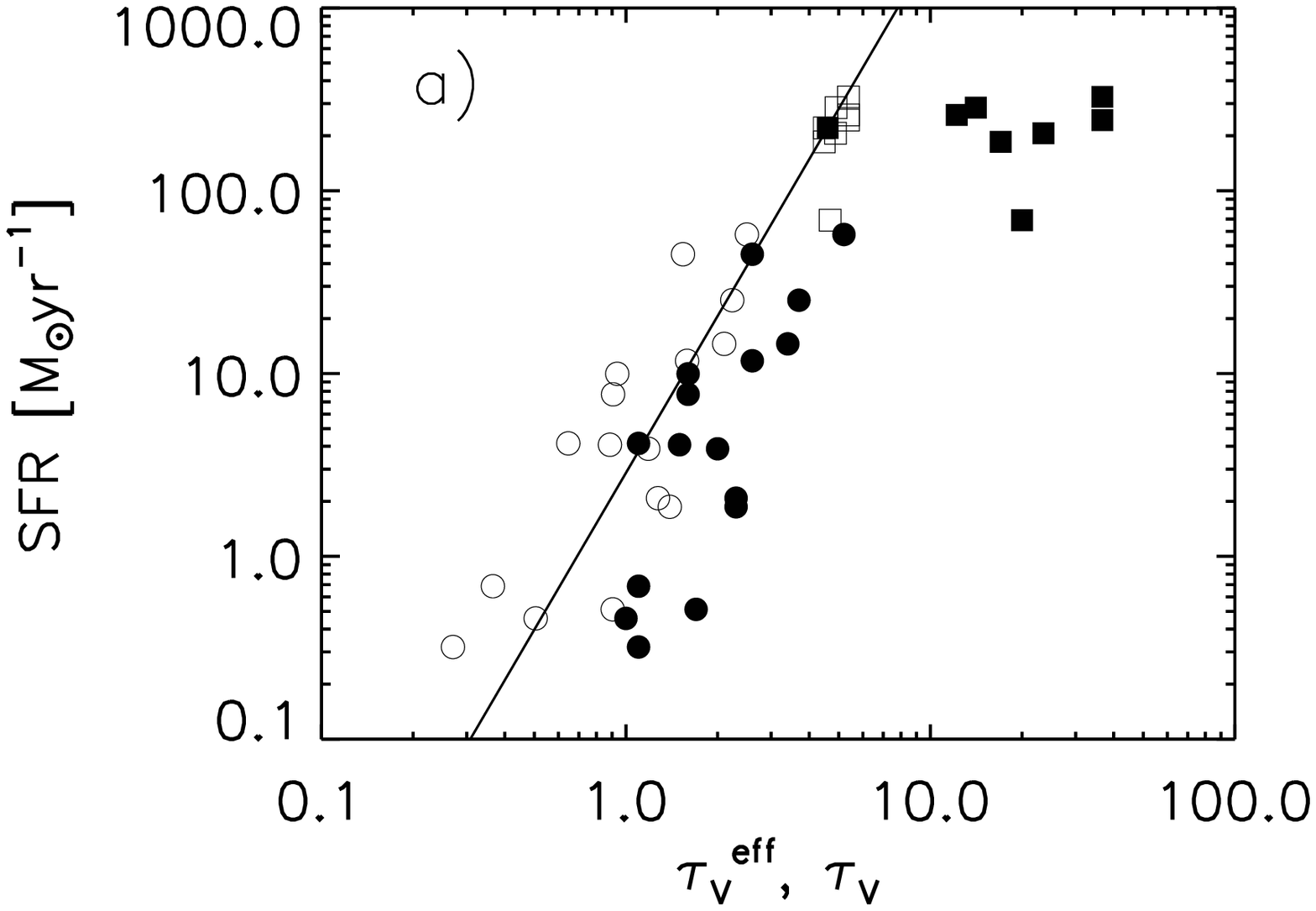}}
  \resizebox{8cm}{!}{\includegraphics{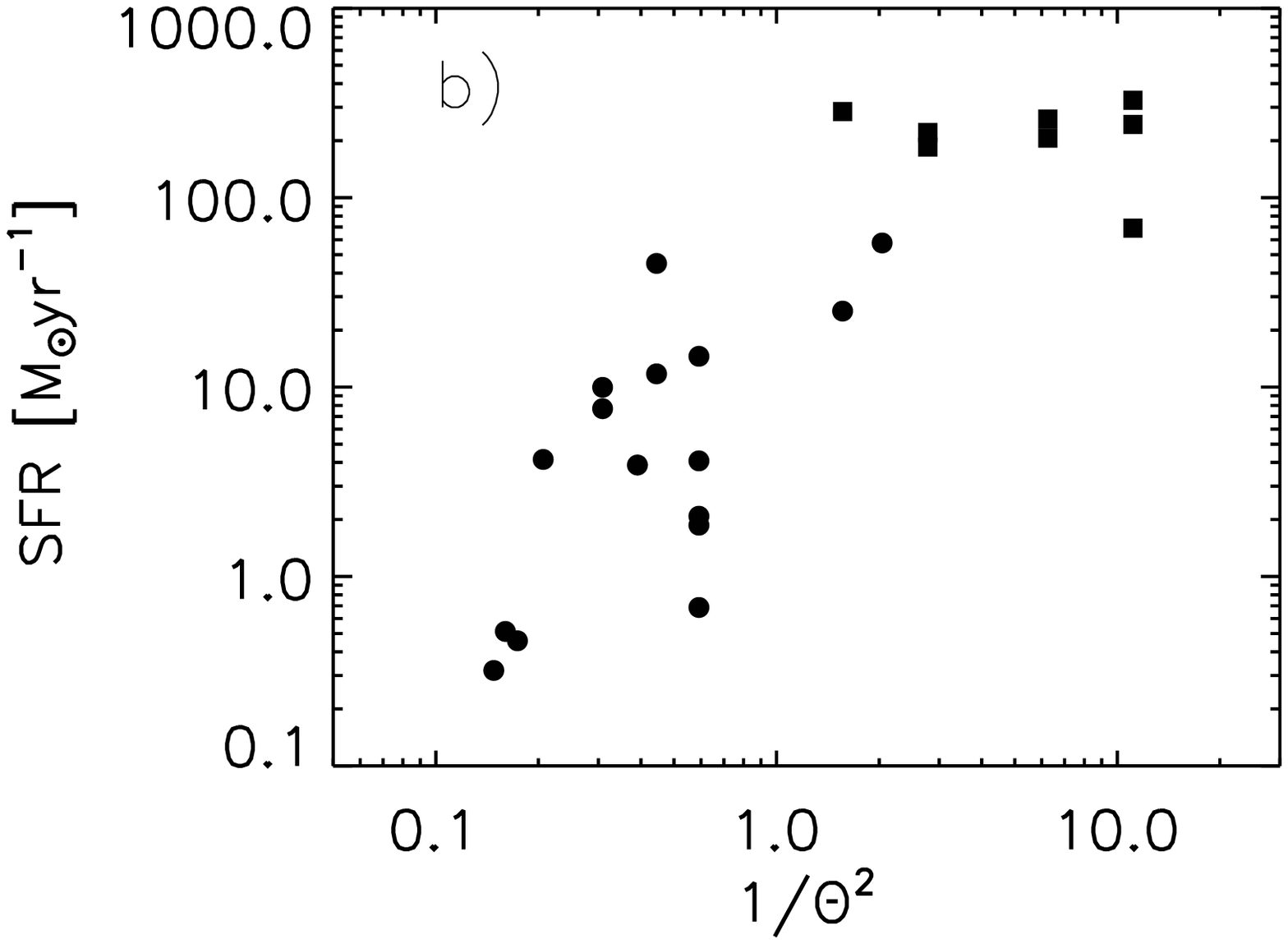}}
  \resizebox{8cm}{!}{\includegraphics{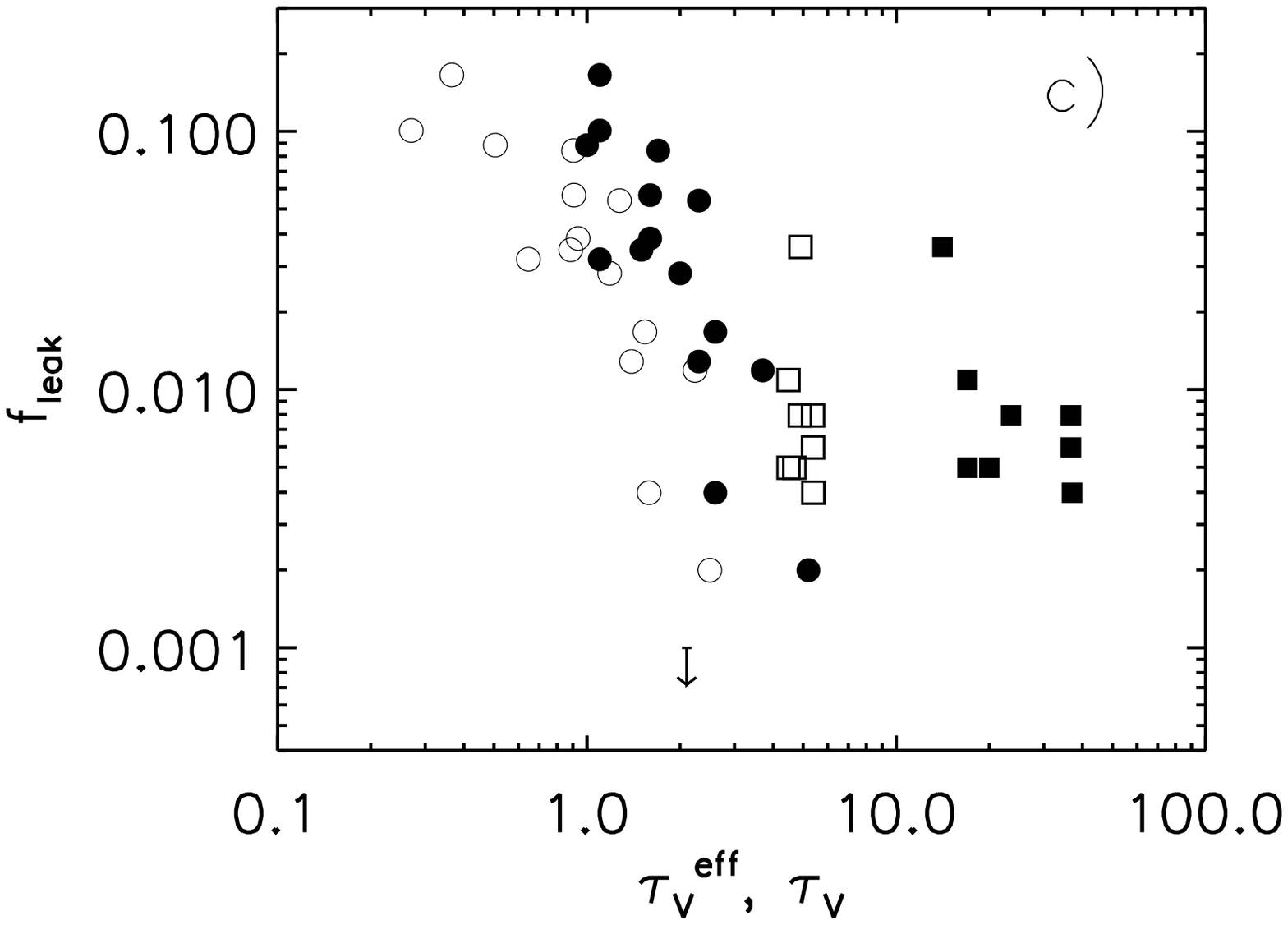}}
  \caption{Relations between the derived quantities. 
Circles and squares indicate UVSBGs and ULIRGs, respectively.
a) SFRs are plotted against the effective $V$-band
optical depth (open symbols) and
$V$-band optical depth (solid symbols).
The solid line is
a power-law relation, SFR $ \propto ( \tau_V^{\mathrm{eff}} )^{3}$.
b) SFRs are plotted against $\Theta^{-2}$, which is proportional
to $\tau_V$ for constant dust mass.
c) Leaking fraction as a function of $\tau_V^{\mathrm{eff}}$ (open symbols)
and $\tau_V$ (solid symbols).
Upper limit on $f_{\mathrm{leak}}$ is indicated with an arrow.
}
\label{bimodal}
\end{figure*}

In Fig. \ref{bimodal}a, the SFR is plotted against
$\tau_V^{\mathrm{eff}}$ and $\tau_V$.
Clearly, we can see a power-law correlation between
SFR and $\tau_V^{\mathrm{eff}}$;
SFR $ \propto (\tau_V^{\mathrm{eff}})^{3} $.
A similar trend has already been shown by Heckman et al. (1998)
and Adelberger \& Steidel (2000)
between $L_{\rm FIR}/L_{\rm UV}$ and
$L_{\rm UV+FIR}$, which are related to the reddening effects
and SFR respectively, although the dispersion of the
correlation they presented was much larger than that shown
in Fig. \ref{bimodal}a.
This trend seems to suggest the Schmidt law for
star formation, in which SFR increases with increasing gas
density. However, note that not $\tau_V^{\mathrm{eff}}$, but 
$\tau_V$ is proportional to 
the gas column density in the starburst regions. 
The discrepancy between $\tau_V^{\mathrm{eff}}$ and $\tau_V$
can be clearly seen at high optical depth $\tau_V \ga 5$; 
as a result, the SFR seems to saturate at $\tau_V \ga 5$. 
This suggests that the mode of starburst activity 
changes at a critical column density corresponding 
to $\tau_V \simeq 5$.

Such bimodal feature can be seen in the relation between
the SFR and the compactness factor as shown in Fig.
\ref{bimodal}b, in which we can also find a critical compactness
factor $\Theta^{-2} \simeq 3$.
Both bimodal features of the SFR against
the column density and the compactness suggest that the gas density
in the starburst regions can directly affect the starburst activity
since in general the geometrical compactness with the high column density
implies the existence of a high density region. 

In Fig. \ref{bimodal}c, we plot $f_{\mathrm{leak}}$ against
$\tau_V^{\mathrm{eff}}$ and $\tau_V$.
There is an anti-correlation between these properties for 
$\tau_V \la 5$, while this anti-correlation breaks for 
larger $\tau_V$. 
Feedback in starbursts with low ISM density, such as 
UVSBGs, may naturally generate regions of porous
ISM that provide pathways for UV photon leakage even from 
central region of starbursts. 
On the other hand, for starbursts with a high density ISM such as ULIRGs, 
leakage can be expected only from the outermost surface of a starburst 
region and may have little to do with the main starburst 
(Goldader et al. 2002).
Thus, the break of anti-correlation between 
$f_\mathrm{leak}$ and $\tau_V$ can be explained by the feedback 
effects on different ISM density. 




\section {Discussion}
In this section, we will discuss the bimodality found in 
the correlations between starburst properties with our SED diagnostics. 
First, we focus on the correlation between the intrinsic surface brightness 
and the size of starburst region, which shows 
the bimodality with a clear limit on the size of starburst regions. Then, 
we investigate the origin of the bimodality in the starburst activity by 
considering the physical processes of starburst regions, such as 
gravitational instability of the disc and feedback effects.  

\subsection{Bimodal starburst intensity}

\begin{figure*}
  \resizebox{\hsize}{!}{\includegraphics{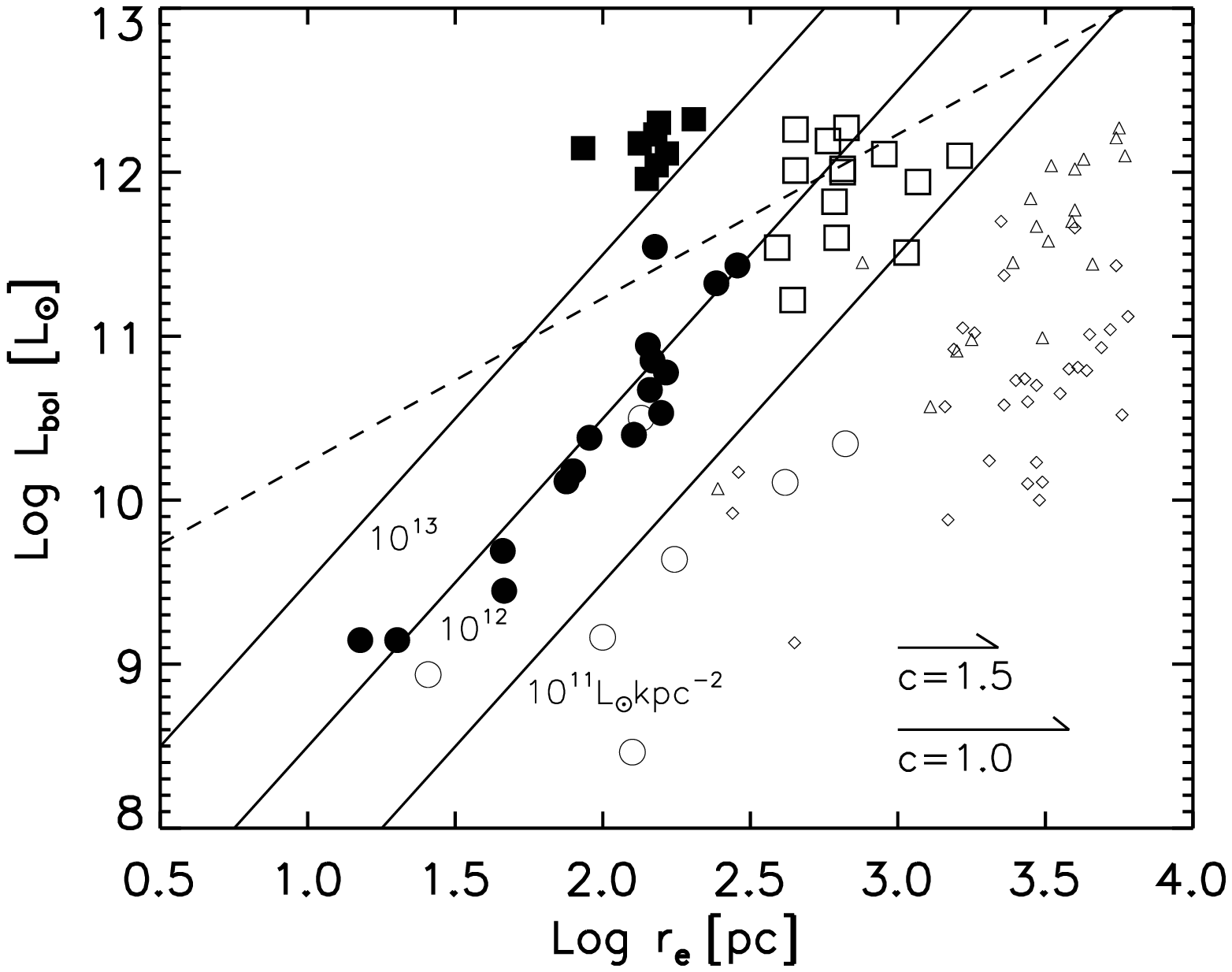}}
  \caption{The bolometric luminosities of starburst galaxies plotted
against effective radii. Filled circles and squares are the
bolometric luminosities and intrinsic 
effective radii derived from the SED of UVSBGs and ULIRGs, respectively.
Open symbols indicate the observed bolometric luminosities and
apparent effective radii at 0.22$\mu$m
(open circles; Meurer et al. 1995), $2.2 \mu$m (squares; Scoville et 
al. 2000).
Small diamonds indicate the $R$-band effective radii and $L_{bol}$
taken from Lehnert \& Heckman (1996) in which 
$L_{bol}$ is estimated from FIR luminosity ($L_{bol}=1.5 L_{FIR}$;
Meurer et al. 1997).
When $R$-band data is not available, we convert the effective
radius measured with H$\alpha$ imaging into $R$-band effective radius
with the average ratio of H$\alpha$ to $R$-band effective radius,
0.6 according to Lehnert \& Heckman (1996).
By using the same ratio of effective radii,
H$\alpha$ effective radii taken from Armus et al. (1990) 
are converted
to $R$-band effective radii, which are plotted against $L_{bol}$
estimated again from $L_{FIR}$ (small triangles).
Dashed line is the limiting luminosity
given in Eq. (\ref{eq_limit}). Two arrows at the bottom left corner indicate
the effect of concentration parameter $c$ on the resulting
$r_e$.
}
  \label{lumrad}
\end{figure*}

In Fig. \ref{lumrad}, 
the bolometric luminosities of starburst galaxies are plotted
against the effective radii.
Filled circles and squares are
bolometric luminosities and intrinsic effective radii (i.e., for
the case without dust) derived from the SED model of
UVSBGs and ULIRGs, respectively.
The bolometric surface brightnesses of UVSBGs are found to be
almost constant at 10$^{12} $L$_{\odot} $kpc$^{-2}$. 
On the other hand, a clear limit on the intrinsic
effective radius can be seen at $\sim 0.3$ kpc.
Moreover, the bolometric surface brightnesses of ULIRGs are an order
of magnitude higher than those of UVSBGs.
This suggests that the mode of starbursts could change at
$L_{bol}\simeq 10^{11}$ L$_\odot$ and $r_e\simeq 0.3$ kpc.

A limiting bolometric starburst intensity $ L_{bol} r_e^{-2} = 2 \times
10^{11}$ L$_{\odot} $kpc$^{-2}$ was proposed by Meurer et al. (1997).
From this intensity limit,
they also proposed a global star formation limit
$ \psi r_e^{-2} \simeq 45 $M$_{\odot}$kpc$^{-2}$yr$^{-1}$.
In Fig. \ref{lumrad}, all open symbols other than open squares
indicate the original data used by Meurer et al. (1997).
For open squares, 
we adopt the effective radii of ULIRGs
observed with the {\it HST}/NICMOS at $2.2 \mu$m taken from Scoville
et al. (2000). These effective radii are 
systematically smaller than those derived from $R$-band images 
(Lehnert \& Heckman 1996; small open diamonds) and from 
images of H$\alpha$ emission (Armus et al. 1990; small open 
triangles),
although the similar class of galaxies is investigated.
Thus, from the observational results it is suggested that
the observed surface brightness depends on the
observed wavelength. 
For UVSBGs, the bolometric surface brightness calculated
with the intrinsic effective radius is higher by a factor of 5 than
the proposed value by Meurer et al. (1997). 
The difference between the intrinsic and observed surface brightness 
is more remarkable for ULIRGs and indicates a bimodal starburst 
activity.

Actually, the limit on bolometric surface brightness we derive
depends on the precise value of adopted
concentration parameter $c$. In Fig. \ref{lumrad},
we adopt $c=2.2$ for all cases.
As discussed in Section 4.4, it is possible that
this concentration parameter could be large for
UVSBGs. In Fig. \ref{lumrad}, we show the
correction factor for the intrinsic effective radius in the case of
$c$=1.0 and 1.5 by arrows at the bottom left corner.
Nevertheless, the derived limit on the surface brightness is
robust for active starburst galaxies, since we confirmed in Section 4.4
that for ULIRGs and relatively luminous
UVSBGs the adopted value $c$=2.2 is reasonably consistent with
observed effective radii.
Note that even if less active starbursts have the smaller value
of concentration parameter, the bimodal feature of
starburst activity shown in Fig. \ref{lumrad} still remains.

In summary, the SED diagnostics find that 1) UVSBGs have 
an intrinsic effective surface brightness 
$ L_{bol} r_e^{-2} \simeq 10^{12}$ L$_{\odot} $kpc$^{-2}$, and 
2) there is a limit on the size of the starburst region of around $ 0.3 $ kpc.
Considering these results,
we will propose a picture of bimodal starburst activity
which consists of mild starburst activity in UVSBGs with a
regulated surface intensity around
$ L_{bol} r_e^{-2} \simeq 10^{12}$ L$_{\odot} $kpc$^{-2}$ and an
intensive mode of starbursts in ULIRGs confined in the region of
$r_e \simeq 0.3$ kpc. In the next subsection, we will discuss the 
physical origin of this bimodality.

\subsection{Origin of bimodal starbursts}

\subsubsection{Disc instability in UVSBGs}
The regulating mechanisms for the mild starburst intensity
have already been proposed by Meurer et al. (1997). Following Kennicutt
(1989), they showed that the gas surface density is
somewhat close to a critical value at which gas discs would be
unstable to self-gravity. According to
Toomre (1964) and Quirk (1972), Kennicutt (1989) proposed the critical
surface density for inducing star formation with the instability is 
\begin{equation}
\Sigma_c = \frac{\alpha \kappa \sigma_g }{\pi G} \; ,
\end{equation}
where $ \sigma_g $ is the velocity dispersion of the gas and $\kappa$ 
is the epicyclic
frequency. The parameter $ \alpha $ is
a constant empirically determined as $ \simeq 0.63 $.
With angular frequency $ \Omega(r) = V(r)/r $,
the epicyclic frequency is given by
\begin{equation}
\kappa = \left( r \frac{d \Omega^2}{d r} +4 \Omega^2 \right)^{1/2} \; .
\end{equation}

While Kennicutt (1989) considered the instability around the flat part
of the rotation curve in disc galaxies,
Meurer et al. (1997) applied it for the nuclear starbursts
near the peak of the rotation curve in inner discs.
We also apply the technique of Meurer et al.\ to interpret the bimodal
trends of SFRs in Fig. \ref{lumrad}.
If we consider that the inner disc to be a rigid rotator with
a angular frequency $\Omega$, the dynamical time of the disc is given by
\begin{equation}
  t_{dyn} = \frac{\pi} {2\Omega} \; .
\end{equation}
With a constant $ \Omega $ for 
the inner disc, the critical surface
density is $\Sigma_c = 2 \alpha \sigma \Omega/(\pi G)$.
With the estimations for the dynamical time and the surface density,
the maximum star formation intensity can be expected as
\begin{equation}
\dot{\ \Sigma_{\ast}} \simeq \frac{\Sigma_c}{t_{dyn}}
= 9 \times 10^{-5} \alpha \sigma
\Omega^2 \; \mathrm{M}_{\odot} \mbox{yr}^{-1} \mbox{kpc}^{-2} \; ,
\end{equation}
where $\sigma$ and $\Omega$ are in units of km s$^{-1}$ and
km s$^{-1}$ kpc$^{-1}$, respectively.

With $\psi(t)/L_{bol}$ from Eq. (\ref{eq_bolsfr}),
the bolometric surface brightness limit can be introduced as
\begin{eqnarray} \label{surfaceb}
L_{bol} r_e^{-2} & \simeq & 3.1 \times 10^5
\alpha \sigma
\Omega^2 \; \mathrm{L}_{\odot} \mbox{kpc}^{-2} \nonumber \\
& =  &
3.7 \times 10^{12}
\left( \frac{\alpha}{0.61} \right) \left( \frac{\sigma}{20 \mbox{km
s}^{-1}} \right) \nonumber\\
& & \times \left( \frac{\Omega}{10^3 \mbox{km
s}^{-1}\mbox{kpc}^{-1}} \right)^2 \; \mathrm{L}_{\odot} \mbox{kpc}^{-2} ,
\end{eqnarray}
where we take $t/t_0=2$ as a typical observable starburst age.

Since most of disc galaxies have a central HI velocity
dispersion $ \sigma_g \simeq 15-20$ km s$^{-1}$
(e.g. Dickey, Hanson \& Helou 1990) and angular frequency $\Omega <10^3$
km s$^{-1}$ kpc$^{-1}$ (e.g. Armus et al. 1990),
this limiting bolometric surface brightness 
($\sim 3.7 \times 10^{12} 
\mathrm{L}_{\odot} \mbox{kpc}^{-2}$) corresponds well 
to the limiting intensity of UVSBGs as seen in Fig. \ref{lumrad}.
Thus, starbursts in UVSBGs can be triggered
by the self-gravitational instability in inner gas discs
as already pointed out by Meurer et al. (1997).
This instability seems to induce only mild starburst activity, 
since the unstable region should have a scale of the disc thickness, 
and the whole gas in the disc cannot concentrate 
into the unstable region due to the centrifugal force from the disc's 
rotation. This means that 
the maximum size of starburst regions is expected to be similar to the 
disc thickness $\sim 0.3$ kpc, and therefore 
consistent with the SED fitting results in which no starbursts 
with $r_e \ga 0.3$ kpc are found.



The surface brightness of ULIRGs are an order of magnitude
larger than that expected from the disc instability. 
Therefore, ULIRGs should be triggered by another mechanism which 
can induce a stronger concentration of gas. 
As indicated by Eq. (\ref{surfaceb}), 
the surface brightness of starburst regions in UVSBGs are 
related to the physical properties of parent galaxies. 
On the other hand, 
the strong mass concentration in the central region of 
ULIRGs should result in the strong self-gravity of starburst region, 
and therefore the starburst region in ULIRGs can be dynamically 
isolated from parent galaxies. 
In such a case, the surface brightness of ULIRGs should be 
controlled by the relative strength of the 
self-gravity and its feedback.

\begin{table}
\begin{center}
Table 3.\hspace{4pt} Velocity dispersions of starburst galaxies\\
\end{center}
\tabcolsep=3pt
\vspace{6pt}
\begin{tabular*}{7cm}{@{\hspace{\tabcolsep}
\extracolsep{\fill}}p{4pc}ccc}
\hline\hline\\ [-10pt]
\multicolumn{1}{c}{name} &
\multicolumn{1}{c}{$\sigma$ [km/sec]} &
\multicolumn{1}{c}{Line$^a$} \\
\hline \\ [-8pt]
\multicolumn{3}{c}{UBSBGs} \\
\hline \\ [-8pt]
IC214  & 124.6 &  HI [21cm] \\
NGC1140&  62.3&   HI [21cm] \\
NGC1569&  23.1  & HI [21cm]\\
NGC1614&  75.0   & CO 2.3$\mu$m\\
NGC1614 &  107.7 &CO(1$\to$2)\\
NGC4194&  41.7   &HI [21cm]\\
NGC4194& 104.0  &  CO 2.3$\mu$m\\
NGC4194&  73.2  & CO(1$\to$2)\\
NGC4385&  36.7  & HI [21cm]\\
NGC5236&  78.2  & HI [21cm]\\
NGC5253&  25.8  & HI [21cm]\\
NGC6052& 127.8  & HI [21cm]\\
NGC6090&  61.4  & HI [21cm]\\
NGC6090&  50.0  &  CO 2.3$\mu$m\\
NGC6090&  55.2  & CO(1$\to$2)\\
NGC6217&  61.6  & HI [21cm]\\
NGC7250&  52.0  & HI [21cm]\\
NGC7552&  62.9  & HI [21cm]\\
NGC7673&  51.1  & HI [21cm]\\
NGC7714&  65.0  & HI [21cm]\\
NGC7714&   106.6 & CO(1$\to$2)\\
\hline \\ [-8pt]
\multicolumn{3}{c}{ULIRGs} \\
\hline \\ [-8pt]
IRAS12112 & 146.4 &CO(1$\to$2)\\
Mrk273    & 71.8 &CO(1$\to$2)\\
Mrk273    &160.0 & CO 2.3$\mu$m\\
IRAS14348 &107.7 &CO(1$\to$2)\\
Arp220    &182.3 &CO(1$\to$2)\\
Arp220    &150.0 & CO 2.3$\mu$m\\
IRAS22491 & 85.6 &CO(1$\to$2)\\
\hline \\ [-8pt]
\end{tabular*}
\vspace{6pt}
\par\noindent
Note: $^a$ Emission lines (HI 21cm; CO(1$\to$2)) and
absorption line (CO 2.3$\mu$m) adopted to derive
the velocity dispersion. See the caption of Fig. \ref{velocity}
for references to the observations.
\\
\label{sig-table}
\end{table}

\begin{figure}
  \resizebox{\hsize}{!}{\includegraphics{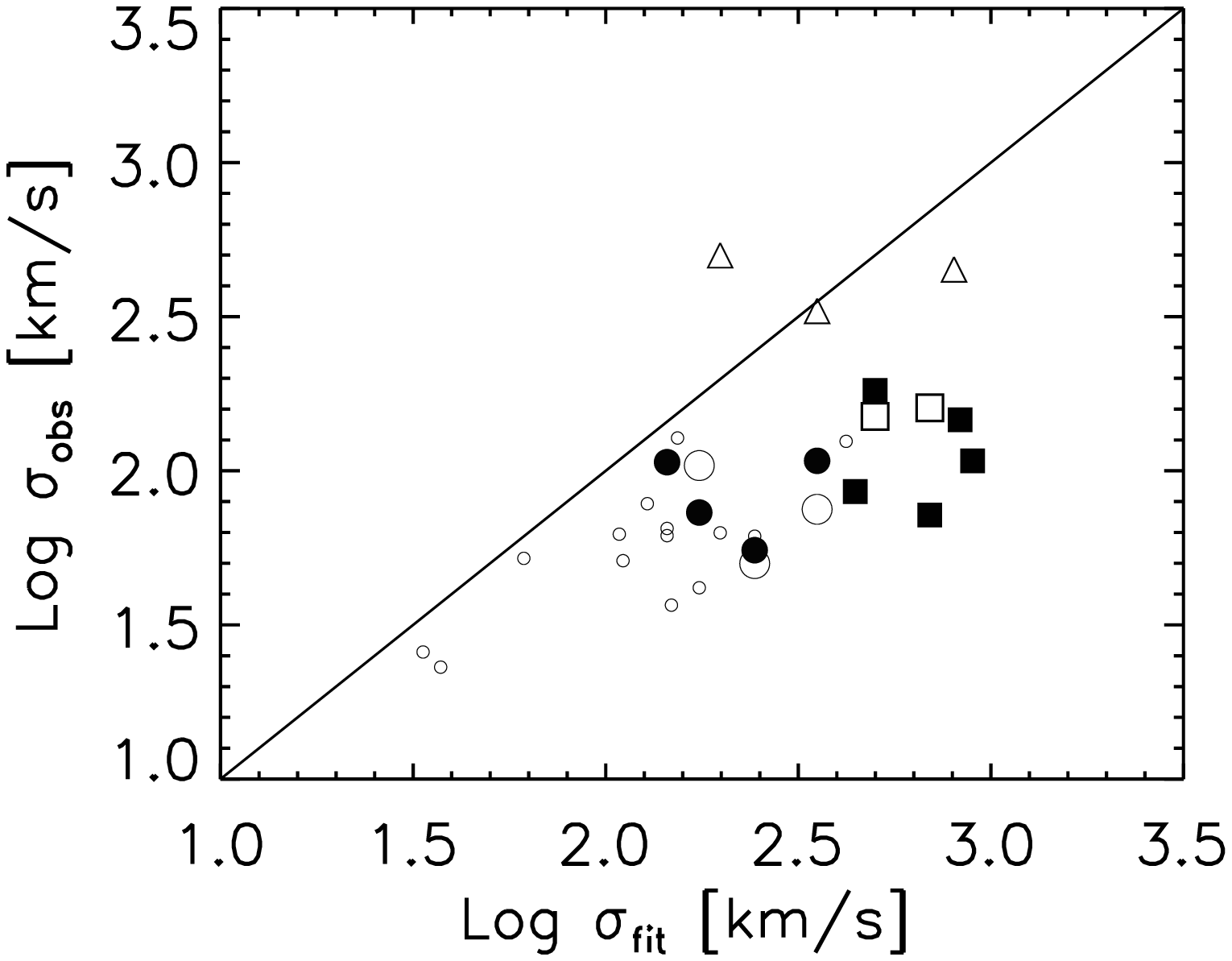}}
  \caption{Comparison of velocity dispersions in starburst regions. 
We estimate the velocity dispersions from the SED fitting results; 
i.e, $\sigma_{fit}=(GM(<r_e)/2r_e)^{1/2}$ where we assume 
$ M(<r_{e})= 2 M_{\ast}(<r_e)$.
Squares and circles indicate ULIRGs and UVSBGs, respectively.
For solid symbols, we adopt the velocity dispersions derived
from the line profile of the CO(1$\to$2) emission 
(Sanders, Scoville \& Soifer 1991),
while HI line width measurements are used for small open circles, which
are taken from LEDA database.
To convert both CO and HI line widths to the velocity dispersion,
we assume that the line profile is Gaussian; i.e., $\sigma$=FWHM/2.35=
W$_{20}$/3.62, where W$_{20}$ is the width at 20\% of the peak.
For large open circles and open squares, we adopt velocity dispersions
derived from the 2.3 $\mu$m CO absorption feature (James et al. 1999;
Shier \& Fischer 1998). For open triangles, we plot terminal velocities,
instead of velocity dispersions, which are 
obtained for Mrk273 (450 km $s^{-1}$), NGC1614 (330 km $s^{-1}$) and
NGC7552 (500 km $s^{-1}$) by Heckman et al. (2000).
}
  \label{velocity}
\end{figure}

\subsubsection{Self-gravity vs.\ feedback in ULIRGs} 
For a starburst with a typical duration of $t_0$,  
the kinetic energy of gas clouds with random velocity $v_g$ can be 
written as $ 1/2 M_g v_g^2 \simeq L_{kin} t_{0} $, where 
$L_{kin}$ is the kinetic luminosity due to feedback. 
Assuming that $ L_{kin} $ is 
proportional to the bolometric luminosity with a constant 
efficiency $f_{kin}$, i.e., $ L_{kin} = f_{kin} L_{bol}$, 
we can write the kinetic energy per a unit mass as  
\begin{eqnarray}
\frac{1}{2} v_{g}^2   & = & \frac{ L_{kin} t_{dyn} }{ M_{g} }
  = f_{kin} \frac{ L_{bol}}{\psi} 
\; .
\end{eqnarray}
By using the relation between 
$L_{bol}$ and SFR $\psi$ in Eq. (\ref{eq_bolsfr}),  
the typical velocity can be written as 
\begin{eqnarray}
v_g  = 454 \left( \frac{f_{kin}}{0.01} \right)^{1/2}
\left( \frac{t}{t_0} \right)^{1/2} \mbox{km s}^{-1} \; .
\label{vgas}
\end{eqnarray} 
Note that the effect of feedback is independent of the mass in  
the starburst regions and also in the host galaxies, 
even though a larger gas supply can induce a more active starburst. 
On the other hand, the gravitational effect 
becomes more prominent in more massive systems.  
This means that too massive starbursts cannot be 
sustained by feedback against strong self-gravity.   

This limiting effect of feedback against self-gravity 
can be estimated by the comparison of $v_g$ with the 
escape velocity $ v_{esc,c}$ 
($ v_{esc,c}^2 \simeq 2 G M(<r_{e}) r_{e}^{-1} $), 
which indicates the depth of the gravitational potential well. 
A critical mass-radius relation is expected for the condition, 
$ v_{esc,c} = v_g $. This relation
can be rewritten to another luminosity-radius relation
by using the mass to light ratio of starburst regions [Eq. (\ref{eq_masslum})].
We can then introduce a limiting bolometric luminosity
\begin{eqnarray} \label{eq_limit}
L_{bol,c} & \simeq &  1.2 \times 10^{2}
\left (\frac{M_{\ast}}{\mathrm{M}_\odot}\right )
 \left ( \frac{t}{t_0} \right)^{-1.8}
\mathrm{L}_{\odot}
\nonumber \\
& \simeq &  1.5 \times 10^{12}
\left( \frac{f_{kin}}{0.01} \right)
\left( \frac{M_{\ast}}{M_{\ast}(<r_e)} \right) \nonumber \\
& & \times \left(\frac{r_e}{1 \mbox{kpc}} \right)
\left( \frac{t}{t_0} \right)^{-0.8}
\mathrm{L}_{\odot}  \;  ,
\end{eqnarray}
for $t_0=100$ Myr,
where we use $ M(<r_{e})= 2 M_{\ast}(<r_e)$.
The total mass within $r_e$ is somewhat uncertain, 
since a realistic distribution of gas and the contribution 
from the dark matter is important. Therefore, we simply assume 
$M(<r_e)=2 M_* (<r_e)$. 
At a typical starburst stage $t/t_0 = 2$,
we take a limiting bolometric luminosity
$ L_{bol,c} \simeq 1.7 \times 10^{12} (r_e/1\mbox{kpc}) $ L$_{\odot}$.
The feedback effect cannot sustain the starburst region 
against self-gravity when the bolometric luminosity exceeds this limit. 
This limiting luminosity is indicated by the dashed line 
in Fig. \ref{lumrad}, which crosses the line of limiting 
luminosity of the starbursts induced by the disc instability 
around 0.3 kpc. 
Fig. 14 shows that all the ULIRGs are above the dashed line 
where the feedback effect cannot sustain the 
starburst region against the self-gravity, while all the
UVSBGs but one are below the dashed line.  
Therefore, 
the mode of starburst in ULIRGs can be different from that in UVSBGs 
as expected. 

The SED diagnostics for ULIRGs indicate
a strong mass concentration in the starburst regions 
as shown in Fig. 13 b). 
Furthermore, the analysis of self-gravity and feedback 
allow us to specify that starburst regions in ULIRGs are too massive to be 
sustained by the feedback effect. 
Such a system in ULIRGs should be dynamically unstable, which can be 
confirmed by the observed kinematical signature. 
With the mass and effective radius derived from the SED fitting, 
we can estimate the velocity dispersion as 
$\sigma_{fit}=(GM(<r_e)/2r_e)^{1/2}$, which is independent of 
the direct kinematical estimation from line observations.  
If a starburst region is  
dynamically stable as a virialized system, 
the observed velocity dispersion should be close to this value. 
In Fig. \ref{velocity}, we compare $\sigma_{fit}$ 
with observed velocity dispersions. 
The observed velocity dispersions are derived from the
emission lines of CO(1$\to$2) and HI at 21cm, and
the absorption line of CO at 2.3$\mu$m.
These velocity dispersions are tabulated in Table 3.
The velocity dispersions by 2.3$\mu$m CO absorption
are most suitable to probe the gravitational potential
in the starburst region, since the effect of bulk motion 
on the CO absorption feature is less important than that 
on gas emission lines. 
Comparing to the HI 21 cm lines, the emission line of CO molecules
seems to be more suitable for our purpose, since these molecules
are more concentrated in the starburst region in general.
However, we also rely on the measurements of HI 21cm line,
since the observation of this line can be found in the 
majority of our sample galaxies, and HI line widths
correlate well with the other kinematical tracers such as
H$\alpha$ and O[II] which are strong in the starburst region
(Kobulnicky \& Gebhardt 2000). 
Note that cool gas is kinematically decoupled with 
the outflowing gas which is indicated by triangles in Fig. \ref{velocity}. 

We can see in Fig. \ref{velocity} that 
the sequence of the velocity dispersion seems to be bimodal, again. 
The deviation of the observed velocity dispersion is most remarkable 
for ULIRGs. This means that, for ULIRGs, 
the whole starburst region is dynamically unstable since 
the observed velocity dispersion is not enough to 
support it against the gravity due to large masses 
derived from the SED diagnostics.  
Such a dynamically unstable state for ULIRGs 
cannot be caused by the local disc instability as 
in UVSBGs, since it requires  
large mass concentration into the whole starburst region.   
Therefore, the dynamical disturbance of a galactic scale is 
necessary to activate ULIRGs. The merging of galaxies 
can be the most plausible trigger mechanism of ULIRGs.

\section {Conclusions}

We develop an evolutionary SED model of starburst regions in which the
stellar and chemical evolution, including dust, are taken into account 
consistently.
We investigate the basic properties of nearby starburst galaxies, such as
the SFR, optical depth and apparent effective radius by using this
evolutionary SED model. We studied the SED of
various starburst galaxies with a SFR of 0.5 -- 300 
M$_\odot$ yr$^{-1}$, classified as UVSBGs or
ULIRGs. Our SED model essentially reproduces all the SEDs of analysed
starburst galaxies over the whole range of starburst activity.
For starburst galaxies having AGN, we find a clear excess of
MIR emission when compared with the best-fitting SED model.
The fitting results
are confirmed by comparing the derived SFRs and optical depths with the
emission line measurements. Irrespective of the degree
of reddening, our model can derive the SFR of starburst regions in a 
unified manner. The apparent effective radii derived from the SED 
fitting are found to be consistent with observations. 
Thus, our model can derive both the SFR and intrinsic effective radius 
from the SEDs. 

The variety of the starburst SEDs is caused
mainly by the compactness of starburst regions,
not from the starburst age.
From the SED diagnostics, we find 
bimodal correlations between the SFR and the compactness/optical
depth; more active starburst regions tend to be more compact and
heavily obscured.

The bimodal starbursts consist of mild state with
a limiting intrinsic surface brightness $ L_{bol} r_e^{-2} \simeq
10^{12}$ L$_{\odot} $kpc$^{-2}$ and intense state
with a characteristic scale of $ r_e \simeq 0.3 $ kpc. 
The mild starburst can be triggered by
the disc instability as proposed by Meurer et al. (1997). 
For ULIRGs, the surface brightness is an order of magnitude larger than 
that of UVSBGs. 
A simple analysis of the feedback allows us to derive a  
critical luminosity over which starburst regions cannot be 
sustained by the feedback against self-gravity. 
This limitation of the feedback causes the bimodality of starburst activity, 
in which the size of starburst region has 
a clear limit of $r_e \simeq 0.3 $ kpc. According to this analysis, 
all of the ULIRG sample is identified as self-gravity dominated starbursts. 
Furthermore, we compare the velocity dispersions inferred from 
the SED fitting results with observations to find that starburst 
regions in ULIRGs are dynamically unstable. 
In order to produce the dynamically unstable starbursts with 
strong mass concentration, a violent trigger mechanism is required, 
rather than the disc instability. 
As suggested by various imaging observations, 
the most plausible trigger mechanism of ULIRGs would be galaxy 
merging which can cause the dynamical disturbance over a galactic 
scale.

\section*{Acknowledgements}
TT would like to thank to N.\ Shibazaki, V.\ Vansevi\v cius, 
T.\ Matsumoto and M.\ Rowan-Robinson for their encouraging supports.
We would like to thank the anonymous referee 
for the sensible comments. 
We are grateful to C.\ Pearson for his careful reading of the draft. 
This work was financially supported in part by a
Grant-in-Aid for the Scientific Research (No.11640230, 13011201 $\&$ 14540220)
by the Japanese Ministry of
Education, Culture, Sports and Science.
This research has been supported in part by a Grant-in-Aid for the
Center-of-Excellence (COE) research.
Also, TT acknowledges the support of PPARC. 
We acknowledge the use of Lyon-Meudon Extragalactic Database
(LEDA). This research has made use of the NASA/IPAC Extragalactic Database
(NED) which is operated by the Jet Propulsion Laboratory,
California Institute of Technology, under contract with
the National Aeronautics and Space Administration.


\begin{thebibliography}{}
\bibitem[]{}
Adelberger K.L., Steidel C.C.\ 2000, ApJ, 544, 218
\bibitem[]{}
Alonso-Herrero A., Engelbracht C.W., Rieke M.J., Rieke G.H., Quillen A.C.\
  2001, ApJ, 547, 129
\bibitem[]{}
Arimoto N., Yoshii Y.\ 1987, A\&A, 173, 23
\bibitem[]{}
Arimoto N., Yoshii Y., Takahara F.\ 1992, A\&A, 253, 21
\bibitem[]{}
Armus L., Heckman T.M., Miley G.K.\ 1990, ApJ, 364. 471
\bibitem[]{}
Bryant P.M., Scoville N.Z.\ 1999, AJ, 117, 2632
\bibitem[]{}
Calzetti D., Bohlin R.C., Kinney A.L., Storchi-Bergmann T., Heckman T.M.\
1995, ApJ, 443, 136
\bibitem{}
Calzetti D., Kinney A.L., Storchi-Bergmann T.\ 1994, ApJ, 429, 582
\bibitem{}
Combes F., Boiss\' e P., Mazure A., Blanchard A.\ 1995, in Galaxies
and Cosmology (Springer-Verlag, Berlin Heidelberg) p. 96
\bibitem{}
Dikey J.M., Hanson M.M., Helou G.\ 1990, AJ, 99, 1071
\bibitem{}
Dwek E.\ 1998, ApJ, 501, 643
\bibitem{}
Fioc M., Rocca-Volmerange B.\ 1997, A\&A, 326, 950
\bibitem{}
Genzel R., Lutz D., Sturm E., Egami E., Kunze D., Moorwood A.F.M.,
Rigopoulou D., Spoon H.W.W.\ et al.\ 1998, A\&A, 498, 579
\bibitem{}
Goldader J.D., Meurer G., Heckman T.M., Seibert M., 
Sanders D.B., Calzetti D., Steidel C.C.\ 2002, ApJ, 568, 651
\bibitem{}
Gordon K.D., Calzetti D., Witt A.N.\ 1997, ApJ, 487, 625
\bibitem{}
Heckman T.M., Lehnert M.D., Stickland D.K., Armus L.\ 2000, ApJS, 129, 493
\bibitem{}
Heckman T.M., Robert C., Leitherer C., Garnett D.R., van der Rydt F.\ 1998,
ApJ, 505, 174
\bibitem[]{}
James P., Bate C., Wells M., Wright G., Doyon R.\ 1999, MNRAS, 309, 585
\bibitem[]{}
Kennicutt R.C.\ 1989, ApJ, 344, 685
\bibitem[]{}
Kennicutt R.C.\ 1998, ApJ, 498, 541
\bibitem[]{}
Kinney A.L., Bohlin R.C., Calzetti D., Panagia N., Wyse R.F.G.\ 1993, ApJS, 
86, 5
\bibitem[]{}
Kobulnicky H.A., Gebhardt K. 2000, AJ, 119, 1608
\bibitem[]{}
Kodama T., Arimoto N.\ 1997, A\&A, 320, 41
\bibitem[]{}
Lehnert M.D., Heckman T.M.\ 1996, ApJ, 472, 546
\bibitem{}
Leitherer C., Heckman T.M.\ 1995, ApJS, 96, 9
\bibitem{}
Lequeux J., Peimbert M., Rayo J.F., Serrano A., Torres-Peimbert S.\ 1979, 
A\&A, 80, 155
\bibitem[]{}
Lutz D., Spoon H.W.W., Rigopoulou D., Moorwood A.F.M., Genzel R.\ 1998, 
ApJ, 505, L103
\bibitem{}
McQuade K., Calzetti D., Kinney A.\ 1995, ApJS, 97, 331
\bibitem[1995]{meu95}
Meurer G.R., Heckman T.M., Leitherer C., Kinney A., Robert C.,
   Garnett D.R.\ 1995, AJ, 110, 2665
\bibitem[]{}
Meuerer G.R., Heckman T.M., Lehnert M.D., Leitherer C., Lowenthal J.\
1997, AJ, 114, 54
\bibitem[]{}
Meuerer G.R., Heckman T.M., Calzetti D.\ 1999, ApJ, 521, 64
\bibitem[]{}
Pagel B.E.J.\ 1997, in Nucleosynthesis and Chemical Evolution of Galaxies
(Cambridge University Press, Cambridge) p. 82
\bibitem[]{}
Pei Y.C.\ 1992, ApJ, 395, 130
\bibitem[]{}
Prugniel Ph., Heraudeau Ph.\ 1998, A\&AS, 128, 299
\bibitem[]{}
Quirk W.J.\ 1972, ApJ, 176, L9
\bibitem{}
Rigopoulou D., Lawrence A., Rowan-Robinson M.\ 1996, MNRAS, 278, 1049
\bibitem{}
Rigopoulou D., Spoon H.W.W., Genzel R., Lutz D., Moorwood A.F.M.,
  Tran Q.D.\ 1999, AJ, 118, 2625
\bibitem{}
Rowan-Robinson M., Efstathiou A.\ 1993, MNRAS, 263, 675
\bibitem{}
Sanders D.B., Scoville N.Z., Soifer B.T.\ 1991, ApJ, 370, 158
\bibitem{}
Sanders D.B., Soifer B.T., Elias J.H., Madore B.F., Matthews K.,
Neugebauer G., Scoville N.Z.\ 1988, ApJ, 325, 74
\bibitem[]{}
Scoville N.Z., Evans A.S., Thompson R., Reike M., Hines D.C.,
Low F.J., Dinshaw N., Surace J.A.\ et al.\ 2000, AJ, 119, 991
\bibitem[]{}
Shier L.M., Fischer J.\ 1998, ApJ, 497, 163
\bibitem[]{}
Soifer B.T., Neugebauer G., Matthews K., Egami E., Becklin E.E., Weinberger 
A.J.,
Ressler M., Werner M.W. et al.\ 2000, AJ, 119, 509
\bibitem[]{}
Storchi-Bergmann T., Calzetti D., Kinney A.L.\ 1994, ApJ, 429, 572
\bibitem[]{}
Sullivan M., Treyer M.A., Ellis R.S., Bridges T.J., Millard B., Donas J.\ 
2000,
MNRAS, 312, 442
\bibitem[]{}
Surace J.A., Sanders D.B.\ 2000, AJ, 120, 604
\bibitem[]{}
Takagi T.\ 2001, PhD thesis, Rikkyo Univ. 
\bibitem[]{}
Takagi T., Arimoto N., Vansevi\v cius V.\ 1999, ApJ, 523, 107
\bibitem[]{}
Toomre A.\ 1964, ApJ, 139, 1217
\bibitem[]{}
Trentham N., Kormendy J., Sanders D.B.\ 1999, AJ, 117, 2152
\bibitem{}
Veilleux S., Kim D.-C., Sanders D.B.\ 1999a, ApJ, 522, 113
\bibitem{}
Veilleux S., Sanders D.B., Kim D.-C.\ 1999b, ApJ, 522, 139
\bibitem{}
Witt A.N., Gordon K.D.\ 2000, ApJ, 528, 799
\bibitem[]{}
Wynn-Williams C.G., Becklin E.E.\ 1993, ApJ, 412, 535


\end{thebibliography}
\end{document}